\documentclass[preprints,article,submit,pdftex,moreauthors]{Definitions/mdpi} 

\firstpage{1} 
\pubvolume{1}
\issuenum{1}
\articlenumber{0}
\pubyear{2026}
\copyrightyear{2026}
\datereceived{ } 
\daterevised{ } 
\dateaccepted{ } 
\datepublished{ } 
\hreflink{https://doi.org/} 

\usepackage{longtable}
\usepackage{rotating}
\usepackage{microtype}
\usepackage{comment}

\providecommand{\UseTaggingSocket}[1]{}

\newcommand{\req}[1]{Eq.\,(\ref{#1})}

\newcommand{\rf}[1]{Fig.~{\ref{#1}}}
\newcommand{\rt}[1]{Table~{\ref{#1}}}
\newcommand{\rsec}[1]{Sect.\,{\ref{#1}}}

\pdfoutput=1 
\renewcommand{\linenumbers}{}
\Title{Science of Nuclear Fusion:\texorpdfstring{\\ \Large{Insights and Ideas}}{ Insights and Ideas}}

\TitleCitation{Science of Nuclear Fusion}

\Author{Johann Rafelski ${}^{1}$\orcidA{} and Andrew J. Steinmetz ${}^{2,1}$\orcidB{}}

\AuthorNames{Johann Rafelski and Andrew J. Steinmetz}

\isAPAStyle{%
\AuthorCitation{Rafelski, J. \& Steinmetz, A. J.}
}{%
\isChicagoStyle{%
\AuthorCitation{Rafelski, Johann and Andrew Steinmetz.}
}{
\AuthorCitation{Rafelski, J.; Steinmetz, A. J.}
}
}
\address{%
${}^{1}$ Department of Physics, The University of Arizona, Tucson, AZ 85721, USA\\
${}^{2}$ School of Physics, Georgia Institute of Technology, Atlanta, GA 30332, USA}

\abstract{Advances in several physics domains open up novel paths to smaller scale, higher energy density opportunities to advance small systems for nuclear fusion. Here we survey both legacy and several novel ``table-top'' approaches which attract current interest. We furthermore address a few related practical and challenging nuclear science topics arising in the context of magnetic confinement and inertial confinement fusion. The contents emphasis includes: By example of solar fusion cycles we draw attention to aneutronic fusion reaction chains. Considering the natural isotopic abundances we assess more carefully the meaning of the term ``limitless energy'' in the context of actual fusion power realizations. We describe achievements in laser-driven proton-boron fusion, and extensions to a self-sustaining and nearly fully aneutronic proton-boron-nitride reaction cycle. We propose another aneutronic option, where the target is a mix of beryllium and light helium isotope; this 3-helium is arguably the most mentioned fusion component in this article. We look in depth at the plasmonic opto-electric field-enhancement for fusion, and at the particle (muon) catalyzed fusion option. We describe problems in harnessing the \emph{dt} fusion for civilian use. We introduce space travel as forthcoming application of aneutronic fusion.
}
\keyword{nuclear fusion; muon catalyzed fusion; laser fusion; inertial confinement; magnetic confinement; relativistic optics; nano-antennas}

\begin{document}
\tableofcontents
\section{Overview}\label{sec:overview}
In this manuscript we introduce scientific aspects of ongoing nuclear fusion for energy research with emphasis on the modern approaches made possible by recent technological advances. We present many novel results embedded into a presentation that makes an effort to also review related prior art. In \rsec{sec:overview} we describe, avoiding use of prerequisite knowledge, the scientific challenges of realization and implementation and addressing a few unfulfilled promises. We explain the reason to search for aneutronic reaction cycles that avoid the primary hazard of fusion energy: the production of energetic neutrons in the technologically most easily achieved approaches. More specific scientific arguments are confined to: \rsec{sec:Fusion}, where we introduce select methods, and \rsec{sec:newFusion}, where we address novel nuclear fusion for energy approaches. These results are made accessible to the general science-literate public in the summary, \rsec{sec:Dis}.
\subsection{Nuclear fusion}\label{ssec:fusion}


\subsubsection{Basic principles}\label{sssec:basics}
The term `nuclear fusion' in its most general sense applies to all light-element nuclear transmutation reactions. In many reactions of interest, the energy released traces back to the exceptionally tight binding of four nucleons in the $\alpha$-particle, a bound state of two protons and two neutrons, $\alpha=ppnn$. The $\alpha$-particle is the nucleus of the helium atom ${}^{4\!}_{2\!}\mathrm{He}^{++}$, first recognized remotely in the solar spectrum and named after the Greek god Helios, who personifies the Sun, our nearest natural fusion reactor and life-sustaining energy source. Other light elements and their isotopes are also relevant to nuclear fusion, both on Earth and in stars.

The redistribution of nucleons in all fusion reactions occurring between two electrically charged nuclei is hindered by their mutual Coulomb repulsion, which in general requires quantum tunneling through this so-called Coulomb barrier or wall. The tunneling probability increases with collision energy, while higher density raises the collision frequency. Collision energy may be increased either by raising the temperature of a thermal plasma or by arranging a non-equilibrium, collective flow of fusible material.

The tunneling probability is governed by relative speed, which at given energy is greatly inhibited by increasing masses of reactants; the root of the reduced mass is therefore found in the tunneling exponent. The product of colliding charges determines the height of the repulsive wall. To amplify the chance of fusion we preferentially seek reactions in which the Coulomb wall is low and/or narrow. In order for a near steady-state thermal burn to be self-sustaining, the energy released by nuclear fusion must exceed the radiative and outflow losses. The often-cited `Lawson criterion' encapsulates this requirement for quasi-static plasma confinement; it has been the primary figure of merit for magnetic-confinement devices, which have remained `20 years away' from a burn condition break even for more than half a century. Our scientific discussion reaches beyond the Lawson triple product of reactant density $n$, confinement time $\tau$, and plasma temperature $T$, which establishes an engineering boundary for one specific usually considered set of reactants. In the present work we will pay attention to gain and loss terms explicitly and quantitatively, without relying on order-of-magnitude criteria, allowing direct comparison among diverse nuclear fusion conceptual solutions.

\paragraph*{\bf Plasma engineering versus nuclear science}
Impressive engineering effort has been devoted to achieve magnetically confined fusible plasma in clever magnetic-field configurations. Such systems are relatively large, as otherwise we could not keep the low density plasma magnetically confined. The technology of magnetic fields and plasma confinement is in policy documents in a misguided manner declared to be `nuclear fusion research'. However, such research is entirely unrelated to nuclear fusion research, including inertial confinement nuclear fusion realization, which is implemented in very small, very high density explosive systems. We show in this work that dilute thermal burn plasma magnetic confinement fusion devices have no future in civilian energy production, as they will be always associated with very costly, if not impossible to control, ultra-fast neutron flux produced in a large plasma volume. General study and the development of conceptual physics principles concerning nuclear fusion reactions, which topics are addressed in this work, differs in principle and in actual context from such dilute and large plasma fusion systems seen on the fringe of the present day very active nuclear fusion research. 

One must clearly see the difference in context: Science vs Engineering. New laser technologies allow development of intense lasers driving several new science paths to nuclear fusion. It is nuclear science with particles colliding and undergoing energy producing reactions, not magnetic and plasma engineering, or the development of powerful lasers, that governs the nuclear science questions: Which nuclear reactions are available in what situation, how much energy they release and in what form, and the rate of neutron production among natural and practically always occurring multi step fusion reactions. Some of the nuclear fusion science, in this broader context, has long been a foundational topic in nuclear astrophysics~\cite{Salas2026Neelima,Nunes:2009,Iliadis:2007,Arnett:1996ev,Rolfs:1988,Clayton:1984}. These insights can be used to advance understanding of science of nuclear fusion for energy production.

\paragraph*{\bf Scope of this presentation}
Therefore we focus this presentation on the nuclear science questions in the context of nuclear fusion for energy production: We treat nuclear fusion reactions quantitatively. Our scientific discussion reaches beyond the engineering boundary for one specific usually considered set of reactants. We furthermore look at nuclear fusion energy deposited only by charged nuclear reaction products in the fusion reaction domain, given that neutrons carry their energy to secondary locations, while the fate of (thermal) X-rays depends on wall enclosure. We study the range of high energy charged particle products to understand better the fraction of energy they deposit in reaction volume. This allows us to compare realistically the heating power to the radiative X-ray power volume loss. The outcome depends sensitively on the nuclear reaction channel, the nuclear plasma composition, density and travel path length, and whether the system operates in thermal or non-thermal conditions associated with directed particle beams.

 \subsubsection{Hydrogen and helium}\label{sssec:HHe}
Several isotopes of hydrogen and helium play central roles in nuclear fusion schemes. The lightest hydrogen isotope is the proton, ${}^{1\!}_{1\!}\mathrm{H}^+\equiv p$. When a neutron $n$ binds to a proton, the stable heavy-hydrogen isotope ${}^{2\!}_{1\!}\mathrm{H}^+\equiv d=pn$ called deuterium is formed. Atomic deuterium D has chemical properties very similar to ordinary hydrogen H; however, even in simple molecules the factor-of-two mass difference between H and D produces measurably different chemistry and vibrational frequencies.

Deuterium ($d=pn$) is a naturally abundant heavy-hydrogen isotope, available in literally unlimited abundance in all water. Its near-universal presence reflects incomplete burn during the Big Bang Nucleosynthesis (BBN) epoch in the expanding Universe~\cite{Rafelski:2023emw,Rafelski:2024fej}, which left a primordial ratio $\mathrm{D}/\mathrm{H}=2.53\times10^{-5}$~\cite{Cooke:2017cwo}, or about one $d$ per $40{,}000$ protons. However, ocean water on Earth is enriched by a factor 6.16 compared to this reference; the measured value is $\mathrm{D}/\mathrm{H}=1.5576\times10^{-4}$~\cite{Meija:2016}, or one $d$ per $6420$ protons. Most light isotopes found on Earth today are also nuclear `ashes' of past cosmological and stellar evolution. Some light elements are, however, not well understood in current nuclear fusion astrophysical chains; they are explained as being produced by cosmic radiation. We will not address here this `origin' question subject to ongoing research using the cosmic flux measurements by the Alpha Magnetic Spectrometer Collaboration with the instrument sited on the International Space Station, see for example~\cite{AMS:2025xoq} for Lithium isotope abundance update.

\paragraph*{\bf Tritium and helium-3}
In the technological race to realize first-generation fusion reactors, the most important hydrogen isotope is the yet heavier tritium T, with nucleus $t=pnn$. While $d$ is naturally stable, $t$ has a half-life of $12.32$~yr (4500 days) and is therefore practically absent as a natural isotope in Earth's crust. We note that $0.46\%$ of all tritium inventory decays every 30 days, transmuting into the light helium isotope ${}^3_2\mathrm{He}^{++}=ppn$ with two protons and one neutron. Consequently, all tritium used in $dt$-fusion must be manufactured, and the slow natural decay of $t$ will, if first generation fusion reactors are implemented, gradually enrich the available terrestrial ${}^3_2$He inventory. Since mixing of material at the Moon surface is not present for billion(s) of years, a relatively large abundance of ${}^3_2$He has been identified in a relatively narrow Moon surface shell. Mining of Moon surface inventory accumulated over billions of years from solar wind is planned today.

\paragraph*{\bf Principal deuterium reactions and energy units}
The importance of ${}^3_2$He is easily recognized considering the principal deuterium-induced fusion reactions 
\begin{align}
&d+d&\to& \quad \label{eq:ddQp}
t (1.011\,\mathrm{MeV}) &+& p (3.022\,\mathrm{MeV})\quad &\mathrm{Q}=&4.0327\,\mathrm{MeV}\\
&&\to& \quad {}^3\mathrm{He}^{++} (0.817\,\mathrm{MeV})&+& n 	(2.452\,\mathrm{MeV})\quad &\mathrm{Q}=&3.2689\,\mathrm{MeV}\label{eq:ddQn}\\
&d+t&\to& \quad \label{eq:dtQ}
\alpha (3.5601\,\mathrm{MeV}) &+& n 	(14.0292\,\mathrm{MeV})\quad &\mathrm{Q}=&17.5893\,\mathrm{MeV}\\
&d+{}^3\mathrm{He}^{++} &\to & \quad \alpha (3.673\,\mathrm{MeV}) &+& p (14.680\,\mathrm{MeV}) \quad &\mathrm{Q}=&18.353\,\mathrm{MeV}\label{eq:dHeQ}
\end{align}
We see in above short list that only ${}^3_2$He leads to charged fusion products: In parentheses we see how the fusion energy yield $Q$ is distributed among reaction products respecting energy-momentum conservation (allowing for the relativistically correct form). When we discuss fusion we use `MeV' as unit of energy, where `M' stands for mega, a million, and `eV' is the energy gained by an elementary charge $e$ falling through a potential of one Volt. The equivalent in the SI unit Joule `J' follows from the value of the elementary charge in Coulomb, $e=1.602\times 10^{-19}\,\mathrm{C}$, which means eV$=1.602\times 10^{-19}\,\mathrm{[VC=VAs=Ws=J]}$. Chemical processes release in each reaction step a few eV, while nuclear processes release a million times more; hence MeV, and its thousandth part `keV' (`k' for kilo), are the natural units of fusion.

\paragraph*{\bf Charged products and multi-alpha nuclei}
Charged particle fusion products such as ${}^{4\!}$He${}^{++} =ppnn=\alpha$ have a short `paper thin' range in matter; their energy is locally shared and heats the nuclear fusion domain. Moreover, the produced He-atom with $\alpha$-particle nucleus can be safely vented or stocked for industrial use, adding to existing natural abundance. This is also the case for sometimes produced bound states of three-$\alpha$'s (carbon nucleus) and yet more rarely produced even heavier multi-$\alpha$ bound states (four-$\alpha$s is the oxygen nucleus, five-$\alpha$s is the neon nucleus).

The bound state of two-$\alpha$'s is the ${}^8$Be-nucleus. This beryllium isotope is unstable; the electric charge Coulomb repulsion overwhelms by a small amount the short range nuclear attraction acting between just two-$\alpha$'s. However, adding one or two neutrons results in the beryllium isotopes ${}^9$Be and ${}^{10}$Be, of interest for fusion, which we discuss in \rsec{ssec:Beryl}. Here we note that ${}^{10}$Be has a natural half-life of 1.387 million years and is extremely scarce, but relatively easily produced in nuclear processes in small quantities.

\subsubsection{Bottling weapon-grade fuel burn}\label{sssec:bottleHBomb}
It is widely assumed that the first accessible generation of nuclear fusion reactors will burn deuterium-tritium ($dt$) fuel, producing an $\alpha$-particle and a neutron via $d+t\to \alpha +n$, \req{eq:dtQ}. In a beam-target fusion experiment, the probability that an incoming particle will induce a given reaction is quantified by its reaction active surface called cross-section $\sigma(E)$. Dimensionally, $\sigma$ has units of area; in nuclear physics one almost always uses the barn
\[
1\ \mathrm{barn} = 10^{-28}\ \mathrm{m}^2,
\]
so that typical fusion cross-sections lie in the range $10^{-3}-10^3\ \mathrm{mb}$ (millibarns, $10^{-3}$ b).

The $dt$ reaction has the largest among all two-body fusion process reaction cross-section, which is about 5 barn. Furthermore this value is found at an accessible reaction energy near and below 65 keV in the CM reference frame (center of momentum, also called center of mass). For comparison, in the conjugate reaction (one neutron in tritium replaced by a proton) $d\,{}^{3}\mathrm{He}$ we find a 6 times smaller reaction cross-section requiring a four times greater CM reaction energy.

\paragraph*{\bf Why 5 barn is a large cross-section}
Before moving forward, let us establish why reaction cross-section, a surface valued at 5 barn ($5\times 10^{-24}\,\mathrm{cm}^2$), is a `large' value. A proton has a classical radial size of $10^{-15}\,\mathrm{m}\equiv 1\mathrm{fm}$. This corresponds to geometric reaction size 0.03 b = 30 mb, the normal scale of strong interaction. Notwithstanding the Coulomb barrier to tunnel through, the `surface size' of reacting $dt$ is 200 times bigger. In the classical billiard ball picture, this implies that colliding balls exchange constituents even though they are not touching; classical balls are several classical reaction sizes apart. 

This demonstrates that the reaction cross-section is not in the realm of classical billiard ball collisions; instead it is a resonant quantum phenomenon originating in quantum wave resonant behavior. This combines with tunneling, which reduces by a factor 1000 the very large strength of this resonance. All of the above applies also to $d\,{}^{3\!}\mathrm{He}$; however, the Coulomb wall is twice as high. Even at the 4 times higher collision energy the quantum tunneling turns out to be 20 times less effective. By combining all factors we arrive at the 5-6 times smaller $d\,{}^{3}\mathrm{He}$ reaction cross-section.

\paragraph*{\bf Energy partition and tritium consumption}
In the $dt$ fusion, the $\alpha$-particle is carrying 20\% of the fusion yield. This energy is deposited locally and helps sustain the fusion plasma. When evaluating the energy balance in a burning plasma, only these 20\% contribute to self-heating; the neutron carries away $\simeq 80\%$ of the $Q=17.59$\,MeV released and escapes the relevant reaction zone. A secondary recovery of this energy is required associated with the production of the fusible tritium. Counting the energy of the neutron, a $1$~GWe (e for electrical power delivery) thermal reactor consumes $\simeq\ 160$~kg of tritium per year, which has to be produced atom by atom, and matched with $\simeq\ 106$~kg of deuterium extracted from sea water. 

A single nuclear fusion reactor consumes many times the annual tritium inventory, which is disappearing with a half-life of 4500 days. Put differently, just to sustain the $dt$-fusion reactor we need a second fusion reactor where for each produced neutron we fabricate at least 1.05 tritons, allowing for 5\% decay and collection loss. Scientifically this is possible. However, all those promising $dt$ fusion must solve the second reactor problem: $n+X\to 1.05 t+Y$ while confining the ultrafast fusion neutrons. Good Luck! Clearly all this requires a very novel idea to allow implementation.

We conclude: rather than the commonly addressed challenge of achieving $dt$-burn, in our opinion the central unsolved challenge of the $dt$-based nuclear fusion reactor energy economy is harnessing the $dt$-reaction energy from the ultra-fast 14 MeV neutrons and simultaneously breeding the replacement tritium fuel. Arguably, solution of the problem of how to harvest the neutron energy while using these penetrating particles to breed replacement tritium is more challenging than scaling up to commercial viability the existing fusion demonstrators.

This matter is further complicated by the fact that $dt$-fusion and thermonuclear weapons share the same fundamental physics and much of the same infrastructure, raising serious fusion weapons proliferation concerns. These considerations motivate the search for alternative fusion schemes, preferably those with minimal or zero neutron production, which insight motivates much of this work.

\paragraph*{\bf Breeding the replacement tritium}
Returning to clarify the nuclear science of $dt$-based nuclear fusion: The neutron-handling solution must harvest neutron energy and simultaneously re-create the burned tritium, ideally in situ, through a second nuclear reaction such as
\begin{equation}\label{eq:Lid}
1)\quad d+t\to \alpha +n +17.58\,\mathrm{MeV}, \qquad\to \qquad 2)\quad n+^{6\!}{\mathrm Li}^{3+}\to t +\alpha + 4.78\,\mathrm{MeV}\,.
\end{equation}
The combined yield of this two step process is $Q_2=22.36$~MeV, and the net reaction burns one $d$ and one ${}^6$Li into two $\alpha$-particles, with neutron and triton as transient intermediaries. Suitable neutron-multiplying ($n+X\to n+n+\tilde X$) or tritium-producing and neutron-preserving ($n+Y\to n+t+\tilde Y$) reactions can compensate losses and even overproduce tritium to start additional reactors. 

Finding the fuel is not at issue: Clearly there is enough abundance of deuterium in sea water, and lithium supply is in plain sight in electrical (car) batteries. In fact we will not even need to separate the natural mix of Lithium isotopes: The high energy neutrons produce in an endothermic process also a tritium with the extra neutron attached to $^{7\!}{\mathrm Li}^{3+}$ (compared to $^{6\!}{\mathrm Li}^{3+}$) released and able to undergo the cycle \req{eq:Lid}.

All this is straightforward to write down; the engineering realization is formidable. The key challenges include:
\begin{enumerate}
\item Reaction cross sections for tritium production by ultra-fast 14\,MeV neutrons on lithium are relatively small and yet important as these are the processes that will overproduce tritium. For final breeding step, neutrons need to be moderated to near 1~MeV; thus fusion produced 14~MeV neutrons must donate $\simeq 90\%$ of their energy to the energy harvesting system before they are useful for tritium breeding.
\item Neutron emission is isotropic, demanding that the breeding and harvesting infrastructure enclose the $dt$ burn reaction volume on all sides practically completely.
\item Fast neutrons penetrate at relevant level many meters of concrete, requiring that the greatly enlarged fusion reactor volume domain be developed. This specifically for magnetic confinement fusion implies the need for a large in size tritium production and collection. The tritium breeding and harvesting volume is larger and at best comparable in size to the neutron energy-harvesting system.
\item Fast neutrons ($E_n \gtrsim 1$~MeV) drive fast-fission reactions in heavy structural materials, producing long-lived radioactive isotopes, the contamination problem that nuclear fusion was supposed to avoid.
\end{enumerate}

\paragraph*{\bf Tritium as a safeguarded material}
An additional strategic difficulty in running a $dt$ reactor economy is that tritium is recognized as an essential component in nuclear weapons programs; therefore, there have been proposals over the years to tighten safeguards and controls over its production. Despite that ``tritium is not even mentioned in any safeguards agreement''~\cite{von1990modern} or the IAEA statute, the proliferation of reactors able to produce tritium on the order of kilograms would certainly reignite such regulatory calls~\cite{dalton2017toward,abdou2021physics}. And while no overarching framework exists, laboratories licensed to store tritium (regulated via national and sub-national agencies) only stockpile tritium on the order of $10\sim100$ grams~\cite{iaea2024}. To list a few examples: Tritium Process Laboratory in Japan (60 g), Tritium Laboratory Karlsruhe in Germany (40 g), and JET experimental tokamak in the UK (90 g). Laser-based fusion facilities store comparably less, e.g., the Laboratory for Laser Energetics in Rochester, NY is licensed to store at most $\sim1.5$ grams~\cite{lle_cryogenic_tritium}.

A 1~GWe (e for usable electric power) $dt$ plant requires $\sim160$ kg/yr, far beyond the amounts in continuous circulation under current regulations. The political and treaty infrastructure for overseeing such inventories at the global scale required for a meaningful $dt$ fusion energy economy does not exist and would require new international agreements of considerable complexity.

These challenges have not been addressed with adequate rigor by many advocates of $dt$-based nuclear fusion. The brute-force approach of drawing lithium walls around a plasma volume fails on multiple counts:
\begin{itemize}
\item Science: Breeding tritium on ${}^7\mathrm{Li}$ occurs at the upper end of the produced neutron energy while once 90\% of neutron energy has been harvested the final one-for-one breeding on ${}^6\mathrm{Li}$ occurs;
\item Technology: The step in-between the two tritium breeding phases: Reactor designers have to show how they safely and efficiently harvest the dominant fraction (70\%) of reactor produced energy, that is 90\% of neutron carried energy before reaching the final breeding-relevant neutron energy range; 
\item Economical viability: Large 15-50\,meter-scale (diameter), meter-thick lithium structures are economically unrealistic and greatly encumbered by the need to remove 70\% of produced energy. These still new to develop structures are growing in size with the size of low plasma density magnetic confinement devices. 
\end{itemize}
Arguably the gigantic support structures that the total low plasma density large fusion system requires assure that a $dt$ fusion reactor is with near certainty more `dirty' than any fission nuclear reactor.

\paragraph*{\bf Is pure deuterium cleaner?}
Among the answers one sometimes hears is that once $dt$ burn is achieved, the reactor will be upgraded to burn pure deuterium at higher temperatures, implying here presumably that $dd$ fusion is `mostly aneutronic'. We discuss quantitatively in \rsec{ssubsec:SecondHeating} why this argument misleads. Consider \req{eq:ddQp}: the deuterium plasma continuously produces tritium as a $dd$ reaction product; this tritium burns much faster than deuterium, so a hot $dd$ reactor effectively runs a $dt$ reactor with in-situ tritium production. For every two $dd$ events, the secondary tritium burn yields on average one 14~MeV neutron, alongside the one 2.45~MeV neutron from the $n+{}^3\mathrm{He}$ branch, so the neutron problem is not eliminated by switching to pure deuterium fuel.

However, the temperatures needed to run $dd$ reactor which creates tritium internally are far and away. On the other hand, some plasma $dt$ nuclear fusion is indeed just around the engineering corner. This must be viewed from the perspective that such a fusion system is harnessing the hydrogen bomb along with the radiation complications which must be solved upfront, and geared to a specific departure point: for a relatively large scale magnetic demonstration reactor ITER (International Thermonuclear Experimental Reactor) the problem of harnessing neutron energy and breeding tritium is clearly different from the same challenge originated in laser driven inertial confinement fusion system where the scale of the source of $dt$ fusion neutrons can be up to a billion times smaller when measured by the volume source of fusion neutrons.

\paragraph*{\bf Inertial confinement: a smaller neutron source}
Inertial confinement (IC) means there is no time for an ignited fusion pellet to fall apart before it is sufficiently burned up. A functional IC reactor would be a small, nano-hydrogen bomb which needs to be fired by lasers at a  rapid rate, more than once a second and preferably at a 0.5 kHz rate, releasing  50 -- 5000 \,MJ in energy per shot, accompanied by a burst of up to $  10^{21}$ ultra-fast 14.1 MeV neutrons.   Laser technology capable to deliver one or more shots per second with and beyond MJ energy content remains elusive today. Speaking more generally, the technological issue is that a small but noticeable fraction of produced reactor power needs to be converted to continuously delivered laser power. In our opinion it is possible to enclose this burst source with a 5-20 meter radius reactor vessel, harnessing energy and breeding the required tritium.

The point of IC laser driven fusion is that the complete reactor fusion core is very much smaller than all current magnetic confinement primary vessels, which then need to be surrounded by a secondary system with a volume nearly 10 times larger than the primary large fusion core, making neither economical sense nor being radiation reduced compared to fission reactors. In our opinion only the inertial confinement $dt$ fusion nuclear reactor with its very small, point like source of neutrons could be economically viable considering the need for neutron handling and tritium breeding. However, required laser technology and pellet cost remain a formidable challenge needing to be overcome.

Given the issues with $dt$-based nuclear fusion one can only wonder why the reader has heard so little about this. In most popular literature the fusion reactor in the Sun is introduced to advance $dt$-based nuclear fusion energy economy. The fallacy of this argument is absence of any relation between these two nuclear fusion paths. To understand this let us next discuss the way our solar reactor works.

 \subsubsection{Solar aneutronic power}\label{sssub:sunN}
The natural fusion reactor we all observe daily is our Sun. The solar reactor works very differently from any $dt$ fusion reactor: It is gravity which confines the thermal reacting solar core and radiation is screened by 1000s of kilometers of dense non-reacting plasma; the question then is: Could there be $dt$ fusion inside the Sun? Maybe at the creation of the Solar system there was some amount of the required ${}^6\mathrm{Li}$, ${}^7\mathrm{Li}$ allowing to breed the tritium inside the core. However, given its totally negligible abundance in the Universe, the small amount of ${}^6\mathrm{Li}$, if any, which was present initially would have been burned long ago. 

More to the point, consider our Sun, made of protons and $\alpha$ particles (24\% by weight), and since it is a 2-3 generation star,  a noticeable abundance (by number)~\cite{Asplund:2021Sun} of oxygen (up to 0.036\%), carbon (up to 0.022\%), still less neon, iron.  We note that all nuclear energy producing fusion reactions are those which do not produce neutrons. Nuclear fusion cycles with much reduced if any neutrons in primary and secondary steps are called aneutronic. It is clear that truly aneutronic is different from near aneutronic: A nuclear fusion reactor working entirely on a chain of truly aneutronic nuclear reactions is the dream that would realize the buzz-term `limitless safe and clean energy'. This is indeed our Sun with a projected remaining lifespan of nearly 5-6 billion years, and a total lifespan of about 10 billion years.

Is an Earthbound fully aneutronic nuclear fusion reactor possible? Addressing, let alone solving this question, is what motivates this work. We note that truly aneutronic fusion could lead to small reactor implementations, suitable for use in transportation solutions, allowing the development cost to be distributed in time and allowing multiple solutions to arise. Think in this context about space travel where aneutronic fusion could be introduced at first just to boost the thrust, first at percent level and growing in time. To see how this works, look at PC computer development and compare with your smartphone, which also has evolved rapidly. Similarly, we foresee space drive fusion reactor turning many design corners and outgrowing the space applications, and becoming source of terrestrial energy.

\paragraph*{\bf Aneutronic burn in the stars}
Can this dream of truly aneutronic fusion energy realized in our terrestrial environment come true? Let us be encouraged by all the visible stars in the sky. In general all visible stars consist of gravitationally confined hydrogen plasma, comprising also the fusion-produced, and primordial, helium, and trace abundances of some other light elements (called `metallicity') typically created in prior stellar generation(s). 

Two known fusion cycles are powering all visible stars, the proton-proton ($pp$) chains, with two versions called `ppe' and `pep', and the catalytic carbon-nitrogen-oxygen (CNO) fusion cycle. These cycles are accompanied by side cycles; quantum mechanics assures that nothing happens at 100\% level. Amazingly, all stellar main and secondary nuclear fusion cycles the authors know about are fully aneutronic! In general these well studied aneutronic paths to fusion energy are present in each star. Which is relevant and/or dominant depends on the mass of the star, and on the amount of (trace) abundance of carbon. We will discuss in \rsec{ssec:SolarFusion} in more detail the nuclear fusion powering the stars.

Insight about purely aneutronic nuclear fusion in stars and our Sun has two vastly different possible follow-up arguments.
\begin{enumerate}
\item On one hand, we can imagine that the solar nuclear fusion environment demonstrates the principle, and thus considering that all natural fusion reactors are aneutronic, we should expect many other aneutronic nuclear fusion fuel cycles awaiting discovery as it is a surprise and near impossible to imagine that just stellar environments would be aneutronic.
\item On the other hand, if indeed the solar nuclear aneutronic power is a unique situation, we can advance the Anthropic Principle, \emph{i.e.}, we must exist to observe. By implication, the laws of nature are fine tuned to avoid adverse conditions allowing intelligent life to exist. In this vein, if the absence of neutron production and emission is a special and unique property of all visible stellar fusion reactors, and our Sun in particular, this assures limitless energy without deadly radiation is available in many stellar systems allowing intelligent life to evolve.
\end{enumerate}
It is clear that a knowledgeable reader could remind us that we can be destroyed by a large meteorite (think dinosaurs), a supernova (SN) explosion nearby (no candidate star on the horizon for tens of millions of years) or half sterilized by a nearby gamma ray burst. The response to this is simple; if the Sun were a significant source of fast neutrons, we would not be here to worry about that.

The reader should note that the Anthropic principle hypothesis is supported by special features of our very long lived solar reactor: the primary reaction in our Sun proceeds by producing deuterons. Thus the $dd$ fusion reaction could ensue, followed by neutrons. Yet there is no accumulation of deuterium; instead, given the specific environment, deuterons fuse with protons and turn into ${}^3\,\mathrm{He}$. The somewhat heavier high metallicity stars running often on the CNO catalytic cycle requiring Carbon inventory are similarly fine tuned as the cycle is self regulating with ultra high temperature dependence of nuclear reaction rate $R_\mathrm{CNO}\propto T^{20}$.

We do not provide the answer in this presentation which of the two options applies to justify the aneutronic solar energy source. The Anthropic principle alternative could be eliminated by creating a different fuel cycle aneutronic reactor on Earth. Therefore, our objective is to draw attention to this exceptional situation and to advance in the direction of resolving the above enigma while reopening to study the science of (aneutronic) nuclear fusion. While stellar conditions are not transferable to the laboratory (the core is near $150\,\mathrm{g/cm^3}$~\cite{Bahcall:2000nu}) our argument concerns the existence of complete aneutronic fusion cycles and not the reproduction of solar conditions.

\subsection{Quasi static plasma}\label{ssec:plasmaF}
\subsubsection{ITER sine termino}\label{sssec:INTiter}
After more than a half century of plasma fusion and reactor development, accompanied by promises that ``fusion energy is around the corner,'' no experimental plasma fusion reactor is operating. An international project, ITER, has been subject to several redesigns~\cite{Aymer:2001,ITR-18-003,ITR-24-005,loarte2025new}, and a price tag arguably a few times the sum of all other fusion work done in the world, ever. We believe that the fusion `gold-rush' replaced decades ago the ongoing science way too early. This prompted the rapid death of most nuclear science, replaced by plasma physicists untrained in the field they were asked to represent, nuclear fusion. The outcome is seen in \rf{fig:FusionPredictions} where we show a collection of prophecies about the expected fusion success date in dependence on when prediction was made (1986--2025), which include all fusion projects.

\begin{figure}[ht]
\includegraphics[width=0.95\linewidth]{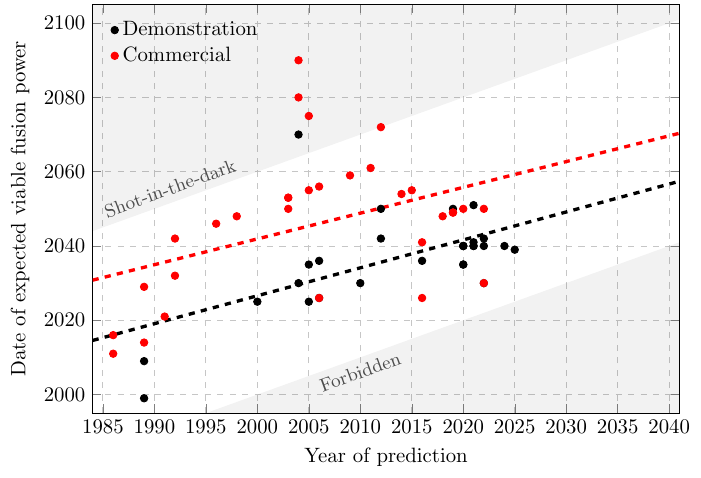}
\caption{Predictions of successful fusion over time (generally magnetically confined plasma fusion). Each point represents a published forecast as compiled by others~\cite{Takeda2023FusionYearsAway,Mohamed2024FusionReadiness,loarte2025new}. Black markers denote predictions referring to demonstration-scale fusion or pilot plants, while red markers denote predictions of commercial fusion power. Dashed lines are linear fits to each set. The upper shaded area (above $y=x+60$) holds `shot-in-the-dark' forecasts more than 60 years ahead; the lower shaded area (below $y=x$) is forbidden, as fusion cannot arrive before the year of the prediction.
}
\label{fig:FusionPredictions}
\end{figure}

The dashed lines in \rf{fig:FusionPredictions} indicate linear fits to the respective data sets. The lightly shaded region below the line $y=x$ is forbidden as successful fusion in the past relative to the year of prediction; the prediction fit will converge to this line when fusion arrives. The lightly shaded region above $y=x+60$ denotes forecasts more than 60 years into the future, which are speculative, ``shot-in-the-dark'' projections rather than fact-based predictions. Trusting the collective wisdom, this \rf{fig:FusionPredictions} implies that a conventional plasma magnetic fusion reactor would not become available in the foreseeable future.

Essential facts about conundrum ITER:
\begin{itemize}
\item 
ITER is a very large quasi steady state magnetically confined low density plasma fusion facility. The details of challenges related to plasma magnetic confinement fusion can be found in books; some go beyond magnetic confinement and consider also inertial confinement, and provide a catalog of the nuclear data. In this context a good text is by Ed Morse~\cite{morse2018nuclear}.
\item
The current intent is to study at ITER fusion processes in a plasma made of abundantly available deuterium $d$ in order to minimize contamination by $t$ burn with high energy neutrons. However, this high energy neutron production is unavoidable as a secondary reaction if $d$ fill is allowed and fusion conditions are achieved, as already explained above. Per unit of produced energy the $d$ fill generates an excess of neutrons that engineers and managers wanted to avoid. We will evaluate chains of hot plasma fusion reactions in \rsec{ssec:ThermalBalance}.
\item
To avoid terminal (to operation) neutron radiation damage, the plasma within ITER can be either run at very low temperature for the deuterium fill, or consist entirely of light hydrogen $p$. In its current context ITER is thus not a thermonuclear experimental reactor but a large and costly plasma experiment.
\end{itemize}

To conclude: It would be good to be sure that we need this highly modified from original design plasma experiment. Novel and promising paths to nuclear fusion, some we describe, do not require any ITER experiments as they either are not operating in near to steady state condition, or if they do, their operational regime is vastly different from that at ITER. Moreover, there is good reason to believe that some of the new ideas we describe in this overview will be successful, upending ITER with design conceived 40 years ago before modern paths to fusion described here were opened up by novel technologies.

One can further argue the case of ITER in two distinctly opposite manners: 
\begin{itemize}
\item ITER has kept in past decades interest in nuclear fusion and contributed to the development of manpower required to start nuclear fusion infrastructure, and 
\item ITER, by binding very large international financial and intellectual resources, has been and remains {\em the primary obstacle delaying success of rising and innovative nuclear fusion approaches}. 
\end{itemize}
This observation is providing an explanation for the data seen in \rf{fig:FusionPredictions}; on one hand optimism about the future of nuclear fusion, on another large international agreement ITER-inertia preventing timely fusion advancements. The ITER project thus lives up to its Latin language meaning: The noun `iter' translates from Latin as being the journey itself; the road (or path) to nuclear fusion requires the noun `via'. ``Iter sine termino'' is a Latin phrase that translates to ``a journey without end'' or ``an endless journey '': at present ITER event calendar is reaching to 2039. It would indeed be to our surprise if unrelated-to-ITER nuclear fusion reactor were not operational before that date.

In view of these long recognized~\cite{lidsky1983trouble} and today well known arguments there is no need to evaluate in an economical and industrial context ITER's fusion power potential. These scientific obstacles  clearly limit ITER's relationship to a viable fusion power generation irrespective of the question if a sustainable nuclear burn could be achieved.

\subsubsection{Alternative plasma fusion approaches}\label{sssec:INTnewPlasma}
Work is ongoing to improve the grandfathered and (arguably as above discussion implies) too large (to fail as well) 1950s quasi-static magnetic confinement technology which ITER represents. Two solutions are explored today for stationary plasma fusion burning. In one approach the static plasma density is considerably increased allowing to reduce reactor size when compared to ITER, thus reducing the size of the source of high energy neutron radiation making it more manageable, while cutting the primary $dt$ fusion reactor construction cost ultimately by 1-2 orders of magnitude.

To achieve this goal, a considerable increase in the strength of the applied magnetic field guiding the plasma is required. Development of strong magnetic fields for fusion has been an objective that has been in the works for decades and the technology is arguably available today such as the compact high-field tokamak SPARC, built around 20\,T high-temperature superconducting magnets~\cite{Creely:2020}. High field and high density plasma burning reactor development does not depend on insights that ITER could have provided, or will provide in some yet not fully determined future.

Another, arguably more promising approach which reduces size and cost, alters furthermore the shape of the confining magnetic field, from the tokamak/ITER style with a pumpkin-like plasma volume to what is a smaller volume stellarator doughnut ring. The implementation of a stellarator has as a technological requirement high precision magnetic fields. We point to the Wendelstein 7-X (\mbox{W7-X}) stellarator at Greifswald~\cite{Grulke:2024}, and a stellarator-based thermonuclear reactor in development in Germany/Bavaria by Proxima Fusion~\cite{Lion:2025}. We stress, however, that none of these approaches has yet reached energy break-even as no magnetically confined plasma, tokamak or stellarator, has ever produced more fusion energy than was supplied to heat it. The record (unbeaten in magnetic confinement for nearly three decades) still belongs to JET, whose 1997 deuterium-tritium burn released a peak $16\,\mathrm{MW}$ of fusion power with scientific gain $Q_\mathrm{sci}=0.63<1$~\cite{Keilhacker:1999,Wurzel:2025}.

\subsection{Paths towards laser fusion}\label{ssec:INTlaserFusion}
\subsubsection{Inertial confinement fusion}\label{sssec:picoICF}
Inertial confinement fusion (ICF) was reportedly conceived even before construction of the first laser in 1960. However, the highest pulse power for a long time was available only in accelerator generated particle beams. With rise of interest in construction of relativistic heavy-ion accelerators in the 1970s this approach was recognized as providing the greatest available energy coupling to a small fusion device; for a history inspired review see~\cite{hofmann2018review}. However, the ICF development shifted toward high-power lasers due to technological advances allowing higher pulse power combined with shorter pulse durations, and recognition of significantly better coupling of the highly pulsed laser power to electrons in the target. Laser driven ICF offers a highly pulsed thermonuclear burn in which a small capsule of fusion fuel is compressed and heated to nuclear fusion ignition conditions. This is explored by two different methods~\cite{Hurricane:2023Review}
\begin{itemize}
\item 
Indirectly driven ICF: by the ablation pressure initiated by an intense driver pulse interacting with vessel walls converting the incident energy into a quasi-isotropic bath of x-rays that symmetrically ablate the fuel capsule; this in essence is a micro-nuclear weapon experiment conducted under controlled conditions.
\item 
Directly driven ICF: A more efficient path is to shine many laser pulses onto fusion pellet so that laser induced pellet ablation of matter compresses the fusible material directly. This is explored at laser fusion facilities where lasers are less powerful.
\end{itemize}
When implemented with deuterium-tritium ($dt$) fuel, ICF shares the physics and associated hazards of neutron-producing fusion in plasma with the key difference that we have a significantly smaller in size neutron source with $14~\mathrm{MeV}$ neutron flux that is highly pulsed. $dt$-ICF remains constrained by the same tritium-handling and neutron-damage issues faced by all $dt$-based fusion concepts. 

However, as noted before, the much smaller size of ICF primary energy and neutron source allows us to imagine practical and economical solutions. Thus, the search for aneutronic variants of ICF remains a critical research frontier.

\subsubsection{Laser driven IC fusion burn}\label{sssec:NIF}
Everybody reading this article will have heard about the spectacular advance in fusion energy production at the National Ignition Facility (NIF) in Livermore, CA in 2022. In a fusion chamber high power lasers flood a relatively large reaction domain with several mega-Joules of energy, creating a uniform spherical field of X-rays which compress a fusion fuel pellet, heating it to conditions of nuclear fusion burn.

The indirect NIF implosion method to achieve nuclear fusion energy is accordingly called indirect inertial confinement (IC) as lasers miss the active fusion domain. Lasers are a rather short-in-time source of X-rays, replacing by their micro-scale action a small nuclear explosion acting as fusion weapon trigger in large devices. The entire burn process arises in dynamically compressed domain before it falls apart, hence the name inertial fusion (IF). 

NIF by design is mimicking, bottling, and miniaturizing nuclear weapons for study of nuclear explosive burn, the primary motivation behind funding and creation of NIF in the first place; the idea has been proven to function~\cite{Zylstra:2022burn,Hurricane:2023Review,Hurricane:2025}. This was the first demonstration that nuclear controlled small scale fusion burn can be achieved in laboratory. The direct fusion burn was reported by Laboratory for Laser Energetics (LLE), University of Rochester~\cite{Manion:2022Omega,Gopalaswamy:2024direct} using the OMEGA laser. We think that reading the abstract of a recently published review will illuminate the scope of this achievement:
\begin{quote}
For many decades, the running joke in fusion research has been that `fusion' is thirty years away and always will be. Yet, these past few years we find ourselves in a position where we can now talk about the milestones of burning plasmas, fusion ignition, and target energy gain greater than unity (scientific breakeven) in the past tense. Fusion is no longer a joke! Yet getting to fusion ignition, the tipping-point of thermonuclear instability resulting in an explosive increase in ion thermal temperature and fusion reaction-rate, and scientific breakeven (target gain, fusion yield/deposited laser energy, in the laser-driven inertial confinement fusion context) has not been easy. In this publication we discuss our present understanding of the physics and technological challenges surrounding ignition and Gain as well as highlight some outstanding problems that still need resolution.\\ -- \emph{Hurricane O. A., et al.}~\cite{Hurricane:2025}
\end{quote}
 
However, in order to adapt ICF scientific-military advance to civilian energy production, we have to overcome considerable nuclear science in regard to control of fusion ultra-fast neutrons along with tritium breeding, and energy harvesting. Independent of this are the unsolved technological challenges: efficient, reliable, powerful MJ-scale lasers firing continuously at Hz scale, and cheap target pellets with a highly precise delivery system to the sustainable X-ray fusion chamber allowing repetitive sustained tiny fusion explosions.

The material science of high fusion neutron flux environment and engineering the associated in-situ tritium production are unsolved scientific and engineering dimensions. However, the ICF path to nuclear energy benefits from nearly point-like geometric source of energetic radiation, circumstances which are far more approachable, promising a considerably lower barrier to economically viable technological solutions for energy harvesting and tritium in situ breeding. And, last but not least, we note that ICF is a method that worked in practice (see quote above) while magnetic confinement failed to deliver in 75 years. We will address the magnetic confinement in depth and show that the promise of infinite clean fusion energy in this context is untenable.

\subsubsection{Antennas for light and fusion}\label{sssec:nanoF}
To compare to a power station, an aneutronic fusion reactor powering a truck, a ship, or used in a space flight fusion engine requires thousands of times smaller power, which, however, must be made available in stand-by on-demand mode. This clearly requires an entirely different scientific and technological solution in comparison with an electrical power fusion reactor. What one can see as a `via' road to such fusion solution uses several frontier technology advances of recent years. 

Today we have available the required technologies: 1) short near and sub-femtosecond lasting coherent high contrast so called relativistic light pulses; and 2) capability to mass produce nano-structures and meta-material, optimized to resonate with such extreme light pulses in the sense of acting as {\em light antennas}, attracting the light pulse energy.

This phenomenon has a well-known historical example in the well-studied Lycurgus cup, see~\rf{fig:LycurgusCup}, a 4th-century Roman glass that appears green in reflection and ruby-red in transmission. Its dichromatic properties arise from Au/Ag nanoparticles $\sim$10--100~nm in size, which resonate at visible wavelengths and selectively scatter and absorb light~\cite{freestone2007lycurgus}. 

\begin{figure}
\centering
\includegraphics[width=0.35\linewidth]{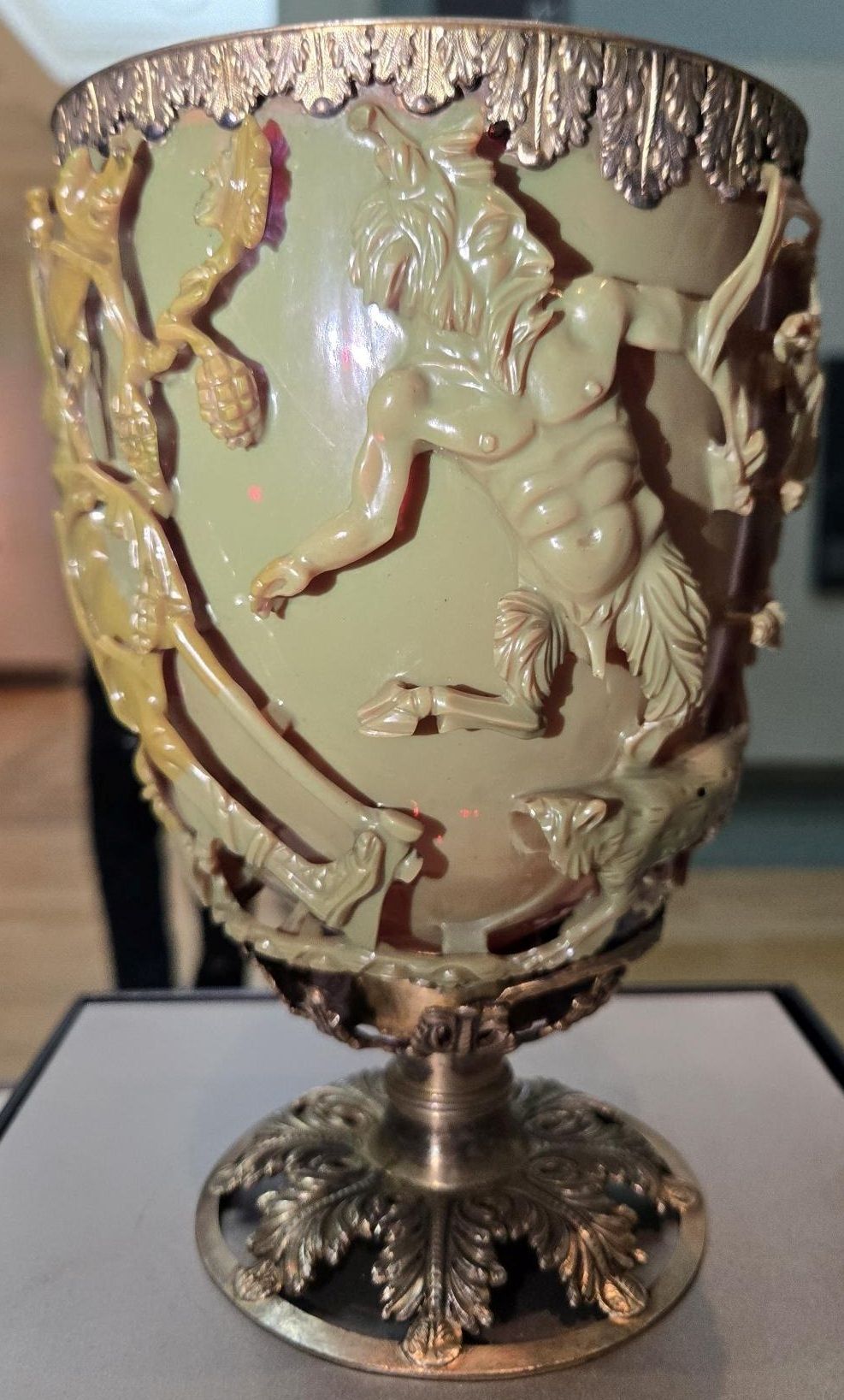}
\includegraphics[width=0.35\linewidth]{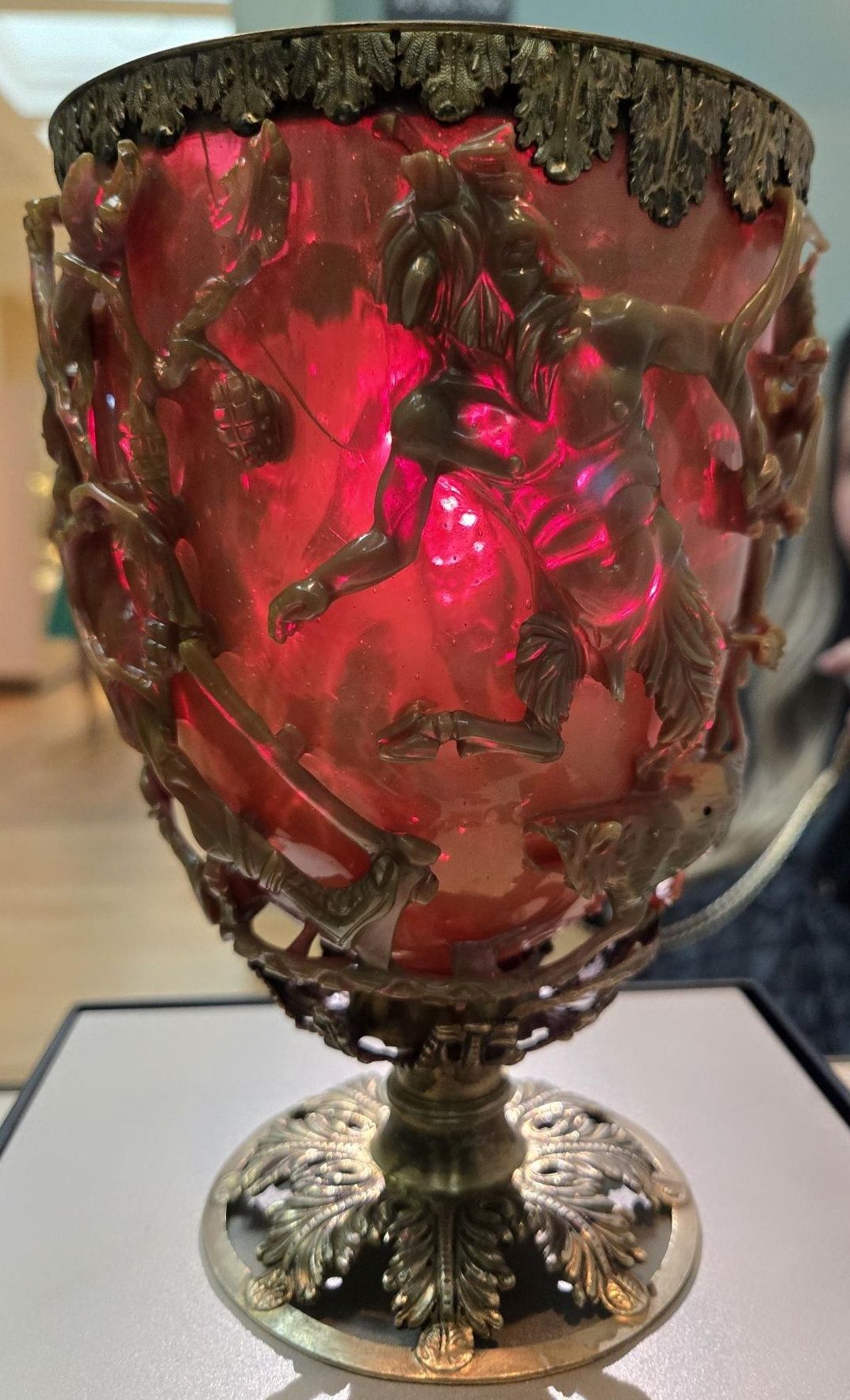}
\caption{The Lycurgus cup under (left) reflected light and (right) transmitted light. Image at the British Museum by author (AJS) under \href{https://creativecommons.org/licenses/by/4.0/}{CC BY 4.0}.}
\label{fig:LycurgusCup}
\end{figure}

When driven near its localized surface plasmon (LSP) resonance, the nanoparticle develops strong near-fields enabling interactions far below the free-space diffraction limit of laser light~\cite{Novotny:2011}, producing local intensity enhancements of $10^4-10^6$. It should be noted that use of antennas for light allows to concentrate laser light to space-time domains well below usual diffraction limit~\cite{Novotny:2007,Novotny:2011} allowing laser light energy to be deposited into small ignition domains. In this manner all of incoming coherent light is delivered to the active fusion centers and constitutes near hundredfold progress compared to NIF where about 1\% of laser energy is used to ignite a single fusion center. 

The use of nano-antennas for light in nuclear fusion was proposed and elaborated by L.~P.~Csernai, N.~Kro{\'o}~\cite{Csernai:2018Radiation,Kroo:2025HighField} and exploits the same principles. It relies on incident light coherent electromagnetic wave coupling to collective oscillations of conduction electrons in nanoscale metallic structures which behave as an optical antenna. Under irradiation by high-contrast, ultrashort laser pulses, resonant nanorods (for example seeded within transparent polymers) can generate: (i) large near-field electric gradients, (ii) acceleration of ions, and (iii) collective electron motion thought capable of transferring correlated momentum to nearby protons.

Such microscopic dynamical mechanisms accelerate particles bypassing the bremsstrahlung losses inherent to thermal plasmas: particle for fusion acceleration occurs on near picosecond time scale in a highly non-equilibrium, non-thermal environment. The related research field is in general referred to as `plasmonic fusion' or perhaps better termed `nano fusion' as plasmonic response of the resonant nano-antenna can be used in multiple ways~\cite{ChGJR2025patent}.

 \subsubsection{Opto-electro-mechanical Fusion}\label{sssec:optoelecmechFusion}
Laser driven nano fusion relies on a transient, highly localized energy deposition to trigger, or better said, to spark, primary nuclear reactions. The research in this field~\cite{Kroo:2025HighField,Biro:2025EPJ,Csernai:2025EPJ,Kedves:2025EPJ} has as its primary objective the demonstration of efficient fusion `ignition' using relativistic laser pulses of nano-scale fusion domains. 

Once this stage is fully understood the next objective is to use the ignition process applying several non parallel laser pulses aiming to achieve a bulk shock burn which ultimately should lead to abundant fusion yield~\cite{Csernai:2026MDPI,Biro:2023Univ}. In aneutronic nano-fusion cycles the burn progresses at a speed determined by the range of produced charged particles, per cycle time, implying near relativistic speed of burn propagation in ordinary matter. There is a lot of undiscovered physics in these still little explored research areas awaiting an in-depth theoretical study, and an experimental exploration with the objective of formation of a shock front bulk nuclear fusion burn. 

How this can be done with success is of course the trillion dollar question; some of the novel ideas have led to patent filing, including by Grayson and Rafelski~\cite{ChGJR2025patent}. Our Arizona solution involves light concentration with the objective of achieving at sub-picosecond scale nuclear fusion ignition by opto-electro-mechanically accelerated motion of nuclear fuel within sub-wavelength (partially) enclosing structures. This approach does not use LSP collective electromagnetic field to achieve motion of fusible material, the method employed by the NAPLIFE collaboration~\cite{Kroo:2025HighField,Biro:2025EPJ,Csernai:2025EPJ,Kedves:2025EPJ}. Instead we use LSP to help focus light to sub-diffraction limit, reaching intensity that will highly ionize a heavy element `pusher' of the fusible material.

The motion of fusible isotopes is due to the collective Coulomb force: Ionized heavy and LSP capable element `pusher', e.g. gold, is chosen to have a high inertia; the also ionized much less inertial i.e. lighter fusible material repels, moving in the opposite direction. Momentum conservation allows a large kinetic energy acquisition and thus the capability of overcoming in collective directed motion the aneutronic energy threshold leading to nuclear fusion albeit in a small target area devoid of electrons, removed by the laser pulse. We already addressed the LSP focusing of light; the Coulomb induced collective motion is today well known: There is a laser-driven classroom demonstration experiment called the Coulomb explosion: A pinhead in vacuum is ionized by a laser shot, and the ionic Coulomb explosion is observed.

To achieve fusion conditions one would want to master such collective Coulomb field directed motion in high density material rather than in the vacuum. In our Arizona approach the first objective is removing most of the electrons from (many) 50--100 nm scale light antennas. These can, for example, be (semi) hollow structures embedded in transparent to light wavelength fusible material. Normal about 3\,$\mu $m light pulse focal point comprises a volume that is more than 20,000 times the volume of a single resonator. This means that whatever we discuss can happen in a synchronous manner to thousands of resonators found in a typical focal volume of a laser pulse. If the available light pulse energy allows the focal point to be made larger keeping the ionization level of the target, even millions of resonators could experience LSP assisted removal of electrons followed by the complex dynamical evolution we studied using the particle-in-cell method. The antenna for light approach uses laser energy effectively avoiding instabilities other methods for direct laser fusion suffer from. 

\paragraph*{\bf Synchronized nano-shell implosion}
One particularly interesting simulation result underpinning the patent filing by Grayson and Rafelski~\cite{ChGJR2025patent} is the implosion of a hollowed heavy nano-shell structure, which, when filled with fusible material, reminds us of more standard inertial confinement fusion (ICF) but miniaturized, and duplicated thousandfold if not millionfold. It is interesting to note that each nano structured antenna is enveloped by laser coherently which is assuring a synchronized dynamical motion. For 90 nm objects synchronized by light velocity this means $ 0.3\times 10^{-15}\,\mathrm{s}$ precision. Each of the many antennas in the path of the light pulse is synchronized to react by light pulse with each antenna acting as the center of ignition of a small spherical explosive burn. Choosing the right density depth profile of nano-antennas could be the key to effective use of laser light for fusion front ignition and effective matter compression.

This assures an extraordinary capability to achieve synchronous collective dynamics of ignited fusible material with prescribed geometric distribution (meta material) nearly impossible to achieve otherwise in the context of ICF, and certainly out of reach for conventional particle beams which miss by about 5 orders of magnitude the time sharpness of laser pulses. This allows to explore many laser beam direct and indirect inertial confinement implosions necessary to achieve large enough nuclear fusion volume capable to retain the fusion energy carried by charged particle fusion products. Such a system offers the required synchronicity that inhibits dynamical implosion instability, a critical advantage of laser pulse opto-electro-mechanically facilitated nuclear fusion. Such a million-fold enhanced particle intensity originates in each of the many spatially distributed fusion sparks synchronized with speed of light and thus allowing shock wave formation; this extreme condition we hope to use to start shock burn in the bulk of the material.

\paragraph*{\bf A fragmented Manhattan Project}
Several fusion companies have been created around such nascent ideas of which our preferred solution was described in more detail above. This domain of nuclear fusion is furthermore attracting investor attention, both private and government. This in many aspects is reminiscent of the Manhattan Project: A rush to convert to praxis the science principles about aneutronic nano-fusion within a highly interdisciplinary context employing relativistic laser pulses, opto-electronic fusion targets and using opto-electro-mechanical resonant devices. However, unlike the national and coherent Manhattan Project, today in nuclear fusion each `player' follows a significantly different path and there is large fragmentation and secrecy surrounding the science basis of the different approaches.

Because fusion reactions driven in their first step by laser accelerated particles can achieve and exceed the aneutronic nuclear fusion reaction threshold, this is the method of choice to advance aneutronic fusion in general. The creation of a truly aneutronic chain is a challenge to which we return several times in this work. We will discuss by example some aneutronic reactions and give an example of aneutronic reaction chain in \rsec{sssec:BNcycles} and \rsec{ssec:Beryl}.
 
It is hard to predict the method allowing the achievement of practical application. However, the clear objective is to overcome the energy threshold and enable aneutronic fusion path. Commercial companies are aiming at aneutronic energy from fusion. However, it is more likely that the primary first application of ongoing work on nano-fusion will be to space propulsion; we comment on this at a few opportunities and give a longer discussion at the very end of this report as an important outlook to near future of nano-fusion. 

\subsection{Particle catalyzed fusion}\label{ssec:INTmuCF}
\subsubsection{The birth of particle physics: Particle catalyzed nuclear fusion}\label{sssec:particles}

The basic principle of particle catalyzed (nuclear) fusion (CF) is transparent to anyone who has explored the question why a bottle of hydrogen gas, which contains about 1:3200 part of HD molecules, does not spontaneously warm up due to production of light helium in spontaneous aneutronic $pd$ fusion reactions. The resolution to this puzzle is that electrons keep nucleons bound at \AA ngstrom separation, which assures stability against fusion on a cosmic time scale.

A 200 times heavier particle, the muon, here just a heavy electron, would bind hydrogen to a 200 times shorter distance. This would allow fusion reactions at a rate ranging between one per microsecond, to one per picosecond, depending on isotopic molecular composition: For different mixes of the three hydrogen isotopes in the molecule there are 6 possible combinations. CF reactions are possible at room temperature and ambient pressure. What controls the fusion rate and yield is the formation speed of `exotic' mesonic hydrogen molecule and self-poisoning of the catalyzing particle, typically by permanent binding to the fusion produced helium. 

One would think that such an insightful application of particle physics to fusion could only be recognized well after the standard model of particle physics was conceived, thus rather recently, but in fact, all of particle physics understanding is very recent. Admiring today's `Mendeleev'-like table, it is not on top of our mind that when one of us (JR) was born, little if any of it was known. Even less appreciated is the fact that the birth of particle physics nearly coincided with the birth of this new idea how nuclear fusion could be facilitated or better said, catalyzed. 

We remind that, arguably, the birth of particle physics begins in 1935: The need to understand nuclear interactions required, as Hideki Yukawa postulated, a particle called the `meson' having the mass of about 200 electron masses to describe the known interaction range. Not observed as a part of matter, such a particle had to be unstable. Photographic emulsion method to detect both the new particle and its decay was developed by Marietta Blau immediately following this; she won in Vienna a science prize in 1937 -- but the world went to war. Today she is widely recognized, e.g. by a street named after her at CERN, for this achievement.

Following the first experimental reports of the observation in photographic emulsion of energy release in a meson transmutation process in 1947, which marks the birth of particle physics, F.~C. Frank~\cite{Frank:1947} has advanced the idea of particle (meson) catalyzed (nuclear) fusion (CF). The surprise here: CF was proposed at the birth hour of particle physics practically in coincidence with the discovery of the pion, and its decay to a heavy electron, the muon. 

\paragraph*{\bf Frank's proposal}
Frank was motivated in his work by the desire to avoid the introduction of a second mesonic particle. What made the connection to fusion possible is that the visible energy release in the today well known charged pion decay to a charged muon (a heavy electron), within the experimental error of the era, could be generated by CF. In the long article printed in Nature~\cite{Frank:1947}, Frank presented in considerable detail what we today call muon catalyzed fusion ($\mu$CF). 

Even so, and despite Frank's scientific prominence, this important invention went straight into the historical dustbin for a few years, the time needed to establish the existence of the not strongly interacting heavy electron, the muon, produced in the observed meson decay. Since Frank wanted to avoid the existence of this decay introducing CF, the nuclear fusion content in his seminal paper was at first thought irrelevant considering that CF was not needed given the confirmed meson decay/transmutation. Frank, busy with his other at the time popular physics, did not adapt his innovative CF thinking to use the relatively long lived muon in practical application to fusion research.

Muon catalyzed fusion was instead described seven years later by Ya.B. Zel'dovich. Like Frank, but in a more specific manner with respect to the properties of muons, Zel'dovich considered a diversity of relevant atomic and molecular processes~\cite{Zel:1954}. In that work Zel'dovich thanks Andrei Sakharov for discussion, but does not cite or mention Sakharov's internal unpublished report of 1948~\cite{Sakharov:1989P}.

\subsubsection{Experimental muon catalyzed fusion breakthrough}\label{ssec:muCFexp}
Finally, three times is the charm, we see $\mu$CF enter public attention due to an announcement in the world news of an experimental discovery, again using the Marietta Blau method, photographic emulsions. J. D. Jackson~\cite{Jackson:2010Rem} recalls:
\begin{quote}
``Luis Alvarez and colleagues discovered muon-catalyzed fusion of hydrogen isotopes by chance in late 1956 (added: see Ref.\,\cite{Alvarez:1957un}). On sabbatical leave at Princeton University during that year, I read the first public announcement of the discovery at the end of December in that well-known scientific journal, The New York Times. A nuclear theorist by prior training, I was intrigued enough in the phenomenon to begin some calculations (added: see Ref.\,\cite{Jackson:2010Rem}).''
\end{quote}
There are several important implicit messages in above quote: Real nuclear fusion discoveries and interpretation involved people trained in nuclear science such as Luis Alvarez, the future 1968 Nobel Prize in Physics recipient. Discoveries should be disseminated in reputable daily press reaching a wide readership and triggering prompt follow-up work, in this case by another (eminent nuclear) theorist, J.D. Jackson. In our opinion the opposite, for example wide reporting in sensational media of `nuclear fusion' processes or discoveries made by unrelated professionals such as plasma engineers or chemists, is in general creating scientific chaos, discrediting a very important for our civilization application of subatomic science.

Jumping forward, Jackson in his seminal work~\cite{Jackson:1957zza} created a no-go theorem: Under most optimal conditions no more than 85 fusions per muon could ever be achieved, which seemed not enough for practical applications. This limit was based on sticking of the muon to the in-fusion produced $\alpha$-particle. In the following two decades continued improvement in understanding of the $\mu$CF processes is seen in many theoretical and experimental works in the Soviet Union; status of the field as of 1979 was summarized by one of us (JR)~\cite{Rafelski:1979ab}. In part stimulated by this widely circulated CERN report, Steven E. Jones and collaborators in early 1980s studied at Los Alamos Laboratory $\mu$CF in dense, very cold, $dt$-liquid achieving 150+ nuclear fusions per muon~\cite{Jones:1983pw,Jones:1986kk,Jones:1986Na}. This result was at first grossly contradicting the theoretical understanding of the epoch; it had placed $\mu$CF near, if not above, the engineering break-even point, depending on the technology available for muon generation, a topic we turn to in \rsec{sssec:muon_production}. Engineering break-even threshold means we could, at a great cost, develop a reactor that would run on empty, not delivering power but sustaining itself.

Jones' results were initially contested without contrary evidence~\cite{jones1993evaluation}; in our reasoning this was solely rooted in scientific disbelief that experiments carried out on a shoestring budget by newcomers in their result could be contradicting an easy to follow theoretical quantum physics no-go theorem. However, this theorem did not, for example, consider that the muon captured and bound deeply to the fusion $\alpha$ could be stripped in the follow-up collisional slowdown process~\cite{Rafelski:1988wq}. Nor was there ever consideration of the possible follow-up nuclear fusion reaction such as $\alpha\mu+t$ which today seems plausible considering a possible reaction resonance in the compound ${}^7\mathrm{Li}$ nucleus~\cite{Aprahamian:2025AA}. 

Jones' $\mu$CF result on maximum achievable fusion yield held its ground and is today the accepted standard. In the following two decades the theoretical no-go $\mu$CF bounds were stretched in order to accommodate Jones' results, but in expert opinion of one of us (JR) there remains a lot more to be understood. The current learned opinions about maximum number of fusions a muon can catalyze are as soft as was the original Jackson no-go theorem. Detailed analysis of experimental results carried out by Jones and his collaborators~\cite{jones1993evaluation} in regard to the surprising speed of tritium catalytic cycles accompanied by limited catalyst poisoning if taken at face value indicate a much higher upper limit maybe above 500 fusions per muon. A review of muon-catalyzed fusion results up to 1996 can be found in Jones's summary chapter~\cite{Jones1997}, which makes clear that the experimental program was far from exhausted at the time of its termination in USA.

\subsubsection{Fusion confusion}\label{ssec:conCF}
The above discussion shows a sudden end of a promising path to nuclear fusion. We did not even attempt to answer the lingering question, could particles other than the muon be of use? Or as we will discuss below, \rsec{sssec:ManyFusions}, could elements with atomic number $Z>1$ be relevant? There is indeed the lingering question, where is an ongoing vibrant international research program on particle catalyzed fusion? 

In the US around 1990 all $\mu$CF funding dried up and researchers moved on to other opportunities and terminating engagement with final reports~\cite{Rafelski:1991DOE}. To understand this unusual situation we need to look at pre-1990 history. The experimental discovery of $\mu$CF, as Jackson recounts~\cite{Jackson:2010Rem}, was accompanied by the characterization in follow-up discussion as `cold' nuclear fusion process, such as the 1987 $\mu$CF Scientific American article by Rafelski and Jones~\cite{Rafelski:1987Jones} and in directly related follow-up review literature~\cite{Rafelski:1990gg}.

However, the term `cold' fusion was adopted by another unrelated research project in April 1989: It was used by electro-chemists measuring heat and calling the often observed caloric imbalance as originating in `cold fusion', a claim made at first without nuclear detection capabilities let alone any relevant nuclear transmutation results. This research claimed rapid and spectacular progress, and achieved in the following even a greater disrepute. Inadvertently, the true `cold' particle catalyzed fusion became victim, at first of a seemingly much more successful `cold' competitor, and later of the reputation problems that ensued surrounding the word `cold'-fusion. In consequence of resulting research landscape, no US funding agency would look at projects involving $\mu$CF for long decades.

Adding insult to injury, because high yielding $\mu$CF approaches require handling of relatively large quantities of tritium, today there is the additional practical barrier as we noted before: any facility handling more than about 100 g of tritium is subject to bilateral and multilateral safeguards inspections. This restricts `table top' scientific study of $\mu$CF, moving the decision process to be about development of a major research program for energy production, based on scant scientific readiness.

\paragraph*{\bf Heavy electrons are not muons}
It is important to remind that catalyzed heavy particle fusion cannot use `heavy electrons' introduced to describe electron dynamics in special materials~\cite{Kuramoto:2000}. Mass is particle inertial resistance to applied force. A heavy electron is a quasi-particle invented to describe how a solid state medium inhibits particle response to applied force. Clearly such heavy electrons cannot exist outside the specific solid state environment and cannot be used to mimic muon catalyzed fusion. On the other hand, if the heavy electron medium consists of the fusible materials, collective material effects inhibiting electron inertial motion, which imply electron enhanced localization, could be of interest to nuclear fusion.

\paragraph*{\bf Ongoing work and historical importance}
Today scientific research in $\mu$CF continues in Japan, see for example Refs.~\cite{Kamimura:2021msf,Yamashita:2022rtu}. A recent experimental study of resonant $dd\mu$ formation is notable~\cite{Toyama:2026}. We argue in \rsec{ssec:MuonCatalyzed} that $\mu$CF approach to fusion deserves further experimental and scientific exploration. One of us (JR) checked that there is related experimental activity beyond Japan at an international research site  licensed to handle a large tritium inventory where also beam of muons is available, i.e. the PSI laboratory near Z\"urich in Switzerland.
 
The historical importance of $\mu$CF lies in following:
\begin{itemize}
 \item[(a)] It provided the first unambiguous demonstration of catalytic nuclear fusion, showing that light nuclei could be induced to fuse rapidly even at ambient temperatures; 
 \item[(b)] It revealed the critical role of non-equilibrium in nuclear burning, highlighting that fusion processes need not rely exclusively on thermal plasmas;
 \item[(c)] should one day a composite dark matter made up of a dark heavy charged particle be identified, by analogy to muonic systems, this path to catalyzed fusion could be of great interest~\cite{Rafelski:1989pz}.
\end{itemize}
Therefore, we suggest the $\mu$CF path is not entirely at its scientific endpoint.

\section{Nuclear Fusion Everywhere}\label{sec:Fusion}
\subsection{Availability of nuclear fusion energy}\label{ssec:Nenergy}
\subsubsection{Relevant nuclear `ashes'}\label{sssec:NuclearAshes}
The present-day elemental composition of the Universe reflects the cumulative outcome of several nuclear cosmological and astrophysical events: the Big Bang Nucleosynthesis (BBN), explosive stellar nucleosynthesis, and the radiogenic enrichment over cosmic time. The present isotopic element distribution is a fossil record of all nuclear activity in the cosmos, shaped by both fusion and fission processes. All stable isotopes may be regarded as the nuclear `ashes' left behind by successive generations of diverse nuclear transmutation processes.

In a hypothetical very distant future, in the limit of complete elemental stellar recycling, the elemental equilibrium distribution could be dominated by the most tightly bound nuclei, nickel and iron. These most-bound per nucleon elements form the boundary between elements capable of yielding energy in fusion (smaller) and fission (larger atomic number). The actual elemental inventory provides a good understanding of candidate nuclei for energy-production schemes.

This situation is addressed in \rf{fig:NuclearAsh}: We present on the horizontal axis the isotope abundances in the Earth's continental crust shown in terms of the number of atoms per Si atom, on log scale. The vertical axis is the atomic number $Z$. The bottom frame presents the lightest elements of interest in fusion energy generation, the top frame presents heavy elements of interest considering nuclear fission. Different stable (and a few unstable) isotope abundances of the same element are connected by horizontal lines showing respective logarithmic abundance.

\begin{figure}
\centering
\includegraphics[width=0.85\linewidth]{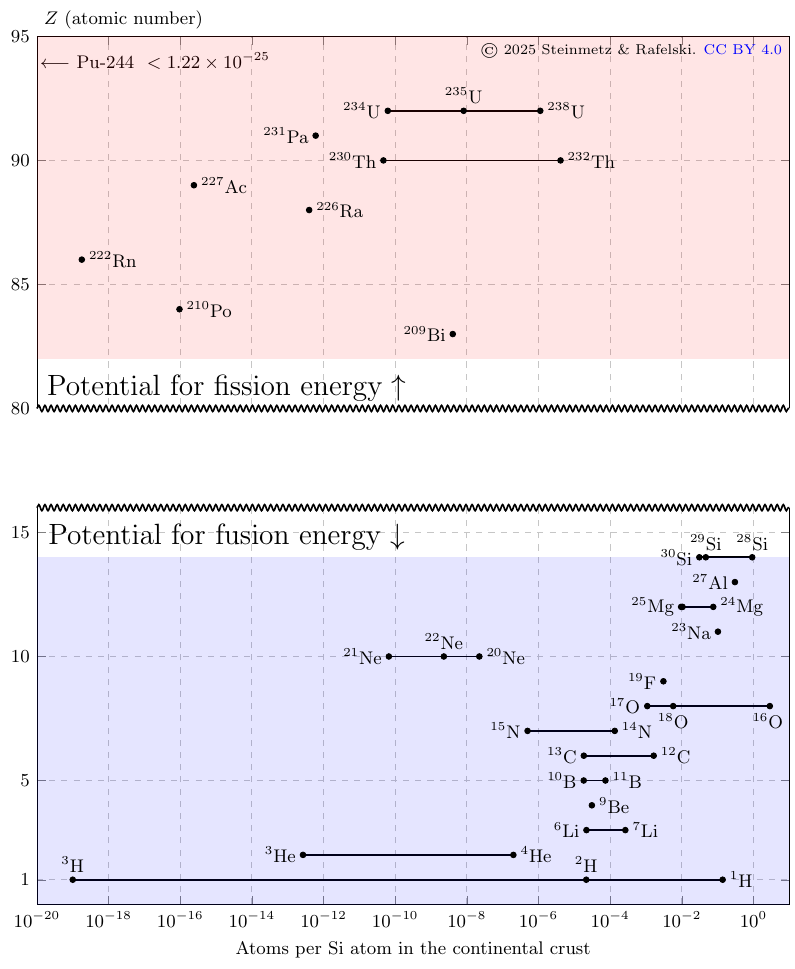}
\caption{Select isotope abundances in the continental crust normalized to silicon (log scale; atoms per Si atom) versus atomic number $Z$ seen on vertical axis, skipping from $Z=15$ to $Z=80$. These two frames suitable for fusion and fission are clearly highlighted (top:fission, bottom: fusion). Data source: Ref.\,\cite{Haynes:2016} for continental crust composition and Ref.\,\cite{Meija:2016} for terrestrial isotope abundances. The isotope abundances of lighter elements in cosmic rays near the Earth were measured by the Alpha Magnetic Spectrometer (AMS) experiment~\cite{AMS:2023anq}. The arrow at top left points to ${}^{244}$Pu, whose abundance ($<1.22\times10^{-25}$) lies below the plotted range.}
\label{fig:NuclearAsh}
\end{figure}

We note that while considering nuclear fusion energy producing reactions we are often interested in least abundant isotopes; their depleted absence indicates that they have been burned in natural transmutation processes. However, some are in low abundance since they are not retained in terrestrial material as is the low abundance of noble gas ${}^{4\!}\mathrm{He}$. Many of the isotopes with depleted abundances seen in the bottom frame have already entered into the discussion of nuclear fusion cycles.

In \rf{fig:NuclearAshUniverse} we show the abundance of all elements inferred for the Universe. Now the vertical axis is the atomic abundance normalized to one atom per million Si atoms, while the horizontal axis is the atomic number. In this global abundance protons and ${}^{4\!}\mathrm{He}$ dominate by many orders of magnitude and the seemingly scarce deuterium is among the most abundant isotopes; this energy reservoir dominates Uranium energy reservoir, allowing for yield difference by a factor $\simeq10^7$.

\begin{figure}
\centering
\includegraphics[width=0.99\linewidth]{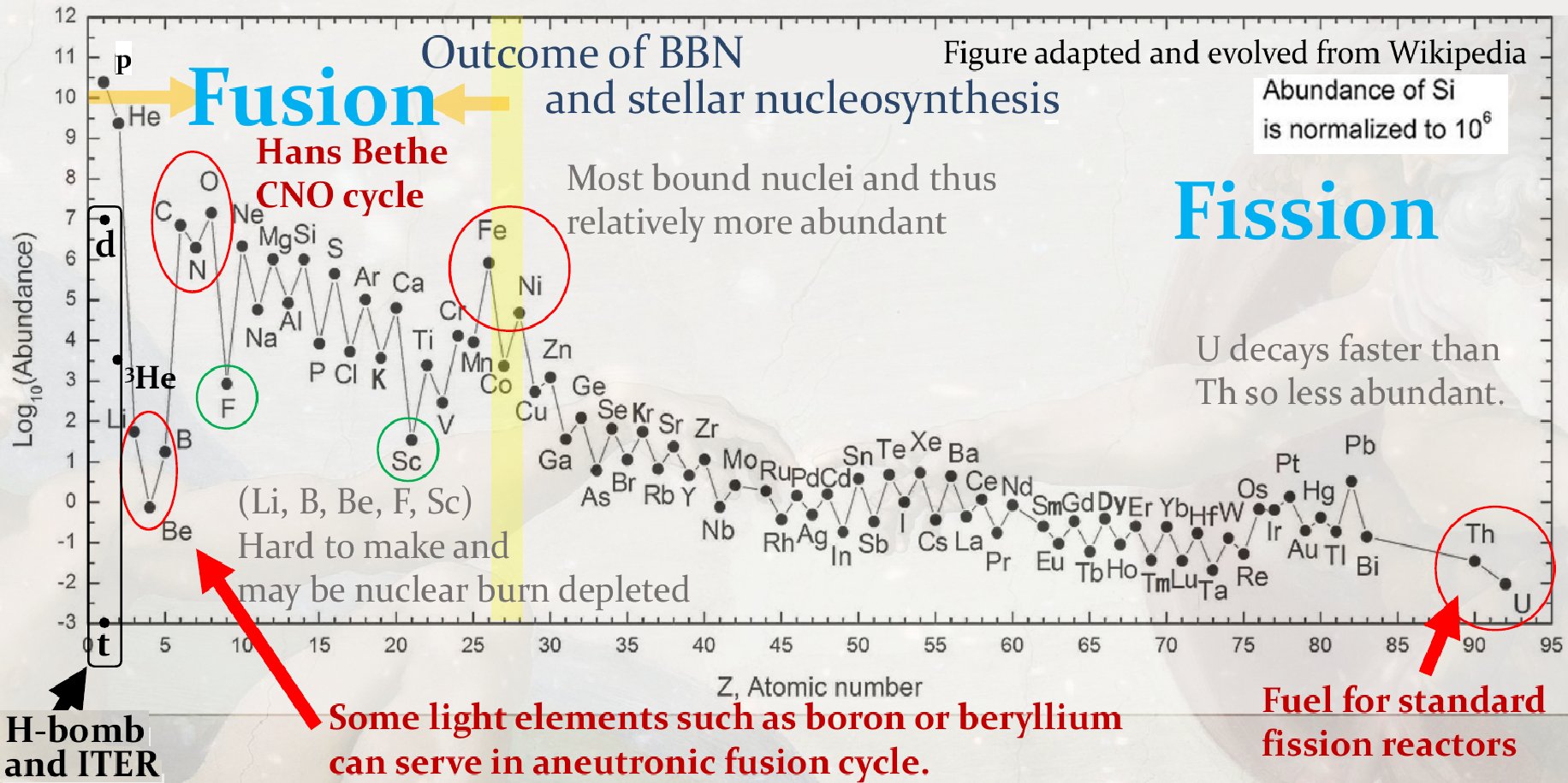}
\caption{Elemental abundances in the Universe normalized to silicon (log scale; atoms per million Si atoms) versus atomic number $Z$. For further insight see comments in the figure drawn from JR presentation at Particles and Plasma meeting in Summer 2025. Adapted from Ref.~\cite{MHzas:2012ABU} under \href{https://creativecommons.org/licenses/by-sa/3.0/}{CC~BY-SA~3.0}; revisions by the authors.}
\label{fig:NuclearAshUniverse}
\end{figure}

Abundance maxima occur at specific atomic nuclei whose synthesis is favored by binding-energy maxima and/or reaction pathways. Light elements are predominantly products of fusion reactions. From the perspective of fuel for nuclear fusion we note many common isotopes with relatively small abundance. Elements heavier than iron are primarily produced by neutron-capture processes in explosive stellar environments. These heavy nuclei can undergo fission releasing energy.

\subsubsection{The (nuclear) interaction length} \label{sssec:InterLength}

Conventional thermonuclear fusion proceeds through equilibrium plasmas, in which reactant nuclei follow a Maxwell-Boltzmann velocity distribution and the fusion rate depends on the thermal average of the microscopic reactivity, $\langle\sigma v\rangle(T)$. In contrast, non-equilibrium or beam-target fusion describes situations in which one population of ions (typically energetic projectiles produced by an accelerator, laser, or collisionless shock) impinges upon a relatively cold and dense target medium that remains near rest.

Because charged-particle fusion at low energies is suppressed by Coulomb repulsion, it is convenient to factor out the dominant exponential tunneling term and define the $S$-factor
\begin{equation}\label{eq:Sfactor}
S(E)\;\equiv\;E\,e^{2\pi\eta_\mathrm{S}}\,\sigma(E)\,,
\quad
\eta_\mathrm{S}=\frac{Z_1 Z_2 e^2}{4\pi\varepsilon_0\hbar v}
=\frac{Z_1 Z_2\alpha}{\beta}\,,
\end{equation}
where $E$ is the center-of-mass energy, $\eta_\mathrm{S}$ the Sommerfeld parameter, $v=\beta c$ the relative velocity, and $Z_{1,2}$ the nuclear charges. By construction, $S(E)$ varies smoothly with $E$ even when $\sigma(E)$ itself changes by many orders of magnitude~\cite{Barker:2002mfp,wang2012reanalysis}.

\begin{figure}
\centering
\includegraphics[width=0.75\linewidth]{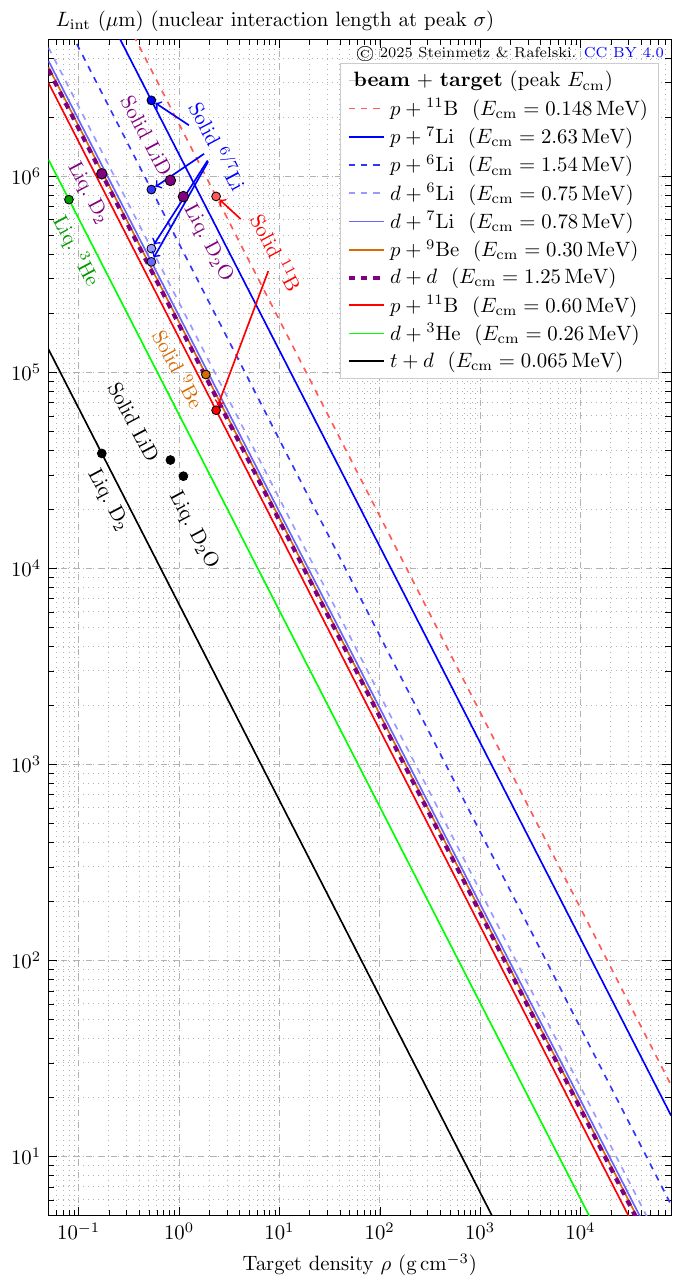}
\caption{Nuclear interaction length $L_{\rm int}$ (in microns) vs.\ target density $\rho$ for thin targets at each channel's peak cross-section $\sigma_\mathrm{peak}$. For every entrance channel $\sigma_\mathrm{peak}$ is summed over all exothermic exit channels at the evaluated CM energy. Cross-sections are obtained from Refs~\cite{Page:2004md,Angulo:1999zz,Xu:2013fha,bosch1992improved,Sikora:2016pB,Otuka:2014wzu,Koning:2019qbo}. Dots mark specific material densities of each target. The $p+{}^{11}\mathrm{B}$ channel enters twice, once for each of its two resonances. The y-axis is extended down to show the interaction length of compressed targets.}
\label{fig:InteractionLength}
\end{figure}

Because the projectiles occupy a narrow energy band rather than a thermal distribution, the relevant quantity for estimating fusion probability is the differential cross-section $\sigma(E)$ evaluated at the projectile energy. For a projectile of known kinetic energy $E$, the mean distance it travels in the target before undergoing a nuclear fusion reaction is characterized by the interaction length
\begin{equation}
 L_{\rm int}
 \;=\;
 \frac{1}{n_{Z}\,\sigma_{\rm fus}(E)} \, ,
 \label{eq:Lint}
\end{equation}
where $n_{Z}$ is the number density of target nuclei and $\sigma_{\rm fus}(E)$ is the fusion cross-section at that energy. Expressing $n_Z$ in terms of the target density $\rho_Z$, atomic weight $A_Z$, and Avogadro's number $N_A$,
\begin{equation}
n_{Z} = \frac{N_A\,\rho_Z}{A_Z}\,,
\end{equation}
allows one to write the product $\rho_Z L_{\rm int}$ as an areal mass
\begin{equation}
 m_{\rm areal}
 \;\equiv\;
 \rho_Z L_{\rm int}
 \;=\;
 \frac{A_Z}{N_A\,\sigma_{\rm fus}(E)} 
 \;,
 \label{eq:m_areal}
\end{equation}
At the peak of the measured or calculated cross-section $\sigma_{\rm fus}(E)$, the corresponding $L_{\rm int}$ describes the characteristic thickness of target material that would yield a fusion optical depth of unity for mono-energetic projectiles. This definition isolates purely nuclear factors from thermal or hydrodynamic effects and provides a convenient metric for comparing different reaction channels under non-equilibrium conditions.

\paragraph*{\bf Comparison across reaction channels}
Figure~\ref{fig:InteractionLength} illustrates $L_{\rm int}$ as a function of target density for several representative aneutronic and light-ion fusion reactions. Each colored curve corresponds to the inverse relation $L_{\rm int} = m_{\rm areal}/\rho$ evaluated at the cross-section's resonance or peak value. The steep inverse dependence ($L_{\rm int}\propto 1/\rho$) means that even dense condensed-phase targets remain effectively transparent for reactions with small cross-sections. For example, the broad resonance of the $p+{}^{11}\mathrm{B}\!\to\!3\alpha$ process ($\sigma\!\approx\!1.2$~b) requires an areal mass of $m_{\rm areal}\!\approx\!15$~g\,cm${}^{-2}$, corresponding to $L_{\rm int}\!\sim\!6.5\times10^{4}$~$\mu$m in solid boron. The same reaction also proceeds through the narrow resonance at $E_\mathrm{cm}=0.148$~MeV, shown as the second $p+{}^{11}\mathrm{B}$ curve. Its measured peak $\sigma\!\approx\!0.1$~b is an order of magnitude smaller and requires $m_{\rm areal}\!\approx\!185$~g\,cm${}^{-2}$ in solid boron. This last value should be read as an upper bound as the resonance width $\Gamma\!\simeq\!6$~keV is comparable to the energy resolution of the measurements, so the true peak is likely sharper and higher shortening the interaction length for a sufficiently narrow beam. By comparison, the familiar $d+t\!\to\!\alpha+n$ channel with $\sigma\!\approx\!5$~b achieves a somewhat shorter $L_{\rm int}\!\sim\!3.7\times10^{4}$~$\mu$m in solid lithium deuteride.

Although the interaction-length concept is strictly applicable to mono-energetic beams, it provides a useful heuristic for visualizing the relative accessibility of different reactions in non-thermal or beam-driven environments where the particle energy distribution is narrow and the target density is well defined.

\subsection{The most accessible path to fusion}\label{ssec:eazyfusion}
\subsubsection{The need to control neutrons in dt-fusion}\label{sssec:ICFneutrons}

In~\rf{fig:NeutronSafety} we see that alpha particles are stopped by thin material barriers such as paper, while beta particles penetrate further, are blocked by sub-millimeter aluminum foil. Gamma rays require dense shielding such as lead to attenuate their flux. Neutrons, being uncharged and highly penetrating, traverse normal shielding materials, must be moderated (energy removal) and ultimately absorbed by appropriate wall material.

The deuterium-tritium ($dt$) fusion reaction,
\begin{align}
 d + t \;\longrightarrow\; \alpha(3.56~\mathrm{MeV}) + n(14.1~\mathrm{MeV}),
\label{eq:dt}
\end{align}
produces a copious flux of $14~\mathrm{MeV}$ neutrons with velocities $v_{n} \approx 0.173\,c$. These neutrons are uncharged and thus highly penetrating: they traverse light shielding materials such as paper and aluminum with negligible attenuation, penetrate even dense high-$Z$ shields like lead, and require dedicated light elemental e.g. hydrogenous moderators and neutron absorbers to capture effectively. In contrast, $\alpha$, $\beta$, and $\gamma$ radiation exhibit far shorter ranges in matter. 

\begin{figure}[ht]
\centering
\includegraphics[width=0.85\linewidth]{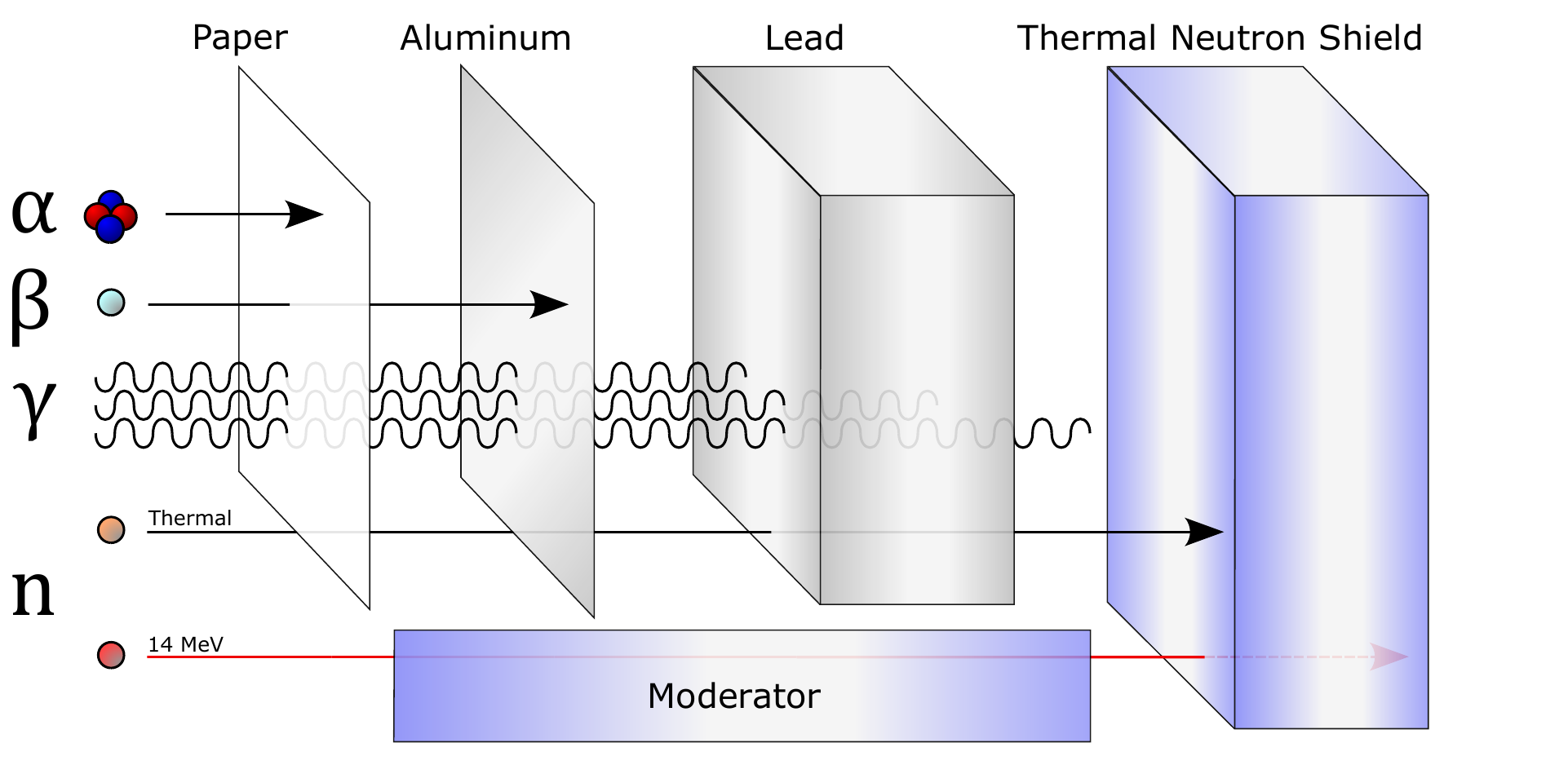}
\caption{Illustration of penetration and shielding of different types of nuclear fusion relevant radiation. Adapted from Ref.~\cite{Ehamberg:2007ABG} under \href{https://creativecommons.org/licenses/by-sa/3.0/}{CC~BY-SA~3.0}; revisions by the authors.}
\label{fig:NeutronSafety}
\end{figure}

The high energy neutron caused hazards are intrinsic to $dt$-based fusion, whether plasma, muon-catalyzed or inertial confinement, both direct- or indirect-drive. This imposes stringent requirements on shielding, handling, and transmutation outcome waste management. For civilian energy production, these considerations motivate the exploration of aneutronic fusion schemes, thereby avoiding the intense neutron flux and its associated consequences. These are typically more difficult to achieve in a thermal environment. Thus in general these require dynamic, non-thermal regimes in their implementation.

\subsubsection{Tritium availability} \label{sssec:ICFtritium}

A critical, looming, and underappreciated limitation of the $dt$-based fusion path is the global scarcity of tritium fuel. Tritium ($t$) is radioactive, with a half-life $\tau_{1/2} \simeq 12.32~\mathrm{yr}=4500~\mathrm{days}$, and is not found in appreciable quantities in nature. Present commercial supplies are obtained predominantly as a byproduct from heavy-water fission reactors of the CANDU type in Canada and South Korea (other countries run these reactors to produce a military supply of tritium). These aging facilities are being decommissioned, and no large-scale civilian tritium production infrastructure exists to replace them. This means that any large scale $dt$ fusion reactor must produce, beyond its own need to continue nuclear burn, a large enough excess of tritium to serve as fuel for the expanding $dt$ fusion power plant infrastructure.

\begin{figure}
\centering
\includegraphics[width=0.75\linewidth]{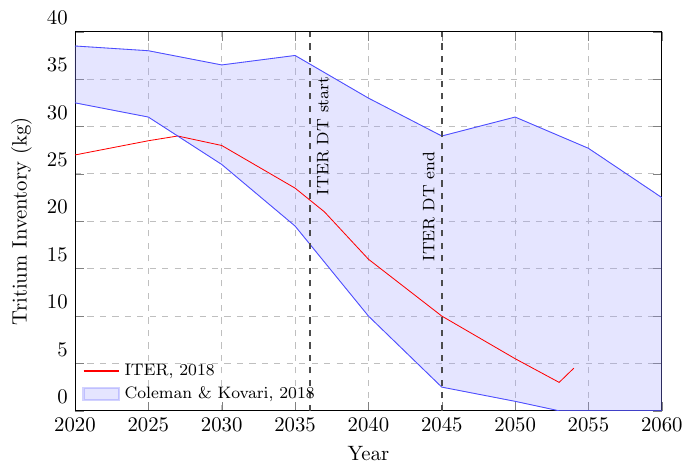}
\caption{Projected global tritium inventories showing the ITER (2018)~\cite{ITR-18-003} estimate (red curve) and supply range (shaded blue) by Coleman \& Kovari (2018)~\cite{coleman2018global} with assumed 2036 start and 2045 end (dashed black lines) of the first deuterium-tritium campaign (DT-1) at ITER~\cite{ITR-24-005}; currently the first tritium fill phase of ITER has been rescheduled to begin in 2039~\cite{loarte2025new}, thus beyond the predictive horizons.}
\label{fig:t_supply}
\end{figure}

As shown in~\rf{fig:t_supply}, the total commercial tritium inventory is presently of order $\mathcal{O}(30~\mathrm{kg})$. It is projected to peak before 2030 and decline thereafter due to radioactive decay, CANDU commercial reactor retirements, and diversion of tritium to other applications. Even without new fusion demand, in consideration of the 4500 days half-life, the tritium supply will decrease sharply over the coming decades. Any experimental plasma burn would further deplete the dwindling supply: For example the ITER experimental program if it progresses to tritium plasma load has an estimated burn rate of $0.9~\mathrm{kg/yr}$.

The fuel consumption rate for an operational $dt$ fusion power plant is daunting: a $1~\mathrm{GW}$ electric ($\sim 2.8~\mathrm{GW_{th}}$) reactor requires $\sim 10^{21}$ $dt$ fusions per second. This corresponds to $\approx 160~\mathrm{kg}$ of tritium per year. At current production costs for artificially made tritium (\$30,000-\$300,000 per gram), the annual fuel bill would be \$5-50 billion per gigawatt of output. No reactor-scale $dt$ fusion economy can be sustained without breeding tritium in situ from lithium via $n+\,^6\mathrm{Li} \rightarrow t+\alpha$ reactions. In June 2022, the magazine \emph{Science} summarized the situation: \emph{``Today there is not enough tritium fuel to initiate one reactor''}~\cite{Clery2022FusionFuel}. The implication is clear that $dt$-ICF fuel cycle is resource-limited well before it is technology-limited by the need to handle high energy neutrons. Both problems will probably have a common solution.

\subsection{Thermal equilibrium \texorpdfstring{$dd$}{dd} fusion}\label{ssec:ThermalBalance}
\subsubsection{Steady state power balance}\label{sssec:PowerBalance}
The suitability of a nuclear reaction for thermal fusion energy in large plasma devices depends on the competition between (i) collisional fusion heating and (ii) radiative and transport losses. For a plasma in thermal equilibrium at ion temperature $T_i$ and electron temperature $T_e$, the volumetric fusion rate between species $i$ and $j$ is
\begin{align}
R_{ij}(T_i)
= \frac{n_i n_j}{1+\delta_{ij}}
\,\langle \sigma v\rangle_{ij}(T_i),
\end{align}
where $n_i$ and $n_j$ are the ion number densities, $\delta_{ij}$ accounts for identical reactants, and $\langle \sigma v\rangle_{ij}$ is the Maxwellian-averaged fusion reactivity. Accurate analytic fits for thermal reactivities are available in the literature for most light-ion fusion channels, notably the Bosch--Hale parametrization and subsequent NACRE compilations~\cite{bosch1992improved,Angulo:1999zz,Xu:2013fha} among others.

If multiple reaction channels $k$ exist (for example $dd\to p+t$ and $dd\to n+{}^3$He), the total rate is
\begin{align}
R_{ij}
= \sum_k R_{ij}^{(k)}, 
\qquad
R_{ij}^{(k)}
= \frac{n_i n_j}{1+\delta_{ij}}
\,\langle \sigma v\rangle_{ij}^{(k)} .
\end{align}

Each channel releases an energy $Q_{ij}^{(k)}$ distributed among reaction products. Only the fraction deposited locally as kinetic energy of charged particles contributes directly to plasma self-heating. Denoting this charged-particle energy by $E_{\rm ch}^{(k)}$, the primary heating power density is
\begin{align}
P_{\rm fus}
= \sum_{ij}\sum_k 
R_{ij}^{(k)}\,E_{\rm ch}^{(k)}\,f_{\rm dep}^{(k)},
\label{eq:powerFusion}
\end{align}
where $f_{\rm dep}^{(k)}$ is the fraction of charged energy deposited before escape or transport losses. 

It is convenient to normalize to the square of the electron density,
\begin{align}
\tilde P_{\rm fus}
\equiv \frac{P_{\rm fus}}{n_e^2},
\end{align}
so that fuel-composition effects enter only through ratios such as $n_i n_j/n_e^2$. For a multi-species plasma with ion charges $Z_i$, charge equilibrium of the plasma requires
\begin{align}
n_e = \sum_i n_i Z_i .
\end{align}

\subsubsection{Radiative losses}\label{sssec:RadLosses}

In a hot fusion plasma the thermal energy can be lost through several radiative channels. We therefore write the total radiative loss power density in the schematic form
\begin{align}
P_{\rm rad}
= \sum_k P_k
= \sum_k \mathcal{Q}_k(T_e,T_i,\{n_j\}),
\end{align}
where the index $k$ labels the relevant radiation processes and $\mathcal{Q}_k$ denotes the corresponding emission functional form. In practice, most radiative channels in a fully ionized plasma are controlled primarily by the electron temperature $T_e$, since electrons both radiate and absorb electromagnetic energy much more efficiently than ions.

The dominant contributions are~\cite{Blumenthal:1970gc}:
\begin{itemize}
\item[(a)] \textbf{Electron--ion bremsstrahlung} ($e+Z\rightarrow e+Z+\gamma$), which scales as $n_e n_i Z_i^2 \sqrt{T_e}$.
\item[(b)] \textbf{Line radiation} from partially ionized ions, important only when the plasma is not fully stripped.
\item[(c)] \textbf{Cyclotron/synchrotron emission}, arising from electron gyration in magnetic fields, with emissivity scaling roughly as $P_{\rm cycl}\propto n_e T_e B^2$.
\end{itemize}

In the high-temperature, fully ionized regime, line radiation is negligible and bremsstrahlung generally provides the dominant radiative sink. While ion cyclotron emission is suppressed by the large ion mass ($P\propto m^{-2}$), it can still compete with bremsstrahlung within sufficiently strong magnetic fields.

Focusing on the leading contribution, the bremsstrahlung power density for Maxwellian electrons may be written in compact form~\cite{Beresnyak:2023nrl}
\begin{align}
P_{\rm brem}
= C_b\, n_e^2\, Z_{\rm eff}\,
\sqrt{T_e}
\left(1 + a_1\frac{T_e}{m_e c^2}\right)
g_B ,
\label{eq:powerBrem}
\end{align}
where $T_{e}$ is written in keV and $C_b=4.83\times10^{-37}\,\mathrm{W\,m^3\,keV^{-1/2}}$ is obtained from the average of the free--free scattering process $e+Z\to e+Z+\gamma$. The parameter $g_B\simeq1.1$ is the velocity-averaged Gaunt factor in the asymptotic limit, and $a_1\simeq2.6$ parametrizes leading relativistic corrections~\cite{morse2018nuclear}.

The effective charge is defined as
\begin{align}
Z_{\rm eff}
= \frac{1}{n_e}
\sum_i n_i Z_i^2 ,
\end{align}
with $n_e=\sum_i n_i Z_i$. 
Because bremsstrahlung scales as $Z_i^2$ while electron density scales as $Z_i$, even modest concentrations of high-$Z$ ions substantially increase radiative losses~\cite{ochs2026bremsstrahlung}.

\begin{figure}
\centering
\includegraphics[width=0.75\linewidth]{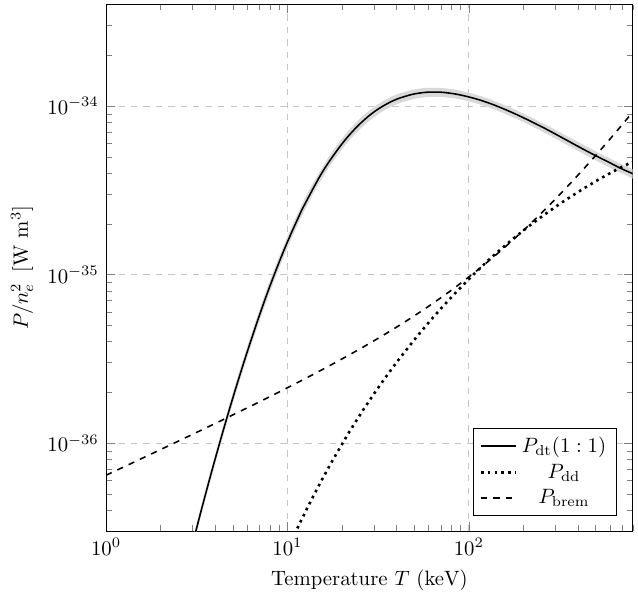}
\caption{
Density-normalized power balance of $dt$ and $dd$ fusion fuels. Each panel shows charged-particle heating $ P_{\rm fus}/n_e^2 $ and bremsstrahlung loss $ P_{\rm brem}/n_e^2 $ as functions of plasma temperature $T$ (1--800\,keV). 
}
\label{fig:PowerBalance_AllFuels}
\end{figure}

Figure~\ref{fig:PowerBalance_AllFuels} illustrates the temperature dependence of these density-normalized heating and radiation terms for two representative fuels. The figure shows conventional neutronic fuels $d+t$ (with 3.5\,MeV $\alpha$-heating) and $d+d$. Note that $d+d$ will naturally occur within a $dt$ fuel mixture as a competing process. All curves are normalized to $n_e^2$ to permit direct comparison independent of density. The intersection of heating and radiation curves marks the temperature at which volumetric charged-particle heating balances radiative losses, highlighting the comparatively favorable low-temperature window of $d+t$.

\paragraph*{\bf Helium-3 spiked deuterium fusion}
We show in~\rf{fig:He3Spike}, as ${}^3$He is added to ``spike'' a deuterium plasma, the fusion heating changes because the $dd$ contribution is reduced by the $[(1-x)/(1+x)]^2$ weighting where
\begin{equation}
x\equiv \frac{n_{^{3}\text{He}}}{n_{d}+n_{^{3}\text{He}}}\,,
\end{equation}
while the $d+{}^3$He fusion channel given by
\begin{equation}
d + {}^3{\rm He} \to \alpha + p
\end{equation}
grows like $x(1-x)/(1+x)^2$. The common denominator arises because at fixed electron density, the doubly charged ${}^3$He displaces two deuterons, $n_e=n_d+2n_{^3\mathrm{He}}$. Meanwhile bremsstrahlung increases monotonically through $Z_{\rm eff}(x)=(1+3x)/(1+x)$. The result is a tradeoff where a modest ${}^3$He fraction can enhance $P_{\rm fus}/n_e^2$ over pure deuterium in the temperature range where the $d+{}^3$He reactivity is competitive, but higher ${}^3$He simultaneously raises radiative losses, so the net margin between heating and bremsstrahlung depends sensitively on both temperature and mixture fraction.

\begin{figure}
\centering
\includegraphics[width=0.75\linewidth]{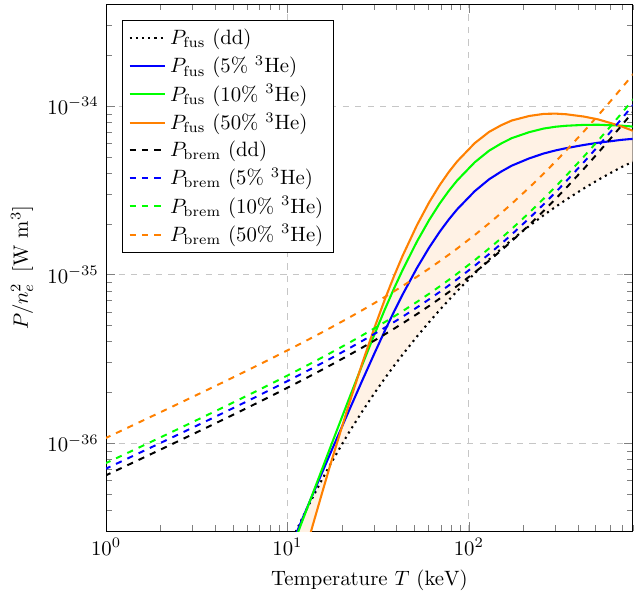}
\caption{Density-normalized fusion self-heating $P_{\rm fus}/n_e^2$ (solid/dotted) and bremsstrahlung losses $P_{\rm brem}/n_e^2$ (dashed) for $d$--${}^3$He mixtures at fixed $n_e$, shown for pure $dd$ and for 5\%, 10\%, and 50\% ${}^3$He by number, with shading indicating the gain relative to pure deuterium.}
\label{fig:He3Spike}
\end{figure}

\begin{table}[ht]
\centering
\caption{Optimal operating point of a $d$--${}^3$He mixture at fixed $n_e$ as a function of the ${}^3$He number fraction $x$. $T_{\rm opt}$ maximizes the heating-to-bremsstrahlung ratio $\mathcal{G}=P_{\rm fus}/P_{\rm brem}$ evaluated with Bosch--Hale cross sections~\cite{bosch1992improved} at $T_e=T_i=T$, $\mathcal{G}_{\max}$ is that maximum normalized to pure deuterium. The last column gives the fraction $f_n$ of released fusion energy carried by neutrons, for the primary reactions alone and in the strong burn-up limit $\chi_t,\chi_3\gg1$ of \rsec{ssubsec:SecondHeating}.}
\label{tab:He3Spike}
\begin{tabular}{cccccc}
\toprule
$x$ & $Z_{\rm eff}$ & $T_{\rm opt}$ (keV) & $\mathcal{G}_{\max}$ & $\mathcal{G}_{\max}/\mathcal{G}_{\max}(0)$ & $f_n$ (primary / full burn)\\
\midrule
0.00 & 1.00 & 153 & 1.17 & 1.0 & 0.37 / 0.36\\
0.05 & 1.10 & 119 & 3.39 & 2.9 & 0.13 / 0.27\\
0.10 & 1.18 & 117 & 4.70 & 4.0 & 0.08 / 0.22\\
0.20 & 1.33 & 116 & 5.74 & 4.9 & 0.04 / 0.14\\
0.35 & 1.52 & 116 & 5.46 & 4.6 & 0.02 / 0.09\\
0.50 & 1.67 & 116 & 4.36 & 3.7 & 0.01 / 0.05\\
\bottomrule
\end{tabular}
\end{table}

As seen in \rt{tab:He3Spike}, the optimum burn temperature is nearly independent of the mixture, $T_{\rm opt}\simeq115$--120\,keV for any $x\gtrsim0.05$, about four times the optimum of an equimolar $dt$ plasma. The gain is strongly front-loaded in $x$ with the maximum $\mathcal{G}_{\max}\simeq5.8$ occurring at $x\simeq0.24$, yet $x=0.05$ already captures about half and $x=0.10$ about three quarters of the attainable improvement before the rising $Z_{\rm eff}$ turns the trend over. The most gain therefore occurs for only a modest helium-3 fraction.

\subsubsection{Secondary heating: \texorpdfstring{$dd$}{dd} example}\label{ssubsec:SecondHeating}
In general fusion daughter products may themselves undergo fusion. This is even more true for fuels with multiple reaction paths; several distinct daughter products may themselves undergo fusion, producing additional heating beyond the primary reaction. An example is deuterium fusion:
\begin{align}
 d + d \to p + t,\qquad
 d + d \to n + {}^3{\rm He},
\end{align}
followed by secondary reactions
\begin{align}
 d + t \to \alpha + n,\qquad
d + {}^3{\rm He} \to \alpha + p .
\end{align}
In the case of an initial $dt$-mixture, the $dt$-fusion reaction would be a competing pathway rather than secondary reaction. The total heating therefore contains both primary and secondary contributions
\begin{align}
P_{\rm tot}
= P_{\rm primary}
+ P_{\rm secondary},
\end{align}
with the latter governed by confinement and burn times of the daughter species.

\paragraph*{\bf Daughter burn-up in a pure deuterium plasma}
In an initially pure deuterium plasma, tritons and ${}^3\mathrm{He}$ nuclei are produced at equal rates by the two $dd$ branches. 
The source rates are
\begin{align}
S_t = b_t R_{dd},\qquad
S_3 = b_3 R_{dd},
\end{align}
with $b_t \simeq b_3 \simeq 1/2$ and
\begin{align}
R_{dd} = \frac12 n_d^2\langle\sigma v\rangle_{dd}.
\end{align}

Each daughter species may subsequently fuse with ambient deuterons before escaping. The evolution equations are~\cite{harms2000principles,morse2018nuclear}
\begin{align}
\frac{dn_t}{dt}
&= S_t
- n_d n_t \langle\sigma v\rangle_{dt}
- \frac{n_t}{\tau_t}, \\
\frac{dn_3}{dt}
&= S_3
- n_d n_3 \langle\sigma v\rangle_{d{}^3\!He}
- \frac{n_3}{\tau_3},
\end{align}
where $\tau_t$ and $\tau_3$ are effective confinement times. In steady state, the burn fractions become
\begin{align}
f_{\rm burn}^{(t)} &= \frac{\chi_t}{1+\chi_t},
\qquad
\chi_t \equiv n_d\langle\sigma v\rangle_{dt}\tau_t, \\
f_{\rm burn}^{(3)} &= \frac{\chi_3}{1+\chi_3},
\qquad
\chi_3 \equiv n_d\langle\sigma v\rangle_{d{}^3\!He}\tau_3.
\end{align}

In the strong burn-up regime ($\chi_t\gg1$, $\chi_3\gg1$), essentially all daughter ions undergo secondary fusion before leaving the plasma. 
The secondary reaction rates then reduce to
\begin{align}
R_{dt} \simeq S_t \simeq b_t R_{dd},\qquad
R_{d{}^3\!He} \simeq S_3 \simeq b_3 R_{dd}.
\end{align}
The corresponding charged-particle heating contributions according to \req{eq:dHeQ},\req{eq:dtQ} are
\begin{align}
P_{\alpha}^{(t)} &\simeq b_t R_{dd} E_\alpha, 
\qquad &(E_\alpha &= 3.56~\mathrm{MeV}), \\
P_{\alpha}^{(3)} &\simeq b_3 R_{dd} Q_{d{}^3\!He},
\qquad &(Q_{d{}^3\!He}&\simeq 18.35~\mathrm{MeV}),
\end{align}
where for the $d+t$ channel, only the energy deposited in the $\alpha$ particle contributes to heating the plasma whereas the $d+{}^3\mathrm{He}$ channel deposits its full $Q$ in charged particles. The incremental charged energy per primary $dd$ reaction is therefore
\begin{align}
\Delta E_{\rm ch}^{(t)}
\simeq b_t E_\alpha
\simeq 1.78~\mathrm{MeV},\qquad
\Delta E_{\rm ch}^{(3)}
\simeq b_3 Q_{d{}^3\!He}
\simeq 9.18~\mathrm{MeV}.
\end{align}

The primary charged heating from $dd$ fusion alone arises from the kinetic energy of charged products in the two branches. Averaging over both channels yields
\begin{align}
E_{\rm ch}^{dd}
\simeq \frac12(4.033 + 0.82)
\simeq 2.43~\mathrm{MeV}.
\end{align}
Including full burn-up of both daughter species yields
\begin{align}
E_{\rm ch}^{dd,\;{\rm full}}
\simeq 2.43 + 1.75 + 9.18
\simeq 13.36~\mathrm{MeV},
\end{align}
more than a five-fold increase relative to primary $dd$ heating alone. This represents an upper bound corresponding to complete confinement and re-fusion of both daughter products. In practice, $\chi_t$ may exceed unity in sufficiently dense plasmas, while $\chi_3$ typically requires higher temperatures because $\langle\sigma v\rangle_{d{}^3\!He}$ peaks at substantially larger $T$ than $\langle\sigma v\rangle_{dt}$. Accordingly, the full enhancement should be regarded as an optimistic limit. Burn-up of tritons and ${}^3\mathrm{He}$ sourced in $dd$ reactions has been observed in tokamaks~\cite{heidbrink1983}. The $d{}^3\mathrm{He}$ fusion products can be diagnosed directly through charged-particle and $\gamma$-ray detection~\cite{sharapov2016}.

\begin{figure}
\centering
\includegraphics[width=0.75\linewidth]{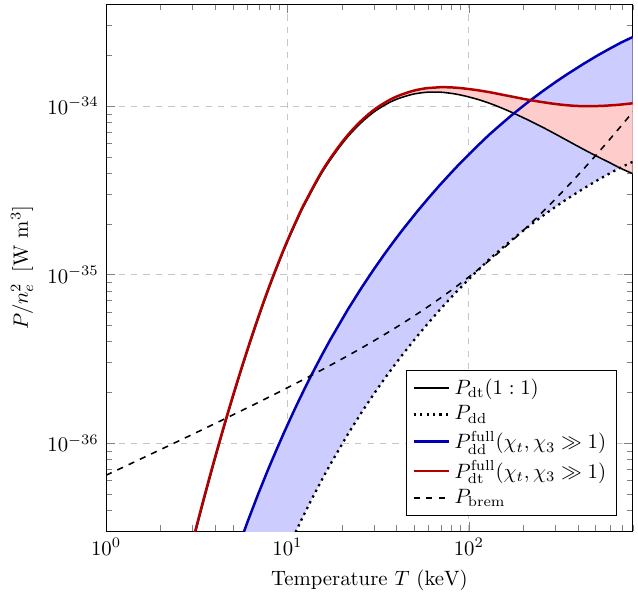}
\caption{
Density-normalized fusion heating for equimolar $dt$, pure $dd$, and fully secondary-burn configurations, compared with bremsstrahlung losses for a hydrogenic plasma ($Z_{\rm eff}=1$). The shaded bands show the gain from full secondary burn: blue between the pure-$dd$ and full-burn $dd$ curves, red between the equimolar-$dt$ and full-burn $dt$ curves. This represents an optimistic situation; see text for cautious remarks.}
\label{fig:DD_boosted}
\end{figure}

\paragraph*{\bf Comparison with baseline $dt$}
Figure~\ref{fig:DD_boosted} compares baseline and fully secondary-burned deuterium-based fuels. The black solid curve shows equimolar $dt$ heating ($n_d=n_t=n_e/2$), and the black dotted curve shows pure $dd$ heating ($n_d=n_e$). The dark blue curve represents the full $dd$ heating in the strong burn-up limit ($\chi_t,\chi_3\gg1$), where both tritium and ${}^3\mathrm{He}$ daughters undergo subsequent fusion before escape; the shaded blue region indicates the additional charged heating relative to primary $dd$. 
The dark red curve shows the corresponding full-burn $dt$ mixture, including concurrent $dd$ reactions within the equimolar $dt$ plasma and complete burn-up of both daughter species; the shaded red region denotes the incremental heating above baseline $dt$. In the strong burn-up limit $\chi_t,\chi_3\gg 1$, daughter species do not accumulate, and the plasma composition remains effectively hydrogenic, so the bremsstrahlung curve (computed with $Z_{\rm eff}=1$) is unchanged.

The fully burned $dd$ curve assumes complete confinement and re-fusion of both daughter species, $t$ and ${}^3\mathrm{He}$, corresponding to the strong burn-up limit $\chi_t,\chi_3\gg1$. This shifts the $dd$ heating curve upward by more than a factor of five relative to primary $dd$ alone. The full-burn $dt$ curve includes additional heating from concurrent $dd$ side reactions within the equimolar $dt$ mixture, again assuming complete secondary burn-up of both daughter species. While this enhancement increases the total charged heating, the baseline $dt$ channel remains intrinsically more favorable at moderate temperatures due to its substantially larger thermal reactivity.

\paragraph*{\bf Two-step reaction accounting for pure $dd$ fuel}
For a plasma that starts as pure deuterium, the secondary-burn network has exactly two steps: The primary step produces tritium and ${}^3$He in (nearly) equal proportion; of reaction products seen in Table~\ref{tab:dd_chain} $n$ escapes, $p$ is inert and both active daughters $t$ and ${}^3\mathrm{He}$ then burn with ambient deuterons in the secondary step. The secondary step reactions produce only inert reaction products $p$, escaping $n$ and $\alpha$-particles, none of which re-enter the reaction chain as further fusible reactants. The reaction network therefore closes after two steps, see Table~\ref{tab:dd_chain} which makes this explicit.

\begin{table}[ht]
\centering
\caption{Two-step reaction network for a pure $dd$ plasma in the strong burn-up limit ($\chi_t,\chi_3\gg1$). $b_t\simeq b_3\simeq 0.5$ are the $dd$ branch fractions. $Q_\mathrm{ch}$ is the energy carried by charged products; only this fraction heats the plasma. The network closes after step~1 because neither $d+t$ nor $d+{}^3\mathrm{He}$ produces a further fusible deuteron or triton from a previously pure-$d$ fuel supply.}
\label{tab:dd_chain}
\begin{tabular}{llccc}
\toprule
Step & Reaction & $Q$ (MeV) & $Q_\mathrm{ch}$ (MeV) & Heating per primary $dd$ (MeV)\\
\midrule
0 (primary) & $d+d\to p+t$ & 4.03 & 4.03 & $b_t\times 4.03 \simeq 2.02$ \\
 & $d+d\to n+{}^3\mathrm{He}$ & 3.27 & 0.82 & $b_3\times 0.82 \simeq 0.41$ \\
\midrule
\multicolumn{4}{l}{Primary charged heating $E_\mathrm{ch}^{dd}$} & $\simeq 2.43$\\
\midrule
1 (secondary) & $d+t\to\alpha+n$ & 17.59 & 3.56 & $b_t\times 3.56 \simeq 1.78$ \\
 & $d+{}^3\mathrm{He}\to\alpha+p$ & 18.35 & 18.35 & $b_3\times 18.35\simeq 9.18$ \\
\midrule
\multicolumn{4}{l}{Secondary charged heating} & $\simeq 10.93$\\
\midrule
\multicolumn{4}{l}{\textbf{Full two-step total} $E_\mathrm{ch}^{dd,\,\mathrm{full}}$} & $\simeq 13.36$\\
\bottomrule
\end{tabular}
\end{table}

The total charged heating per primary $dd$ reaction in the strong-confinement limit is therefore $E_\mathrm{ch}^{dd,\,\mathrm{full}}\simeq 13.36$~MeV, a factor of $\simeq 5.5$ above the primary-only value of 2.43~MeV. This is the enhancement depicted in Fig.~\ref{fig:DD_boosted}. Two points deserve emphasis. First, the $d+{}^3\mathrm{He}$ secondary branch contributes by far the largest share (9.18~MeV) because it deposits its full $Q$ in charged particles; the $d+t$ branch deposits only the 3.56~MeV $\alpha$-particle energy, while the 14.1~MeV neutron escapes. Second, and most important for the message of \rsec{sssec:bottleHBomb}, the secondary $d+t$ reaction produces a 14~MeV neutron for every two primary $dd$ fusions, so a pure $dd$ plasma operating in the strong-burn regime carries a similar neutron hazard to an explicit $dt$ plasma.

The claim that $dd$ fusion is `clean' is therefore incorrect. Each primary $dd$ reaction of Table~\ref{tab:dd_chain} is accompanied on average by one neutron, half of these at 14~MeV, so the extractable charged energy is $E_\mathrm{ch}^{dd,\,\mathrm{full}}\simeq13.36$~MeV per neutron produced, compared to 3.56~MeV per neutron for $dt$. Pure deuterium improves this ratio, but it does not remove the 14~MeV neutrons nor the shielding and activation burden they impose.

\subsection{The aneutronic fusion reactor powering the Solar System}\label{ssec:SolarFusion}
\subsubsection{Solar model}\label{sssec:Solar}
Our nearest example of a stable aneutronic fusion reactor is the Sun itself, which has sustained life on Earth for billions of years. The Sun's composition is dominated by primordial hydrogen and helium left over from Big Bang Nucleosynthesis, with only trace amounts of heavier nuclei (principally carbon, nitrogen, and oxygen) incorporated from earlier generations of stars.

In its core, the Sun produces energy by converting hydrogen into helium-4 via nuclear fusion. Two well-established reaction pathways operate under solar-core conditions:
\begin{enumerate}[label=(\arabic*),nosep]
\item the \textbf{proton-proton ($pp$) chains}, which dominate in stars of $\approx1.3$ solar mass or less, and
\item the \textbf{carbon-nitrogen-oxygen (CNO) cycle}, a catalytic loop that requires the presence of heavier nuclei and becomes increasingly important in more massive or metal-rich stars. The CNO cycle is only possible in relatively recent stellar populations enriched with ``recycled ashes'' from previous generations of stellar nucleosynthesis.
\end{enumerate}

The confining force that sustains fusion in the Sun is provided by its own gravity, which counterbalances the outward radiative pressure generated by the fusion reactions. This hydrostatic equilibrium allows the solar core to operate in a quasi-steady state for billions of years.

The present-day solar reactor generates a luminosity
\begin{align}
L_\odot \simeq 3.8\times10^{26}~\mathrm{W},
\end{align}
and has done so without interruption for approximately $4.6$ billion years. Our planet exists within the narrow range of conditions made possible by the continuous operation of our ``local'' aneutronic stellar core reactor.

\subsubsection{The proton-proton (\texorpdfstring{$pp$}{pp}) chains}\label{sec:PPChains}

In~\rf{fig:PPChain}, we show branching ratios and detailed sub-chains of the so called ($pp$) fusion cycle: 
In our Sun the dominant fusion cycle begins with the production of $d$ by weak interactions; the primordial deuterium was burned long ago. On the other hand our Sun does produce the long lived (4500 days half-life) yet heavier hydrogen isotope ${}^{3}_{1}\mathrm{H}^+\equiv t=pnn$ called tritium. $t$ is produced in half of $dd$ fusion reactions. The other half of $dd$ reactions produce the stable light isotope of helium ${}^{3}_{2}\mathrm{He}^{++}\equiv ppn$ accompanied by emission of a relatively energetic free neutron $n$. However, in a star any produced $d$ rapidly fuses with the abundant light $p$ to form ${}^{3}_{2}\mathrm{He}^{++}$. Because any $d$ that becomes available is burned in this way upon production, the {\em two deuteron} $dd$ fusion reactions will occur extremely rarely. This assures the absence of neutron producing process at this fusion step and it turns out this is the case for the two entire fusion cycles that have been explored.

The produced ${}^{3}_{2}\mathrm{He}^{++}$ can accumulate and once a significant fraction is achieved, two ${}^{3}_{2}\mathrm{He}^{++}$ can meet and fuse, producing an $\alpha$ and two protons, with production of an $\alpha$ closing the fusion cycle. This last burn-step needs to overcome a relatively large Coulomb wall between the two helium nuclei and thus clearly requires extreme temperature conditions along with a large abundance of ${}^{3}_{2}\mathrm{He}^{++}$. Another side chain is the reaction of a newly produced $d$ with residual ${}^{3}_{2}\mathrm{He}^{++}$ producing $\alpha$ and a $p$. Very recently, a very high abundance of ${}^{3}_{2}\mathrm{He}^{++}$ was observed in a spectacular solar flare event~\cite{Bucik:2025}. One can only wonder how ${}^{3}_{2}\mathrm{He}^{++}$ from the interior of the Sun got to be a surface flare. On this note, the moon surface is a well known depository of solar wind carried ${}^{3}_{2}\mathrm{He}^{++}$, which we have not found on Earth. This demonstrates the need for further study of ${}^{3}_{2}\mathrm{He}^{++}$ based fusion reaction (cycles) and inventory.

\begin{figure}[ht]
\centering
\includegraphics[width=0.95\linewidth]{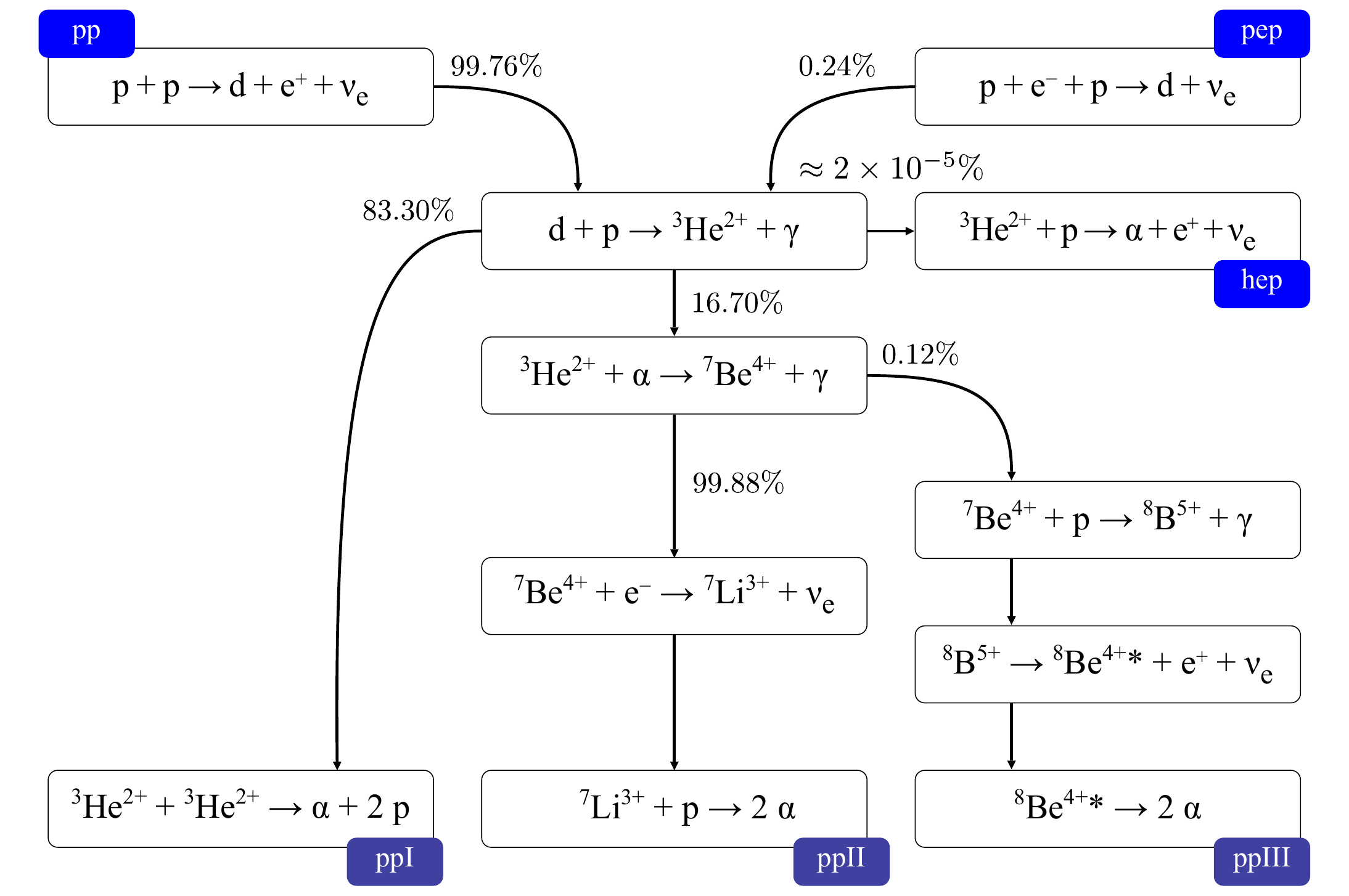}
\caption{The proton--proton (\textit{pp}) chains that power main-sequence stars. Charge states are shown explicitly as a reminder that stellar matter is fully ionized with every species a bare nucleus. The asterisk marks the excited state of ${}^{8}$Be. Adapted from Ref.~\cite{SzamPPCycle:2012} under \href{https://creativecommons.org/licenses/by-sa/2.5/}{CC BY-SA-2.5}; revisions by the authors.}
\label{fig:PPChain}
\end{figure}

Figure~\ref{fig:PPChain} shows the dominant {ppI} branch (83.30\%) where two ${}^3$He nuclei fuse to ${}^4$He; the {ppII} branch (16.70\%) proceeding through ${}^7$Be electron capture; and the rare {ppIII} branch (0.12\%) involving ${}^8$B decay~\cite{Adelberger:2010qa}. Also shown are the less frequent \textit{pep} and \textit{hep} channels. Together these cycles sustain stellar energy generation and dictate the spectrum of solar neutrinos.

The net reaction of the $pp$ chain converts four protons into one helium-4 nucleus
\begin{align}
4p + 2e^- \;\longrightarrow\; \alpha + 2\nu_{e} + 26.7~\mathrm{MeV},
\end{align}
of which approximately $26~\mathrm{MeV}$ per $\alpha$-particle is released as usable thermal energy; the remainder is carried away by neutrinos. The $pp$ chain proceeds through both weak and strong interactions. The initial step $(99.77\%)$ involves the slow weak interaction
\begin{align}
p + p \;\xrightarrow{\ \mathrm{weak}\ }\; d + e^{+} + \nu_{e},
\end{align}
which converts two protons into a deuteron, emitting a positron and an electron neutrino. A small fraction $(0.23\%)$ of stellar deuterium is produced via electron capture. Subsequent steps are governed by the strong interaction: the newly formed deuteron fuses with another proton to produce ${}^{3}\mathrm{He}$, and two ${}^{3}\mathrm{He}$ nuclei can then combine to form ${}^{4}\mathrm{He}$.

This multiplicity of side chains reflects the probabilistic nature of quantum mechanics: while the ppI branch dominates in solar conditions, the less probable ppII and ppIII channels contribute to the production of solar neutrinos and, in rare cases, to the synthesis of heavier light nuclei.

\subsubsection{The carbon-nitrogen-oxygen (CNO) cycles}\label{sec:CNOcycle}

The carbon-nitrogen-oxygen (CNO) cycles constitute a family of catalytic aneutronic fusion processes in which hydrogen is converted into helium through a network of proton captures and $\beta^{+}$ decays operating on nuclei of C, N, and O. First identified by Hans Bethe (recognized by the 1967 Nobel Prize in Physics), the CNO cycles operate alongside the proton-proton ($pp$) chains in main-sequence stars and become the dominant energy-generation mechanism in stars above $\sim 1.3\,M_{\odot}$ due to their much steeper temperature dependence.

\begin{figure}[ht]
\centering
\includegraphics[width=0.75\linewidth]{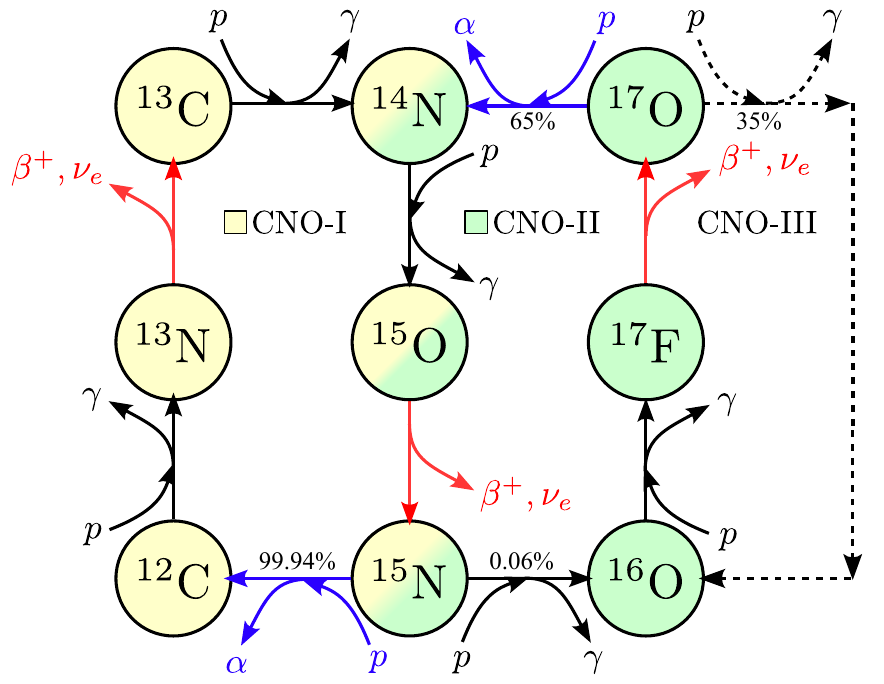}
\caption{The carbon-nitrogen-oxygen (CNO) ``bicycle,'' showing the catalytic hydrogen-burning cycles CNO-I (yellow) and CNO-II (green). The small upper branch through ${}^{17}$O(p,$\gamma$)${}^{18}$F opens the path to the hotter CNO-III sequence (not depicted in detail) and further non-cyclic processes. Black arrows denote proton captures $(p,\gamma)$, red arrows $\beta^+$ decays, and blue arrows the $(p,\alpha)$ reactions that close each cycle; dashed arrows mark the branch toward CNO-III.}
\label{fig:CNO_Bicycle}
\end{figure}

At its core, the CNO mechanism functions as a catalytic loop: four protons are consumed and one ${}^{4}\mathrm{He}$ nucleus is produced, while the catalyst ${}^{12}$C is restored at the end of each cycle,
\begin{align}
4p \;\longrightarrow\; {}^{4}\mathrm{He} + 2e^{+} + 2\nu_{e} + \gamma + \text{(catalyst restored)}.
\end{align}
The principal reaction sequence, known as CNO-I, cycles material through
\begin{equation}
{}^{12}\mathrm{C} \to {}^{13}\mathrm{N} \to {}^{13}\mathrm{C} \to {}^{14}\mathrm{N} \to {}^{15}\mathrm{O} \to {}^{15}\mathrm{N} \to {}^{12}\mathrm{C} 
\end{equation}
Two steps in the process occur through $\beta^{+}$ decay
\begin{equation}
{}^{13}\mathrm{N}\rightarrow {}^{13}\mathrm{C} \text{ and }
{}^{15}\mathrm{O}\rightarrow{}^{15}\mathrm{N}\,,
\end{equation}
which introduce characteristic delays of 2--10~minutes, requiring the cycle to operate in a quasi-steady state. Under present solar conditions, CNO contributes only about $1.7\%$ of the Sun's helium production, but it dominates energy generation in stars with hotter cores~\cite{krane1991introductory}.

Figure~\ref{fig:CNO_Bicycle} illustrates the structure of the CNO ``bicycle,'' which consists of two tightly coupled catalytic loops. The CNO-I sequence closes when ${}^{15}\mathrm{N}$ captures a proton, restoring the catalyst ${}^{12}\mathrm{C}$ and producing an $\alpha$ particle. A competing channel, ${}^{15}\mathrm{N}(p,\gamma){}^{16}\mathrm{O}$ provides the entry point to the CNO-II sequence, which cycles material through ${}^{16}\mathrm{O}$, ${}^{17}\mathrm{F}$, and ${}^{17}\mathrm{O}$. At the next branch point, ${}^{17}\mathrm{O}$ may either return to the CNO-I loop through ${}^{17}\mathrm{O}(p,\alpha){}^{14}\mathrm{N}$, or proceed upward via ${}^{17}\mathrm{O}(p,\gamma){}^{18}\mathrm{F}$, thereby initiating the hotter CNO-III chain. Thus, the relative strengths of the $(p,\alpha)$ and $(p,\gamma)$ reactions at ${}^{15}$N and ${}^{17}$O govern how nuclear flow is partitioned among CNO-I, CNO-II, and CNO-III.

The branching between these cycles is set by the ratio of thermal reaction rates $\langle\sigma v\rangle(T)$, which we evaluate at the solar core temperature $T_{\odot}\simeq1.5\times10^{7}\,\mathrm{K}$ ($T_{9}\simeq 0.015$). Using NACRE~II rates~\cite{Xu:2013fha}, the ${}^{15}$N branch overwhelmingly favors the CNO-I branch
\begin{equation}
\frac{\langle\sigma v\rangle({}^{15}\mathrm{N}(p,\gamma){}^{16}\mathrm{O})}
 {\langle\sigma v\rangle({}^{15}\mathrm{N}(p,\alpha){}^{12}\mathrm{C})}
 \Bigg|_{T_{9}=0.015}
 \approx 5.9\times10^{-4},
\end{equation}
so that only ${\sim}0.06\%$ of cycles proceed into CNO-II at solar core temperatures. In contrast, the branching at ${}^{17}$O is significantly less asymmetric. Taking the median rates reported in ETR25~\cite{Iliadis:2026hww,iliadis_2025_17610211}, the branching occurring when the catalyst becomes ${}^{17}\mathrm{O}$ yields
\begin{equation}
\frac{\langle\sigma v\rangle({}^{17}\mathrm{O}(p,\gamma){}^{18}\mathrm{F})}
 {\langle\sigma v\rangle({}^{17}\mathrm{O}(p,\alpha){}^{14}\mathrm{N})}
 \Bigg|_{T_{9}=0.015}
 \approx 0.54,
\end{equation}
so that approximately $35\%$ of the reacting ${}^{17}$O proceeds through the $(p,\gamma)$ channel, the remaining $65\%$ returning to CNO-I. This non-negligible ``leakage'' allows flux into the CNO-III sequence, especially in stars with hotter cores. While CNO-I dominates in the Sun, CNO-II becomes increasingly important at higher temperatures, and CNO-III provides an upward pathway for material to cycle toward heavier isotopes in a network responsible for hydrogen burning in intermediate and high-mass stars.


\subsubsection{The triple-alpha process}\label{sec:TripleAlpha}

The triple-$\alpha$ process ignites at core temperatures $T \gtrsim 10^{8}~\mathrm{K}$ and marks the transition from hydrogen burning to helium burning in stellar evolution~\cite{Suno:2016fjb}. Unlike the $pp$ chains and the CNO cycle, which convert hydrogen into helium, the triple-$\alpha$ process fuses three ${}^{4}\mathrm{He}$ nuclei (alpha particles) into a single ${}^{12}\mathrm{C}$ nucleus
\begin{align}
3\,{}^{4}\mathrm{He} \;\longrightarrow\; {}^{12}\mathrm{C} + \gamma + 7.27~\mathrm{MeV}.
\end{align}
The reaction proceeds in two steps. First, two ${}^{4}\mathrm{He}$ nuclei combine to form unstable ${}^{8}\mathrm{Be}$, which has a lifetime of only $\sim 10^{-16}~\mathrm{s}$. Second, during this brief interval, if a third ${}^{4}\mathrm{He}$ collides with ${}^{8}\mathrm{Be}$, the three-body system can pass through the resonant Hoyle state~\cite{Hoyle:1954zz,Chernykh:2007zz,Epelbaum:2011md} of ${}^{12}\mathrm{C}$ and decay to its ground state with the emission of a gamma ray. This aneutronic process is responsible for the first major buildup of carbon, adding to relatively small abundance created in the Universe after BBN.

\subsection{Plasma screening and heavy-nuclear catalysis}\label{ssec:PlasmaScreening}
\subsubsection{Heavy nuclear catalysts}\label{sssec:HeavyCatalyst}

Heavy highly-charged nuclei sitting inside a dense bound electron cloud are capable of lowering the Coulomb wall seen by nearby colliding light nuclei~\cite{Grayson:2025kva}. This polarization screening occurs with electron densities near the $Z=79$ gold ion and is distinctly different from the textbook weak-field Salpeter screening regime. This situation is depicted in \rf{fig:AuCatalyst} where two light nuclei approach one another inside the dense cloud of electrons which a heavy, highly charged nucleus gathers around itself in a hot plasma. Within that cloud the Coulomb barrier separating the light reactants is measurably lowered and they tunnel more readily. The heavy nucleus then acts as a catalyst, and being far more massive than the plasma is hot, $m_\mathrm{Au}\gg T$, it does not move on the timescale of the reaction. The high-$Z$ context is a counterpart of the self-consistent screening by light nuclei studied for Big Bang nucleosynthesis (BBN)~\cite{Grayson:2024uwg}, where the correction to thermonuclear reaction rates was found to be small, but very strongly dependent on the nuclear charge $Ze$.

\begin{figure}
\centering
\includegraphics[width=0.40\linewidth]{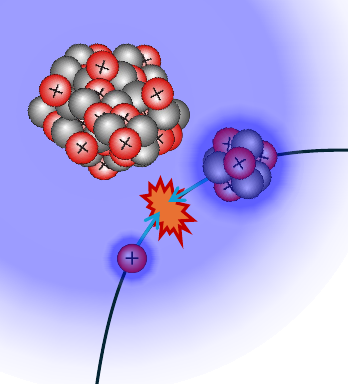}
\caption{Depiction of heavy nuclei catalyzing fusion reactions. A proton and a boron nucleus collide within the large screening cloud of electrons generated by a stationary heavy nucleus such as Au ($Z=79$). Reprinted from Ref.~\cite{Grayson:2025kva} under \href{https://creativecommons.org/licenses/by/4.0/}{CC~BY~4.0}.}
\label{fig:AuCatalyst}
\end{figure}

Gold is the natural candidate for study for heavy nuclear catalysis as it is already present in plasmas of current interest to nuclear fusion:
\begin{itemize}
 \item[(i)] In the gold films used as targets in laser-fusion experiments;
 \item[(ii)] in the hohlraum walls of indirect-drive inertial confinement fusion, see \rsec{sssec:NIF};
 \item[(iii)] and introduced deliberately as the plasmonic nanoparticles of the NAPLIFE targets, see \rsec{sssec:nanoF} and \rsec{ssec:Plasmonics} below.
\end{itemize}

Whether such a dense electron cloud of a heavy gold nucleus matters depends on where in the potential the reaction is actually sampled. In a thermal environment nuclear fusion reactions occur near the Gamow energy arising from the competition between tunneling probability and probability of collision at given relative energy
\begin{equation}\label{eq:GamowE}
E_G = \left(\frac{(\pi T\,Z_1Z_2\,\alpha)^2\,\mu_\mathrm{r}}{2}\right)^{1/3}\,.
\end{equation}
Here $\mu_\mathrm{r}$ is the reduced mass of the two colliding light nuclei.

\begin{figure}[ht]
\centering
\includegraphics[width=0.72\linewidth]{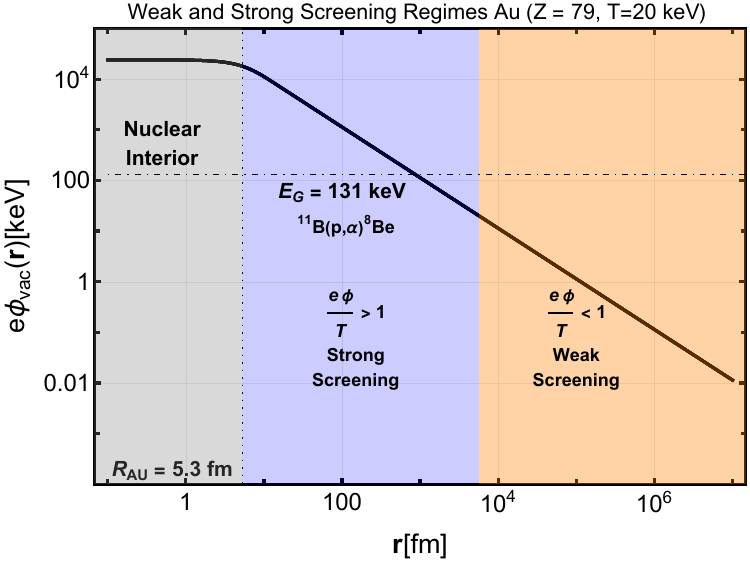}
\caption{Vacuum potential energy $e\phi_\mathrm{vac}(r)$ of a gold nucleus ($Z=79$) in a $T=20$~keV plasma. In the orange region where $e\phi/T<1$, weak-field Debye-H\"uckel screening applies. In the purple region $e\phi/T>1$ and the screening corrections become nonlinear. The gray band is the nuclear interior, $R_\mathrm{Au}=5.3$~fm, bounded by the dotted vertical line. The solid curve is $e\phi_\mathrm{vac}(r)$; the dash-dotted horizontal line marks the Gamow energy $E_G=131$~keV of ${}^{11}\mathrm{B}(p,\alpha){}^{8}\mathrm{Be}$. Reprinted from Ref.~\cite{Grayson:2025kva} under \href{https://creativecommons.org/licenses/by/4.0/}{CC~BY~4.0}.}
\label{fig:ScreenRegimes}
\end{figure}

\subsubsection{Potential-dependent screening mass}\label{sssec:ScreeningMass}
Ordinary Debye-H\"uckel plasma screening assumes the electromagnetic potential energy $V(r)\equiv e\phi(r)$ is everywhere small compared to the plasma temperature $T$, $V(r)/T\ll1$, so that the induced charge density is linear in $\phi$~\cite{Barker:2002mfp}. In some situations this approximation is not sufficient. The solution is to solve the Poisson equation self-consistently, keeping the full nonlinear dependence of the induced charge on the potential. Additionally, for strong-field screening it is necessary to use relativistic Fermi--Dirac statistics for the electrons rather than the Boltzmann approximation; see Ref.~\cite{Birrell:2024bdb} for an improved mathematical treatment of FD limits.

Writing the unit-less potential $\Phi\equiv V/T$ and introducing a potential-dependent screening mass $m_s(\Phi)$, the rescaled Poisson equation takes the compact form
\begin{equation}\label{eq:PBscreen}
-\nabla^2\Phi(r) + \frac{m_s^2(\Phi)}{(\hbar c)^2}\,\Phi(r) = P_\mathrm{ext}(r)\,,
\end{equation}
which reduces to Debye-H\"uckel theory when $\Phi\ll 1$ and to the Thomas--Fermi atomic model of electron cloud as $T\to 0$. In the ultrarelativistic, strong-field limit relevant near the nuclear surface the screening mass itself becomes~\cite{Elze:1980er,Kodama:2002xjg}
\begin{equation}\label{eq:screenmass}
m^2_{s,\mathrm{FD}}(\Phi) \approx \frac{4\alpha T^2}{3\pi}\left[\pi^2 + \Phi^2(r)\right]\,,
\qquad
m/\Phi T\ll 1\,,
\end{equation}
a weak-field (Debye) term plus a nonlinear correction that grows with the local potential rather than saturating exponentially as the older Boltzmann treatment predicts. Over the range probed by tunneling, see \rf{fig:ScreenMassFD}, the simplified Boltzmann statistics overestimates the screening mass, and hence the reaction-rate enhancement, by orders of magnitude relative to the correct Fermi--Dirac result~\cite{Grayson:2024uwg,Grayson:2025kva}.

\begin{figure}[ht]
\centering
\includegraphics[width=0.72\linewidth]{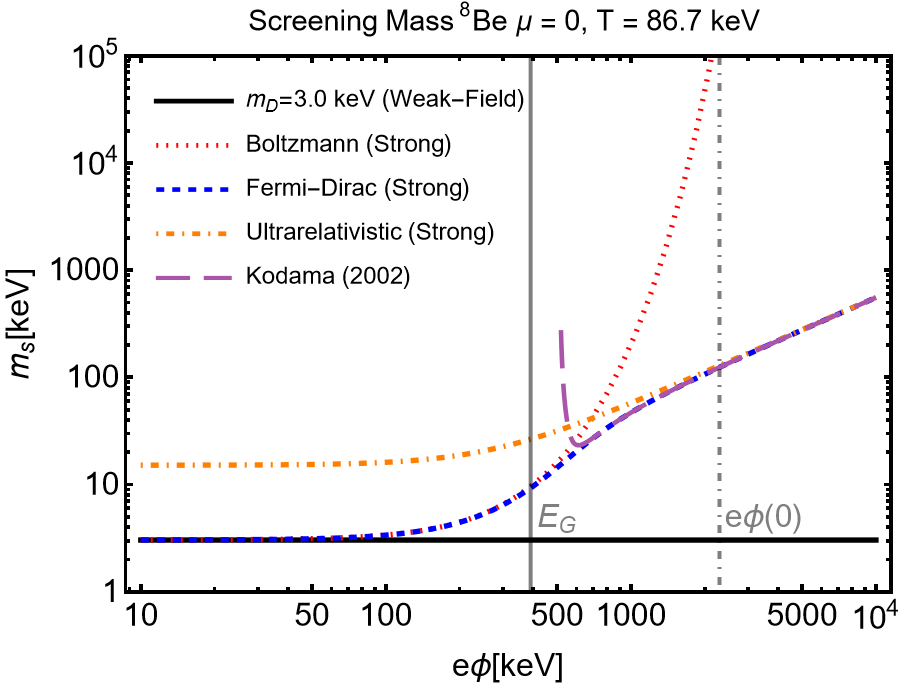}
\caption{Effective screening mass $m_s$ around a finite-sized $^4$He (\,$^8$Be\,) nucleus at the (illustratively chosen) BBN temperature $T=86.7$~keV, as a function of the local potential $e\phi$. The self-consistent Fermi--Dirac result (blue dashed) tracks the weak-field Debye value $m_D$ (black) at small $\Phi$ but rises only as a power law at large $\Phi$; the traditionally used Boltzmann approximation (red dotted) instead diverges exponentially, over-predicting screening in the tunneling-relevant window bounded by the Gamow energy $E_G$ and the on-surface potential $e\phi(0)$ (vertical gray solid and dash-dotted lines, respectively). The purple long-dashed curve labeled Kodama (2002) follows Ref.~\cite{Kodama:2002xjg}. Reprinted from Ref.~\cite{Grayson:2024uwg} under \href{https://creativecommons.org/licenses/by/4.0/}{CC~BY~4.0}.}
\label{fig:ScreenMassFD}
\end{figure}

\subsubsection{Strong field fusion rate enhancement}\label{sssec:FusionEnhancement}
Reaction-rate enhancement in the weak field limit follows the Salpeter form~\cite{Salpeter:1954nc}, in which the fractional change in reaction rate is set by the induced potential evaluated at the origin
\begin{equation}\label{eq:salpeter}
\mathcal{F}_\mathrm{sc} \equiv \frac{R_\mathrm{sc}}{R_\mathrm{vac}} \sim \exp\!\left(\frac{e\phi_\mathrm{ind}(0)}{T}\right)\,.
\end{equation}
We can then apply this same self-consistent machinery to the fully ionized gold catalyst of \rsec{sssec:HeavyCatalyst}, embedded in the $\sim$keV electron plasma. Although \req{eq:PBscreen} is nonlinear, the strong-screening and weak-field (Debye) contributions to the induced potential near the origin turn out to be additive, so the weak-field part can be subtracted at each temperature to isolate a strong-screening contribution $e\phi_\mathrm{ind}(0)\simeq14.2$~keV that is independent of plasma temperature. This is close to the zero-temperature potential shift of the bound-electron cloud of a gold atom
\begin{equation}
 e\phi_\mathrm{HF}(0)=\alpha^2Z^{4/3}a_\mathrm{TF}m_e\simeq14.7\,\mathrm{keV}
\end{equation}
obtained from the Hartree--Fock screened-Coulomb potential shifts tabulated in Ref.~\cite{garrett1967potential} ($a_\mathrm{TF}\approx1.598$ at $Z=79$). For two light reactants of charge $Z_1,Z_2$ tunneling near a heavy catalyst of charge $Z_3$, the full multi-body enhancement factor becomes
\begin{equation}\label{eq:goldF}
\mathcal{F}_\mathrm{sc} = \exp\!\left[\frac{1}{T}\left(Z_1Z_2\,\alpha\, m_{s,\mathrm{FD}}(r_G) + Z_3\frac{Z_1+Z_2}{2}\,\alpha\, m_{s,\mathrm{FD}}(r_G) + \frac{Z_1+Z_2}{2}\,\phi_\mathrm{ind}(r_G)\right)\right]\,,
\end{equation}
evaluated at the Gamow turning point $r_G$. Notably, the classical turning point $r_{E_G}$ belonging to this energy lies at internuclear distances of order femtometers. The controlling parameter is then whether the ratio $Ze\phi(r_{E_G})/T$ exceeds unity at that turning point. See \rf{fig:ScreenRegimes} for gold where the weak-field condition $e\phi/T\ll1$ which underlies Debye-H\"uckel theory holds only beyond $\sim10^4$\,fm, whereas the Gamow energy $E_G=131$\,keV of the ${}^{11}\mathrm{B}(p,\alpha){}^{8}\mathrm{Be}$ reaction at $T=20$\,keV falls deep inside the region where $e\phi/T>1$ denoting the strong-screening region. \rf{fig:GoldEnhance} shows the resulting enhancement for several aneutronic and near-aneutronic reactions as a function of plasma temperature. Because catalysis grows with the charge of the reacting pair, $p+{}^{11}\mathrm{B}$ benefits the most, while $d$-based reactions are enhanced more modestly.

\begin{figure}[ht]
\centering
\includegraphics[width=0.72\linewidth]{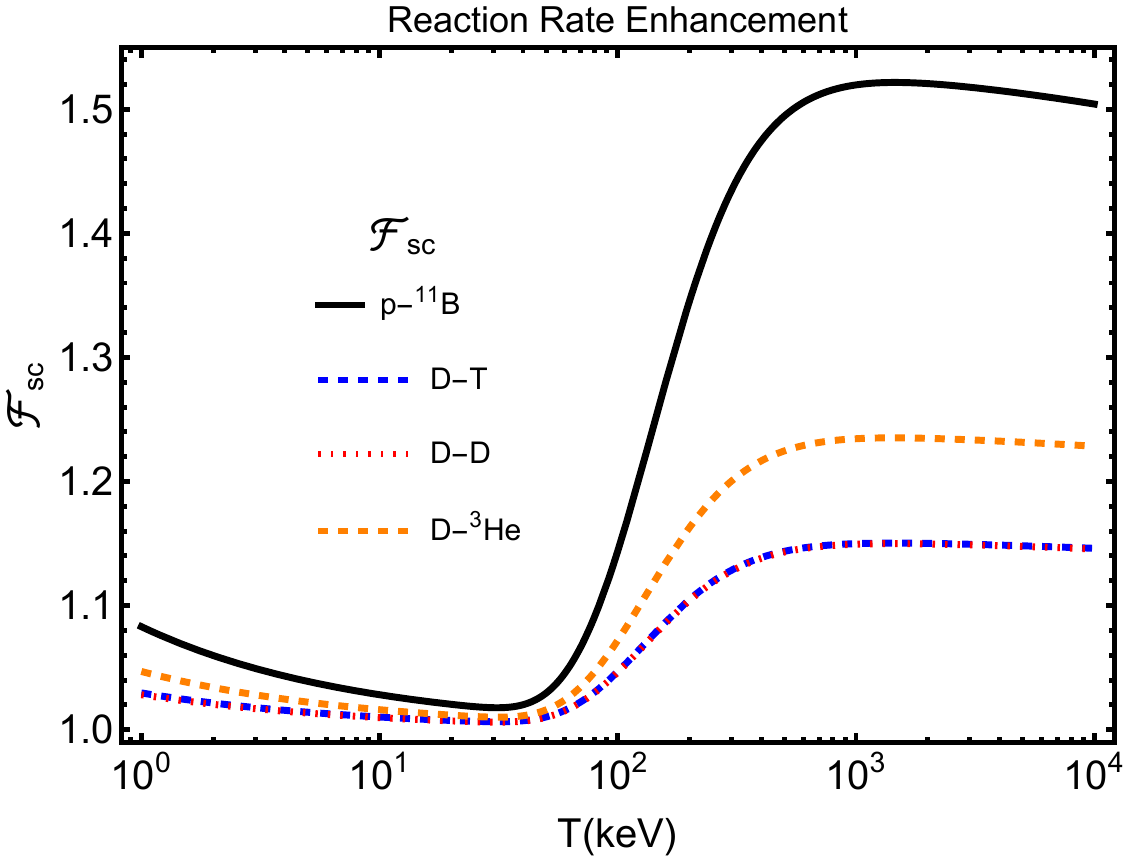}
\caption{Reaction-rate enhancement factor $\mathcal{F}_\mathrm{sc}$ [\req{eq:goldF}] near a fully ionized gold catalyst, as a function of plasma temperature, for $p$-${}^{11}$B (black solid), $d$-$t$ (blue dashed), $d$-$d$ (red dotted), and $d$-${}^3$He (orange dashed). All curves rise slowly at low $T$, where the temperature-independent strong-screening term of \req{eq:screenmass} dominates, then climb sharply above $\sim100$~keV as thermal electrons increasingly penetrate the gold polarization cloud, before flattening once $T$ approaches $m_ec^2$. Reprinted from Ref.~\cite{Grayson:2025kva} under \href{https://creativecommons.org/licenses/by/4.0/}{CC~BY~4.0}.}
\label{fig:GoldEnhance}
\end{figure}

An enhancement factor of order unity may seem modest, but note that \req{eq:goldF} presumes thermal reactants, whose kinetic energy carries them only into the outer reaches of the gold cloud. Laser-accelerated, non-thermal ions penetrate much further, up towards the nuclear surface, where the enhancement is instead governed by
\begin{equation}
 \mathcal{F}_\mathrm{sc}=\exp\left(\frac{\langle Z\rangle e\phi_\mathrm{ind}(R)}{T}\right)\,,
\end{equation}
with $e\phi_\mathrm{ind}(R)\simeq14$~keV and where $\langle Z\rangle$ is the average charge of the reactants. A more in depth study of screening effects in different fusion environments seems indicated and may provide an explanation for the long-standing ``anomalous screening'' seen in low-energy astrophysical $S$-factor measurements~\cite{Grayson:2025kva}.

\section{High-tech Fusion}\label{sec:newFusion}
\subsection{Particle beams and fusion}\label{ssec:exploreLas}
\subsubsection{Aneutronic \texorpdfstring{{$d+^{3\!}\mathrm{He}$}}{d+3He} fusion}\label{sssec:d3Hefuesion}
Let us look more closely at $d+^{3\!}\mathrm{He}\to \alpha+p$ aneutronic fusion which yields 18.35 MeV: 80\% (14.68 MeV) of the fusion yield is carried by a highly energetic proton, and 20\% (3.67 MeV) by the $\alpha$-particle. The reaction cross-section peaks at $E_\mathrm{CM}=250\,\mathrm{keV}$ at a value of 0.85 barns which is significantly lower compared to the $dt$ 5 barn cross-section at 4 times lower reaction energy. The scientific reasons for this are well understood: The Coulomb wall is twice as high with Helium, and the resonant reaction peak is, due to greater electromagnetic interaction, now inside ${}^3\mathrm{He}$, also pushed up.

The thermal burn of $d+^{3\!}\mathrm{He}$ would be accompanied by $dd$ fusion which is `neutronic'. Thus only a non-thermal, non-confined directed particle motion induced fusion reaction can be considered. This would be a ${}^3\mathrm{He}$ particle beam hitting a $d$ target; this often is referred to as `inverse' geometry as the heavier particle is in motion. This allows maintaining control of the inventory of the much rarer ${}^3\mathrm{He}$ and allows better control of forward projected charged particle fusion products, especially useful in space propulsion applications.

\paragraph*{\bf Direct conversion of charged-product energy}
In literature one often notes a fusion company, Helion Energy, which promises direct harvesting of (electrical) power from the motion of the produced charged particles. Abundant fusion reactions are expected in collective plasma cloud collisions. Both promises have steep technological challenges to overcome. What is important to note is that a fusion reactor system with a highly effective conversion of electrical power to reactant particle beam followed by an equally highly effective conversion of fusion produced kinetic charged particle energy into the electrical power could possibly solve the fusion problem, provided that fusion conditions are indeed achieved.

The nuclear fusion energy output in such a system implies amplification by about a factor 20. Therefore in principle a device providing a net electrical gain in each cycle, i.e. gain per injected particle energy balance $\Delta Q_c>0$ can be achieved -- we keep in mind that we assumed that electronic stopping power slowing down fusion reactants can be technologically controlled. If not, any directed particle energy will end warming up the electron cloud with marginal fusion outcome.

\paragraph*{\bf Cycle gain estimate}
A fusion cycle yield is derived from the following sequel steps:
\[
\text{Electricity}\to\text{particle beam}\to\text{fusion}\to\text{electricity.}
\]
The amplification factor per cycle is
\[
\frac{\text{fusion energy out}\,+\,\text{beam energy}\,-\,\text{beam losses}}{\text{beam energy}}
\]
Loss is predominantly due to heating of reactants and energy used in mitigation strategy. Considering particle beams, given the required size of a beam-target system and speed of light, each cycle lasts as little as $\tau_c$ = 20 nanoseconds (6 meters at speed of light in vacuum). Because there can be many cycles per second, a small generated power excess converts to a large power (energy per time) number:
 \begin{align}
 P_c=\frac{\Delta Q_c}{\tau_c}\,, \qquad P= N_cP_c\,.
\end{align}
$P_c$ is the power produced when one injects one particle every $\tau_c$ into the reaction volume. For a larger injected particle number $N_c$ the produced power $P$ scales up. A continuous particle beam comprising a current $j$ has $N_c=j\tau_c$ and thus power $P$ produced is
 \begin{align}
 P= N_cP_c\,, \qquad P=j\Delta Q_c\,.
\end{align}
Clearly the last result is independent of realization and applies to any system that allows the fusion of reactants without large heat loss.

A ${}^3\mathrm{He}$ double charged ion current, with a magnitude $qj= (q/e) 5 \mathrm{A}$ with ${}^3\mathrm{He}$ and a total kinetic energy of 625 keV/particle would drive a fusion target with a beam power of 3.1 MW. Each ${}^3\mathrm{He}+d$ fusion reaction releases 18.35 MeV, thus total fusion and particle energy in each step is about 19\,MeV counting the energy of the particle beam. This energy one must harness as effectively as possible in a direct conversion to electricity. Assuming per full cycle losses of not more than 35\% thus $\Delta Q_c=12.5\,\mathrm{MeV}$, such a fusion reactor system would generate 125\,MW electrical power output; the amplification factor 20 is here simply the ratio of 12.5/0.625=20. As a caveat, to assure small stopping power losses of ${}^3\mathrm{He}$, one has to move away (laser pulse shots) electrons in the $d$-target reaction zone very efficiently, to assure the 35\% loss is realistic.

\paragraph*{\bf Why $dt$ does not work in this scheme}
The reader can easily play with preferred performance numbers to see why anyone who was able to demonstrate effective electricity generation from charged fusion products $p$ and $\alpha$ can plan to build a ${}^3\mathrm{He}$ particle beam based power plant. On first sight, the reader will want to do the same for $t+d$ fusion reaction which has a 7 times bigger reaction cross-section at 4 times lower energy. Clearly much simpler technological demands arise.

However, 80\% of energy yield is carried away by difficult to control ultra fast neutral neutrons. Therefore our amplification factor would be, all other factors being comparable, around 1, allowing for the losses we discussed, while cost of energy would be inflated by ultrafast neutron handling accompanied by tritium breeding and handling. This said: A Helion style $dt$ reactor will be much easier to build compared to ${}^3\mathrm{He}+d$. However the actual usable energy will be derived from the ultrafast 14 MeV neutron pulses. The specific advantage of Helion harvesting charged particle kinetic energy into electricity is manifestly absent.

\subsubsection{Laser-driven two target proton--boron (pB) fusion}\label{sssec:LULIpB}
The proton--boron $p+{}^{11}\mathrm{B}$ nuclear fusion reaction system has been explored in nuclear laboratories using proton particle beams thoroughly, seen its potential importance in aneutronic fusion~\cite{Stave:2011zz,Sikora:2016pB}. In a series of experiments at the Laboratoire pour l'Utilisation des Lasers Intenses (LULI) in 2013 these nuclear science insights were used in experiments which have demonstrated that high-intensity, short-pulse lasers can drive $p+{}^{11}\mathrm{B}$ fusion in a novel non-equilibrium regime, an effort in which one of us (JR) participated~\cite{Labaune:2013dla}.

The scheme employs a two-pulse sequence:
\begin{enumerate}[nosep]
\item A long ($\sim$ns) laser pulse generates a hot, expanding plasma from a boron target, expelling electrons and preparing a strongly ionized medium.
\item A short ($\sim$ps) pulse delivers a beam of energetic protons into this plasma before thermal equilibration, initiating $p+{}^{11}\mathrm{B}$ fusion reactions.
\end{enumerate}
This setup, depicted in~\rf{fig:PBFusion}, has successfully produced highly intense signature $\alpha$ particles of the pB reaction. This success has led to the issue of a patent~\cite{labaune2014production}. The alpha-flux yield from this reaction is shown in~\rf{fig:YieldpB} over various experiments. Nuclear data acquired over the last decade continue to refine the understanding of laser driven $p+{}^{11}\mathrm{B}$ fusion reactions~\cite{sciscio2025laser}.

\begin{figure}[ht]
\centering
\includegraphics[width=0.95\linewidth]{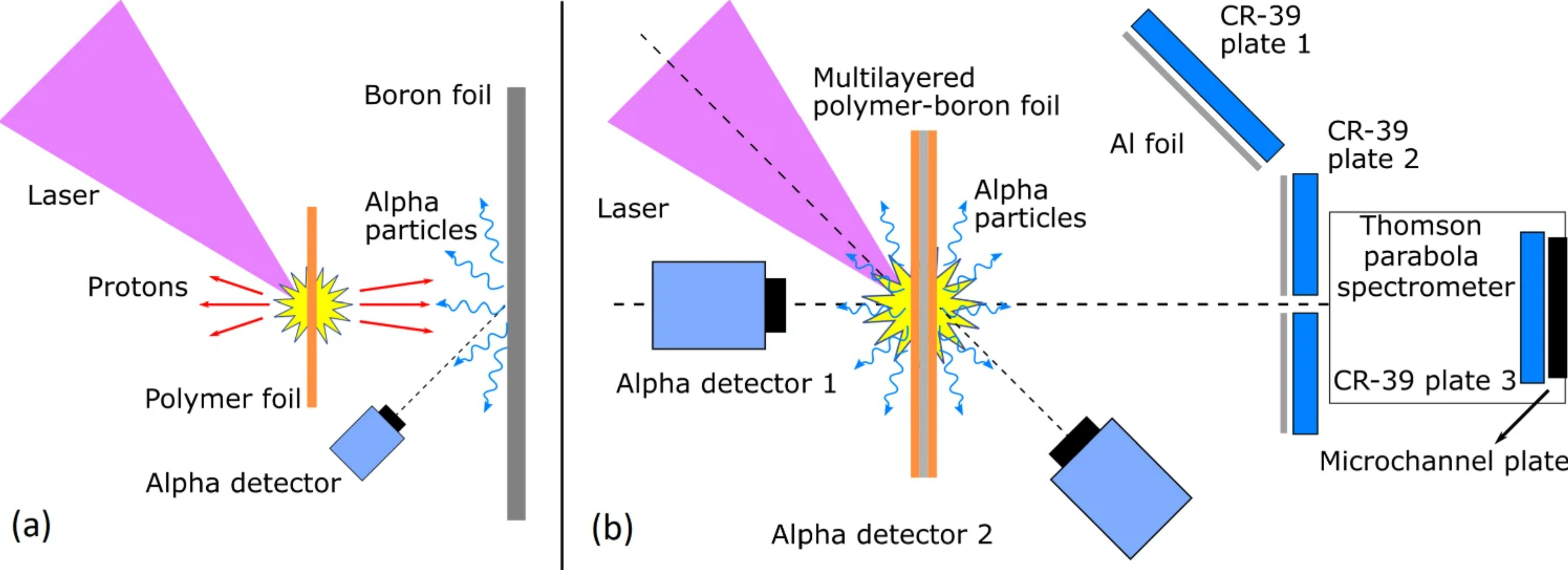}
\caption{Laser-driven pB fusion in (a) pitcher-catcher geometry, where proton acceleration occurs from a thin polymer foil, and (b) single-target configuration containing thin boron layers. The sequential nano- and pico-second laser pulses generate plasma and proton beams. Reprinted from Ref.~\cite{Aladi2025Alpha} under \href{https://creativecommons.org/licenses/by/4.0/}{CC~BY-4.0}.}
\label{fig:PBFusion}
\end{figure}

\begin{figure}
\centering
\includegraphics[width=0.75\linewidth]{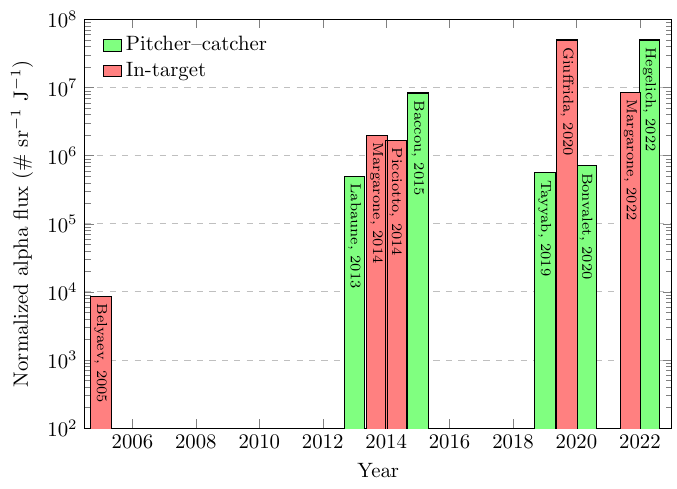}
\caption{Normalized alpha flux (per joule of laser energy) vs.\ year for p${}^{11}$B experiments. Green bars: pitcher--catcher geometry; red bars: in-target geometry. Experiments shown~\cite{Belyaev:2005pB,Labaune:2013dla,Margarone:2014pB,Picciotto:2014pB,Baccou:2015pB,Tayyab:2019pB,Giuffrida:2020pB,Bonvalet:2021pB,margarone2022target,Hegelich:2023pB}. Adapted from Refs.~\cite{mehlhorn2022path,margarone2022target} under \href{https://creativecommons.org/licenses/by/4.0/}{CC~BY-4.0}; revisions by the authors.}
\label{fig:YieldpB}
\end{figure}

The $p$-$B$ fusion experiments were carried out not with isotopically pure ${}^{11}\mathrm{B}$ but with a natural mix which contains 19.9\% ${}^{10}\mathrm{B}$. Experimental work over the past three decades has identified new resonant features in the 
\begin{equation}\label{eq:10BptoLi}
p+{}^{10}\mathrm{B}\to {}^7\mathrm{Li}+\alpha +1.46\mathrm{MeV}
\end{equation}
reaction system~\cite{Kolk:2022isp,Wiescher:2017vhs,Spitaleri:2017spt}. There is a new broad resonance near $E \sim 600~\mathrm{keV}$ mirroring the case of ${}^{11}\mathrm{B}$, as well as a low-energy close to threshold resonance with resonant energy known to within 10\,keV, that is $E \sim 10\pm10~\mathrm{keV}$. This exceptional situation could impact novel paths to fusion we are considering.

These resonances in reaction \req{eq:10BptoLi} are of interest also from astrophysical perspective since the fusion reaction involves production of ${}^7\mathrm{Li}$ whose abundance is in tension with BBN models. They affect also reaction rates and nucleosynthesis pathways in stellar interiors. In applied fusion contexts, especially laser-driven or plasmonic schemes, they determine the proton energy distributions required to maximize $\alpha$-particle yield in interactions with natural boron isotope mix. 

\subsubsection{Boron-nitride fusion cycles}\label{sssec:BNcycles}

Laser-initiated fusion in boron-nitride (BN) targets exploits the coupling between the primary proton--boron reaction and a network of secondary nuclear processes that unfold within a single nano-structured medium~\cite{Labaune:2016mqp}. BN can form cage-like or buckyball-type nanostructures analogous to C$_{60}$~\cite{Golberg:1998BNfull,Oku:2004BnNn}, providing a dense and spatially confined environment for charged-particle interactions.

The primary aneutronic fusion reaction,
\begin{equation}\label{eq:pB11_primary}
p + {}^{11}\mathrm{B} \to 3\,\alpha + 8.7\ \mathrm{MeV},
\end{equation}
is initiated by a picosecond laser pulse that accelerates protons into the BN lattice. A synchronized nanosecond pulse produces a dense BN plasma~\cite{Labaune:2016mqp}. These experiments showed that use of BN enhances observed reaction rates.

\begin{figure}[ht]
\centering
\includegraphics[width=0.75\linewidth]{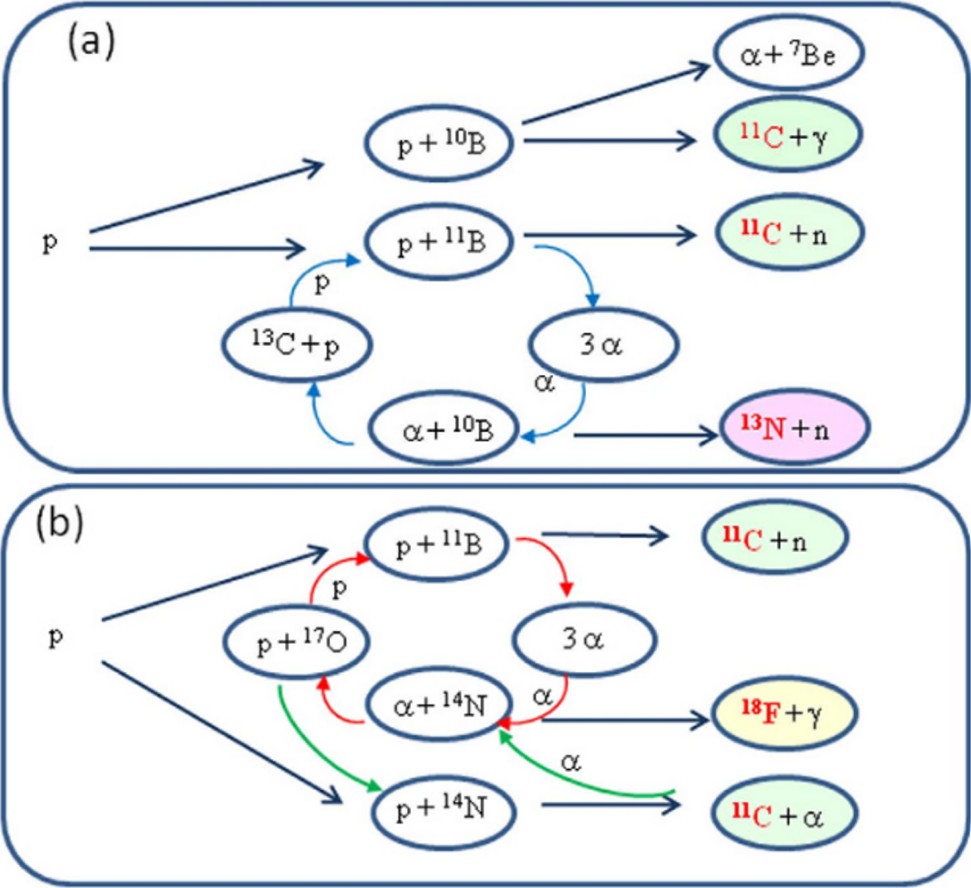}
\caption{
Schematic of the laser-driven proton-boron fusion network in (a) natural boron (B) target and (b) boron-nitride (BN) target. Colored arrows trace the reaction cycles; shaded ovals mark final states containing a radioactive nucleus (red labels). Figure reprinted from Ref.~\cite{Labaune:2016mqp} under \href{https://creativecommons.org/licenses/by/4.0/}{CC~BY-4.0}.}
\label{fig:BNcycles}
\end{figure}

As illustrated in~\rf{fig:BNcycles}~(a), the primary proton reaction seeds a cascade of secondary processes. We see production of neutrons in the secondary $\alpha+{}^{10}\mathrm{B}$ reactions which can be avoided by using mono-isotopic ${}^{11}\mathrm{B}$ target. Boron is routinely isotopically separated as ${}^{10}\mathrm{B}$ is a widely needed neutron absorbing material. The production of neutrons in $p+{}^{11}\mathrm{B}\to {}^{11}\mathrm{C}+n$ reaction has a threshold requiring proton energy in excess of $-Q=2.8\,\mathrm{MeV}$ and thus can be 100\% avoided by using proton beams with lower energy. On the other hand, this leads to the inadvertent discovery of effective radioisotope production, which was patented by Labaune and Rafelski~\cite{Labaune2019US10217538B2}.

Production of neutrons in the secondary reaction
\begin{equation}
 \alpha+{}^{11}\mathrm{B}\to {}^{14}\mathrm{N}+n+0.15\,\mathrm{MeV}\,,
\end{equation}
can only be avoided by mixing the mono-isotopic ${}^{11}\mathrm{B}$ target with another ingredient, which has at available $\alpha$ energy a strong aneutronic reaction cross-section. This step is illustrated in~\rf{fig:BNcycles}~(b): One of us (JR) being a theoretical member of the experimental collaboration~\cite{Labaune:2016mqp} suggested the use of a boron-nitride (BN) target and in the end JR had to procure BN out of his pocket to allow his proposal to proceed. Though the reaction 
\begin{equation}\label{eq:aN}
 \alpha+{}^{14}\mathrm{N}\to {}^{17}\mathrm{O}+p-1.192\,\mathrm{MeV}\,,
\end{equation}
is endothermic, it allows reaction cycle by converting the produced $\alpha$-particle, see \req{eq:pB11_primary}, back into protons. Reactions of protons with ${}^{14}\mathrm{N}$ require photo-capture and thus are also aneutronic. Thus we conclude that use of BN target alters the reaction path towards practically fully aneutronic nuclear fusion process.

The use of the BN target has two initial objectives:
\begin{itemize}
 \item[(a)] to increase the reaction rate by allowing energetic fusion $\alpha$ to react with ${}^{14}\mathrm{N}$ in order to generate energetic protons, thus supporting reaction cycling driven by $p$;
 \item[(b)] and to withdraw the $\alpha$ inventory from the secondary neutron generating reaction in their unavoidable collisions with ${}^{11}\mathrm{B}$.
\end{itemize}
Inadvertently, this target choice had a third consequence as BN is often a natural nano-structured material. The presence of such a structure (see next \rsec{ssec:Plasmonics}) may have created ultra high yield fusion events, which a decade ago were not evaluated in fear of allowing into the data sample an unknown experimental error. Selecting `normal low yield' reactions, the use of (natural isotope mix) BN, was a success with considerable increase in the observed $\alpha$ yield, verifying the proposed fusion material.

\subsubsection{Reactions of interest for multi-component fusion materials}
Table~\ref{tab:aneutronic_reactions} summarizes the principal first step reaction channels of current experimental interest. Before we proceed we should remember: (a) There are many other fusion reactions including aneutronic reactions involving collisions of protons with light elements; (b) The aneutronic $p + {}^{7}\mathrm{Li}$ does not have an active above energy threshold resonance that decays into an energy yielding fusion channel before neutron removal is probable. Thus the aneutronic channel in this reaction is unreachable in two body reactions, while the reachable reaction channels are accompanied by neutron production. We defer further discussion of this often overlooked issue to another venue.

\begin{table}[ht]
\centering
\caption{A select set of fusion reactions, along with the lowest resonant fusion reaction energy $E_\mathrm{rCM}\pm\Gamma/2$ here $\Gamma/2$ is half width in CM frame, the total energy release $Q$ and $Q_\mathrm{ch}$ is the energy fraction carried by charged products, $f_n = 1 - Q_\mathrm{ch}/Q$ is the neutron energy fraction. Reactions marked $\star$ produce no primary free neutrons. For $p+{}^{11}\mathrm{B}$ we list two resonances, the narrow low-energy one and the dominant broad resonance. In bottom section, instead of the lowest resonance, we show $\alpha$-energy threshold for channel activation considering that there are several resonances of significance driving these reactions.}
\label{tab:aneutronic_reactions}
\begin{tabular}{lccccc}
\toprule
Reaction & Products & $E_\mathrm{rCM}\pm\Gamma/2$ [MeV] & $Q$ [MeV] & $Q_\mathrm{ch}[\%]$ & $f_n$ \\
\midrule
\phantom{ $\star$ }$ d + d \to $ & ${}^{3}\mathrm{He}+n$ &$0.4\pm3$ &3.27 & 25 & 0.75 \\
\phantom{ $\star$ }$\phantom{d + d} \to $ & $t+p$ & $0.4\pm3$ &4.03 & 100 & 0 \\
\phantom{ $\star$ }$ d + t \to $ & $\alpha+n$ & $0.05\pm0.037$ &17.59 & 20 & 0.80 \\
\midrule
 $\star$ $d + {}^{3}\mathrm{He} \to $ & $\alpha+p$ & $0.25\pm0.14$ &18.35 & 100 & 0 \\
 $\star$ $p + {}^{6}\mathrm{Li} \to $ & $\alpha+ {}^{3}\mathrm{He}$ & $1.6\pm 0.25$ &4.02 & 100 & 0 \\
 $\star$ $p + {}^{7}\mathrm{Li} \to $ & $2\times\alpha$ &$ -0.33\pm0.037$ &17.35 & 100 & 0 \\
 $\star$ $p + {}^9\mathrm{Be} \to $ & ${}^{8}\mathrm{Be}+d$ & $0.29\pm 0.06$ & 0.56 & 100 & 0 \\
 $\star$ $p + {}^{11}\mathrm{B} \to $ & $3\times\alpha$ & $0.148\pm 0.003$ &8.68 & 100 & 0 \\
 \phantom{ $\star$ }$\phantom{p + {}^{11}\mathrm{B} \to}$ & \phantom{$3\times\alpha$} & $0.61\pm 0.15$ & & & \\
 $\star$ $ {}^{3}\mathrm{He} + {}^{3}\mathrm{He} \to $ & $2\times p+\alpha$ & non-res &12.86 & 100 & 0 \\
\midrule
 $\star$ $\alpha + {}^{14}\mathrm{N} \to $ & ${}^{17}\mathrm{O}+p$ & $E_\alpha>3.1$ & -1.192 & 100 & 0 \\
 $\star$ $\alpha + {}^{14}\mathrm{N} \to $ & ${}^{16}\mathrm{O}+d$ & $E_\alpha>5.0$ & -2.841 & 100 & 0 \\
 \phantom{$\star$} $\alpha + {}^{14}\mathrm{N} \to $ & ${}^{17}\mathrm{F}+n$ & $E_\alpha>6.6$ & -4.735 & 5.5 & 0.95 \\
\bottomrule
\end{tabular}
\end{table}

The top section of \rt{tab:aneutronic_reactions} lists most relevant fusion reactions, starred in middle section are aneutronic cases appearing often in literature. In the bottom section of this table we indicate the use of the fusion $\alpha$-particle to produce secondary $p,d$ in scattering on ${}^{14}\mathrm{N}$. Arguably, there is no other isotope available to regenerate these fusion cycle maintaining particles $p,d$ by fusion produced $\alpha$-particles constrained by the energy available in fusion reaction. As indicated, neutron production threshold is above the energy available in fusion produced $\alpha$-particles. Even though these reactions are endothermic, only a small fraction of fusion energy is consumed, and sustaining a fusion cycle is naturally of greater importance. 

\subsubsection{Beryllium based aneutronic fusion cycle} \label{ssec:Beryl}
We draw attention to arguably the lowest energy yield fusion process with aneutronic reaction products in first step easily fully confined 
\begin{equation}
^9\mathrm{Be}+p\to\! ^8\mathrm{Be}+d + Q \,,\qquad Q=0.5592\,\mathrm{MeV}\,.
\label{eq:9Bep-8Ben}
\end{equation}
This reaction has a lot of interesting physics, as was recently discussed~\cite{Rafelski:2025uio}. The nucleus ${}^9\mathrm{Be}$ can be seen to consist of a two-$\alpha$-particle molecule bound by a neutron orbiting with angular momentum $J=3/2^-$. A relatively low energy slow grazing proton picks up this fast orbiting neutron.

This is, in every regard, the inverse of usual nuclear reaction models: instead of using up a deuteron we make one, and the incoming particle that induces the fusion reaction moves more slowly than the target constituents that participate. Moreover, it is a neutron moving out, which is not subject to the requirement to penetrate a Coulomb wall. Moreover, this neutron is relatively weakly bound else we could not gain energy attaching it to form deuteron.

The weak binding of the neutron in ${}^9\mathrm{Be}$ with separation energy $E_s^{(n^8\mathrm{Be})}=-1.6654$\,MeV implies for the static wave function a long-range tail. The energy gain arises because attaching the neutron to a proton gains $E_d^{(np)}=-2.22457$\,MeV.

\paragraph*{\bf Secondary reactions on beryllium}
The produced deuteron has a good chance to engage with another ${}^9\mathrm{Be}$. Among possible reactions is the following
\begin{equation}
{}^9\mathrm{Be}+d\to\! ^{10}\mathrm{Be}+p + Q \,,\qquad Q= 4.5877\,\mathrm{MeV}\,.
\label{eq:9Bed-10Bep}
\end{equation}
The reaction product, ${}^{10}\mathrm{Be}$
is a molecular state of two $\alpha$-particles glued together by two valence neutrons which are both in the $J=3/2^-$ orbit. This reaction~\req{eq:9Bed-10Bep} could be induced by a $d$ produced in primary reaction~\req{eq:9Bep-8Ben}. The produced proton can restart the cycle of reactions. 

This branch of ${}^9\mathrm{Be}+d$ reaction cycle is clearly aneutronic. However, there is another reaction branch
\begin{equation}
{}^9\mathrm{Be}+d\to\! ^{10}\mathrm{B}+n + Q \,,\qquad Q= 4.3621\,\mathrm{MeV}\,,
\label{eq:9Bed-10Bn}
\end{equation}
which prevents us from calling the two step beryllium cycle aneutronic. Moreover, reactions such as 
\begin{align}
&{}^9\mathrm{Be}+d\to\! ^{7}\mathrm{Li}+\alpha + Q \,,\qquad Q= 7.1521\,\mathrm{MeV}\,,\\
&{}^9\mathrm{Be}+d\to\! ^{8}\mathrm{Be}+t + Q \,,\qquad Q= 4.5927\,\mathrm{MeV}\,,
\label{eq:9Bed-7Lia-8Bet}
\end{align}
do not produce a proton in the final state and thus interrupt the repetitive reaction cycle. Clearly a pure Be is neither a self-sustaining fusion cycle nor at two-step level aneutronic.

\paragraph*{\bf Beryllium-${}^{3}\mathrm{He}$ fusion chain}
As we saw how a beryllium target can be made into a source of deuterons and we want to further exploit this: Among the `star' marked aneutronic reactions on Table~\ref{tab:aneutronic_reactions} we see the case of $p + {}^9\mathrm{Be}$ we have just discussed. This reaction produces a deuteron. Presence of produced weakly bound deuteron in general induces neutron generating reactions as we saw by example of the secondary reaction on ${}^9\mathrm{Be}$. However, the deuteron produced in $p + {}^9\mathrm{Be}$ reaction is just at resonant energy for the reaction $d + {}^{3}\mathrm{He}$ shown in the first starred line of Table~\ref{tab:aneutronic_reactions}. Therefore mixing with ${}^9\mathrm{Be}$ in the target enough ${}^{3}\mathrm{He}$ assures that the second step involving the produced $d$ will resonantly favor fusion on ${}^{3}\mathrm{He}$ which always produces a (high energy) proton. Moreover the primary and secondary protons will not react with in target ${}^{3}\mathrm{He}$.

We conclude that a target mix of ${}^9\mathrm{Be}$ with ${}^{3}\mathrm{He}$ appears to be aneutronic and self-sustaining for very many cycles per incoming proton. This is probably the simplest of all imaginable aneutronic fusion reactor designs: Combination of a proton beam with a target comprising a mix of ${}^9\mathrm{Be}$ with ${}^{3}\mathrm{He}$. Of course there are unsolved problems: we have a relatively high energy 15 MeV proton produced in the secondary $d+{}^{3}\mathrm{He}$ reaction. It is very desirable to add another component to the ${}^9\mathrm{Be}$, ${}^{3}\mathrm{He}$ target allowing protons to be multiplied, `splitting' energy so that the cycle closing reaction would benefit from several protons at convenient energy. To facilitate reactions without stopping power sweeping of electrons by a laser shot is an option.
 
\subsection{Laser plasmonic resonance}\label{ssec:Plasmonics}
\subsubsection{Antennas for light: the NAPLIFE approach}\label{sssec:Anten}
High-intensity laser interactions with metallic nano-inclusions provide a unique mechanism for driving electromagnetic energy into sub-wavelength volumes~\cite{kroo2016plasmonic}. When resonant, these nanostructures support localized surface plasmon polaritons (LSPPs). These are collective electron oscillations that concentrate optical fields far beyond the diffraction limit~\cite{Schuck:2005,Novotny:2007,Novotny:2011}. In the context of aneutronic fusion, such field localization raises the possibility of accelerating light ions near the plasmon to nuclear reaction energies within the available femtosecond timescales. The NAnoPlasmonic Laser Induced Fusion Energy (NAPLIFE) collaboration, led from the Wigner Research Centre for Physics in Budapest, has developed and experimentally pursued this approach over the past decade~\cite{Biro:2025EPJ,Csernai:2025EPJ,tamas2025nanofusion}.

The NAPLIFE concept targets $p+{}^{11}\mathrm{B}$ fusion, which is aneutronic in its primary channel ($p+{}^{11}\mathrm{B}\to 3\alpha + 8.7$~MeV; see \rsec{sssec:LULIpB}), by embedding resonant gold nanorods into a hydrogen-rich urethane dimethacrylate (UDMA)--triethylene glycol dimethacrylate (TEGDMA) co-polymer matrix. The nanorods are sized to match the longitudinal plasmon resonance to the 800~nm wavelength of a Ti:Sa laser. Irradiation by $\sim 40$~fs pulses at intensities up to $\sim 10^{17}$~W/cm${}^2$ launches intense localized surface plasmons, which in turn accelerate the surrounding protons to multi-hundred keV energies via ponderomotive and space-charge forces~\cite{Papp:2022PRXEnergy,Papp:2023FrontPhys}. In particular, NAPLIFE experiments at the ELI-ALPS facility have demonstrated proton acceleration up to 225~keV. This exceeds the requirement to react at the 150~keV (in CM frame) resonance in the $p+{}^{11}$B cross-section~\cite{Kroo2024pB,Kedves:2025EPJ}.

A key feature distinguishing the NAPLIFE scheme from conventional ICF is the explicitly non-thermal character of every step: laser absorption, plasmon excitation, ion acceleration, and nuclear reaction, all proceed on femtosecond-to-picosecond timescales before any thermalization can occur~\cite{Csernai:2025EPJ}. This non-thermal configuration avoids the Rayleigh--Taylor instabilities that plague symmetric compression in ICF, and it allows the process to be scaled to very low laser energies (mJ range) using table-top laser systems~\cite{Biro:2023Univ}.

\paragraph*{\bf Electron sweeping and reduced stopping power}
A relativistic light pulse whose leading edge cuts through the polymer target accelerates the conduction electrons of the gold nanorods to relativistic speeds, sweeping them away from their equilibrium positions. Left behind is a highly charged, spatially ordered arrangement of ionized gold ions and surrounding polymer nuclei, nearly preserving the original nanorod geometry~\cite{Papp:2022PRXEnergy}. While electrons are displaced on femtosecond timescales, the heavier nuclei resist initially and remain nearly stationary on this timescale. This is the method in which inertial confinement physics plays out: the nuclear motion is delayed long enough for the collective Coulomb fields to be created and be able to deliver useful work~\cite{Zweiback:00Coul}.

An important secondary benefit of electron displacement is the dramatic reduction of the material stopping power. In regions temporarily depleted of electrons, the energy loss per unit path length for ions is reduced by a factor of $\sim 10^5$ relative to fully screened matter. This enhancement allows the first fusion-produced $\alpha$-particles to travel much farther and trigger secondary nuclear reactions in the surrounding BN or boron-rich target material. How long such a situation lasts depends on duration of the pulse and electron transport within the target and is to some degree controllable: fast electron return localizes the nuclear burn domain.

The resulting collective motion of the nuclear components creates a compression wave through the target, limited by the balance of kinetic energy against inter-nuclear Coulomb repulsion. This compressed state is far denser than anything achievable in a thermal plasma of comparable pressure, since the counteracting bremsstrahlung from bound electrons is absent. Under these conditions, particle-in-cell (PIC) simulations by the NAPLIFE collaboration demonstrate collective proton acceleration to MeV-scale energies at 30~mJ laser input. This regime is entirely inaccessible without nanoplasmonic enhancement~\cite{Papp:2023arXiv}.

The energy balance for a table-top nano-fusion scheme is favorable compared to NIF-class ICF. With 200~mJ laser pulses repeated at 300~Hz, the total optical input power is $\sim 60$~W. If the $p+{}^{11}\mathrm{B}$ fusion yield per shot provides a gain factor of 50--2000, the system delivers 3--120~kW of output power. The required laser and target technologies are available today at modest cost, offering a clear development pathway from laboratory demonstration to an applied system. A more conservative present-day estimate, based on the $\alpha$-yield scaling shown in Fig.~\ref{fig:YieldpB}, suggests that NAPLIFE-type experiments are already in the regime where the onset of measurable fusion is established; the challenge now is to increase the yield per shot by several orders of magnitude through improved nanorod density, target design, and simultaneous ignition geometry~\cite{Csernai:2026MDPI,Csernai:2024Universe}.

\subsubsection{Nuclear fusion signatures}\label{sssec:FusionSigs}

A central question in nanoplasmonic fusion research is whether the locally enhanced fields can accelerate ions to fusion-relevant thresholds, and how one can distinguish genuine nuclear events from other laser-matter interaction products. The NAPLIFE collaboration has employed a multi-diagnostic approach~\cite{mcnamee2025contaminant,Csernai:2023PRE}: CR-39 nuclear track detectors, Thomson parabola ion spectrometry, Raman spectroscopy, and laser-induced breakdown spectroscopy (LIBS) have all been applied to polymer samples containing resonant gold nanorods~\cite{tamas2025nanofusion,Kroo2024pB}.

\paragraph*{\bf Alpha-particle detection}
The most direct fusion signature is $\alpha$-particle production from the $p+{}^{11}\mathrm{B}\to 3\alpha$ reaction. In NAPLIFE experiments, CR-39 plastic track detectors placed behind the irradiated targets show clear, high-density $\alpha$-tracks in nanorod-seeded samples, while control samples without nanorods exhibit negligible activity~\cite{Kroo2024pB}. Thomson parabola spectra corroborate this finding, revealing MeV-scale ion emission preferentially in nanoparticle-doped targets. Contaminant-free $\alpha$ detection using Thomson parabola spectrometry has also been reported independently in pB laser experiments elsewhere~\cite{mcnamee2025contaminant}. The 150~keV resonance in the $p+{}^{11}$B cross-section (see \rsec{sssec:LULIpB}) is directly implicated by a measured dip in the backward proton emission at this energy in NAPLIFE experiments, providing a nuclear-spectroscopic fingerprint of the fusion process~\cite{Kroo2024pB}.

\paragraph*{\bf Deuterium production}
A secondary, less direct but intriguing signal is the reported production of deuterium in nanorod-seeded polymer samples. This was first detected by Raman spectroscopy of the crater walls, where C-D vibrational modes appear only in irradiated, nanorod-containing samples~\cite{Csernai:2023PRE}. Subsequently, in-situ LIBS measurements of the backward-emitted plume confirmed the D$_\alpha$ spectral line at 656.1~nm exclusively in nanorod-doped targets, with the D/(2D+H) ratio reaching 4--8\% in selected events~\cite{Kroo:2024SciRep}. The deuterium yield scales linearly with laser pulse energy up to $\sim 25$~mJ, and increases with nanoparticle concentration. 

The NAPLIFE collaboration attributes this to the plasmonic near-field enhancement enabling $p+p\to d+e^++\nu_e$ (the weak-interaction branch of the $pp$ chain) or to $d$-formation via inelastic nuclear processes induced by the accelerated protons. However, the interpretation remains open: On one hand the expected rate of nuclear reactions would need to be greatly enhanced; on another the alternative chemical, photochemical, or plasma-chemistry pathways cannot yet be entirely excluded~\cite{Kroo:2025HighField}. Deuterium production detected by spectroscopic identification should therefore be regarded as a tantalizing hint of near-nuclear-threshold processes, not as a standalone fusion proof. To demonstrate nuclear fusion experimentally one must detect classic nuclear fusion products using nuclear detection methods.

\paragraph*{\bf Femtoscopy as a future fusion diagnostic tool}
The NAPLIFE collaboration has proposed adapting the Hanbury Brown--Twiss (HBT) intensity interferometry technique, established in relativistic heavy-ion physics~\cite{Baym:1997ce} for measuring the size and duration of collision fireballs, to the nano-fusion context~\cite{Csernai:2024Universe}. By correlating the momenta of two emitted $\alpha$-particles or deuterons, one can extract the space-time extent of the emission source. Applied to NAPLIFE-type experiments, this femtoscopy approach would provide a model-independent measurement of whether the nuclear emission is consistent with a short in time, spatially localized fusion burn, or with a diffuse, thermalized source, providing a key distinction for validating the non-thermal, nanoplasmonic ignition mechanism. 

\subsubsection{Crater volume and energy deposition}\label{ssec:CraterVol}
\begin{figure}[ht]
\centering
\includegraphics[width=0.95\linewidth]{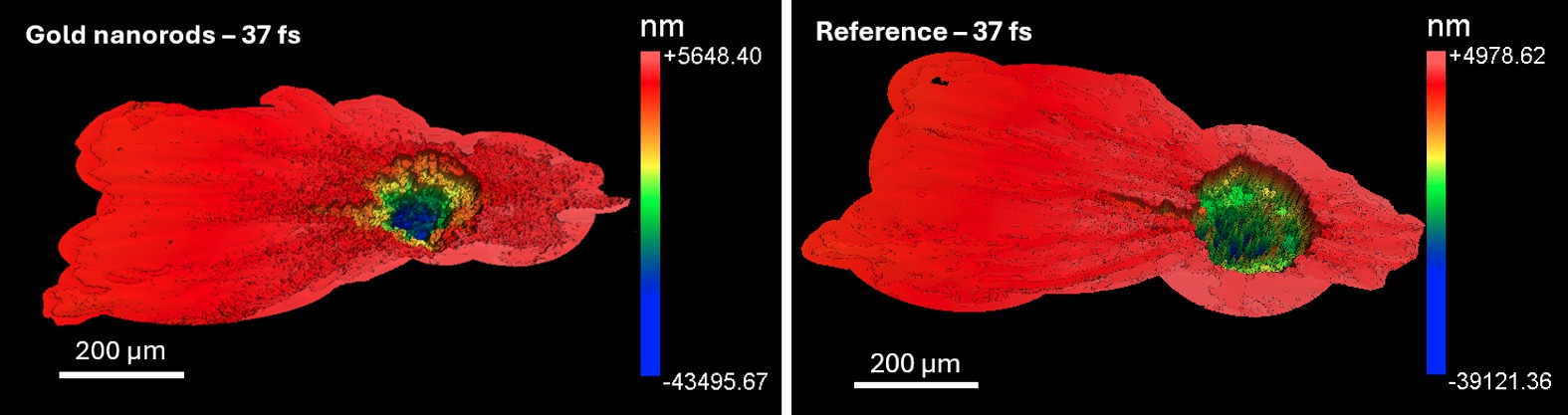}\\[0.5em]
\caption{Laser pulse created crater volumes at constant pulse energy of UDMA-TEGDMA polymer without (left inset) and with (right inset) embedded gold nano-particles. Image courtesy of NAPLIFE Collaboration~\cite{NagyneSzokol:pc}.}
\label{fig:NanorodCraters}
\end{figure}

Ultrashort, high-intensity laser shots produce craters in the irradiated polymer samples. An interesting experimental observation is that crater volumes differ dramatically between `clean' samples and those seeded with resonant gold nanorods. At fixed pulse energy ($\sim$25--27.5~mJ), nanorod-seeded samples exhibit crater volumes up to an order of magnitude larger than unseeded controls~\cite{Kroo:2025HighField,szokol2024pulsed}, the example shown in figure~\ref{fig:NanorodCraters} is for a 37\,fs duration laser pulse.

Two mechanisms contribute to this enhancement, operating on different timescales:
\begin{enumerate}[label=(\arabic*),nosep]
\item 
\textbf{Coulomb explosion.} 
Rapid electron depletion by the laser pulse leaves positively charged regions in the polymer. The resulting space-charge fields explosively accelerate ions outward on sub-ps timescales, well before thermal equilibration. PIC simulations indicate that this mechanism can account for a significant fraction of the crater volume enhancement~\cite{Papp:2022PRXEnergy,Papp:2023arXiv}.
\item 
\textbf{Fusion-generated energetic ions.} 
If $p+{}^{11}\mathrm{B}$ or secondary reactions occur with even modest probability, MeV-scale $\alpha$-particles deposit energy locally and enlarge the crater. The $\alpha$-range in UDMA polymer at 3~MeV is $\sim 20\,\mu$m, sufficient to affect the crater walls~\cite{Kroo2024pB}.
\end{enumerate}

Plasmon-enhanced energy deposition is well established; Coulomb-driven expansion is quantitatively consistent with PIC models; fusion-driven energy deposition, while supported by $\alpha$ signatures, requires additional control experiments. Crater volume is a valuable macroscopic observable for establishing near-field energy concentration, but must be interpreted alongside nuclear diagnostics to draw unambiguous conclusions about the contribution of nuclear fusion yield to this observable. However, the strong enhancement of the laser pulse activity confirms the operational premise of substantial resonant field enhancement near to nano-antennas for light.

\subsubsection{PIC simulations and simultaneous-ignition geometry}\label{sssec:NAPLIFE_PIC}
Particle-in-cell (PIC) simulations have been central to interpreting NAPLIFE experiments and guiding future target design. Papp et al.~\cite{Papp:2023arXiv} simulated a UDMA/TEGDMA target containing a single resonant gold nanorod irradiated by a 30~mJ, 40~fs pulse. Without nanoantennas, the laser with these parameters is not capable of accelerating protons to $p+{}^{11}$B fusion energies. With the resonant nanorod, the near-field enhancement in numerical simulations drives collective proton acceleration to multi-MeV energies within the first 100~fs, with an energy distribution extending to $\sim 3$~MeV. The nuclear reaction yield from this distribution is consistent with experimentally measured $\alpha$-flux levels at 25~mJ~\cite{Kroo2024pB}.

A key challenge for scaling to net-energy gain condition is simultaneous ignition across a macroscopic target volume. The NAPLIFE collaboration has proposed a `time-like' volumetric ignition geometry in which the laser pulse duration and ionization-front propagation speed are matched so that all nanorod sites within a $\sim 100\,\mu$m thick target ignite simultaneously~\cite{Csernai:2025EPJ,Biro:2025EPJ}. This is analogous to simultaneous detonation in a solid explosive, and is designed to avoid premature hydrodynamic expansion that limits fusion yield in conventional ICF. The monochromatic, linearly polarized laser beam carries `mechanical' electromagnetic energy that can be transferred to nuclear motion without passing through a thermal intermediate~\cite{Csernai:2025EPJ}, thereby avoiding the Carnot-efficiency limitation that burdens all heat-cycle energy conversion.

The Coulomb explosion also provides natural pressure amplification: the electrostatic field energy in a fully ionized nanorod cluster scales roughly as $Z^2$ times the laser deposition, creating transient pressures far exceeding direct laser illumination. Combined with the $\sim 10^5$ reduction in stopping power in electron-depleted regions, these effects make the nanoplasmonic target qualitatively distinct from a conventional ICF fuel pellet.

The immediate experimental priority of the NAPLIFE collaboration is statistically unambiguous, shot-by-shot reproducible $\alpha$ detection with confirmed absence of contaminants~\cite{mcnamee2025contaminant}. This requires dedicated runs at ELI-ALPS with controlled target purity, laser contrast $\gtrsim 10^{10}$, and systematic variation of nanorod density and alignment. Directed nanorod orientation during polymerization could increase proton-acceleration efficiency by up to an order of magnitude~\cite{Kedves:2025EPJ}.

\subsection{Muon-catalyzed fusion}\label{ssec:MuonCatalyzed}
\subsubsection{True ``cold fusion''}\label{sssec:CF}
We now explore muon-catalyzed fusion ($\mu$CF) which is the original and only experimentally verified form of cold nuclear fusion. To appreciate why a catalyst is needed, consider the simplest possible scenario: a sealed vessel of heavy hydrogen gas D$_2$. Could the deuterons in D$_2$ molecules spontaneously tunnel through the Coulomb barrier and fuse? Superficially this resembles spontaneous nuclear fission. The answer is no: the inter-nuclear separation in D$_2$ is $\sim 0.74$~\AA, far too large for quantum tunneling to proceed at any measurable rate on a cosmic timescale. A fundamentally different approach is needed to reduce the barrier width.

Two very different strategies have been proposed, and both acquired the popular label `cold fusion.' The first, which we return to below, relies on replacing electrons with muons to shrink the molecular radius by a factor of $\sim 200$. The second, which enters the public eye more often, involves loading heavy hydrogen isotopes into solid metal lattices (e.g.\ titanium or palladium hydrides) and attempting to exploit lattice vibrations to compress the hydrogen isotopes to fusion-relevant separations. If this worked and were verified by nuclear-physics diagnostics, it would indeed be transformative. Unfortunately, the nuclear evidence reported in 1989~\cite{Jones:1989dv} for this approach (the Fleischmann--Pons experiment) remains contested, and many contemporaneous claims were made without adequate nuclear detection capability. This line of research continues today under the name LENR (Low Energy Nuclear Reactions or Lattice Enabled Nuclear Reactions); the large potential payoff continues to attract interest, and science legitimately advances by trial and error.

Returning to the first strategy: negative muons $\mu^-$ are elementary particles with the same charge as the electron but $m_\mu/m_e\simeq 207$ times heavier. In a muonic atom, the $\mu^-$ occupies atomic orbitals that are $m_\mu/m_e\simeq 207$ times smaller than those of an electron (with appropriate reduced-mass corrections). A muonic deuterium atom $d\mu^-$ is therefore $\sim 200$ times more compact than ordinary atomic hydrogen. When two such muonic hydrogen atoms combine into a muonic molecule ($dd\mu^+$, $dt\mu^+$, or $tt\mu^+$), the inter-nuclear separation is correspondingly reduced, making the Coulomb barrier narrow and allowing spontaneous fusion to proceed on nanosecond-to-picosecond timescales. This is $\mu$CF, muon-catalyzed fusion ~\cite{Rafelski:1987Jones,Rafelski:1990gg,Rafelski:1990in}.

\subsubsection{Catalytic cycle}\label{sssec:muCFcycle}
The $\mu$CF mechanism exploits a simple but profound analogy with chemical catalysis: the muon plays the role of a `heavy electron', temporarily binding to two hydrogen isotope nuclei, bringing them close enough to fuse, and then in principle being released to repeat the process. Because the muon is a lepton with no strong-force interactions, it does not itself participate in the nuclear reaction. However, its presence reduces the width of the Coulomb barrier, catalyzes the fusion, and is then free to catalyze further events before it decays.

\begin{figure}
\centering
\includegraphics[width=0.9\linewidth]{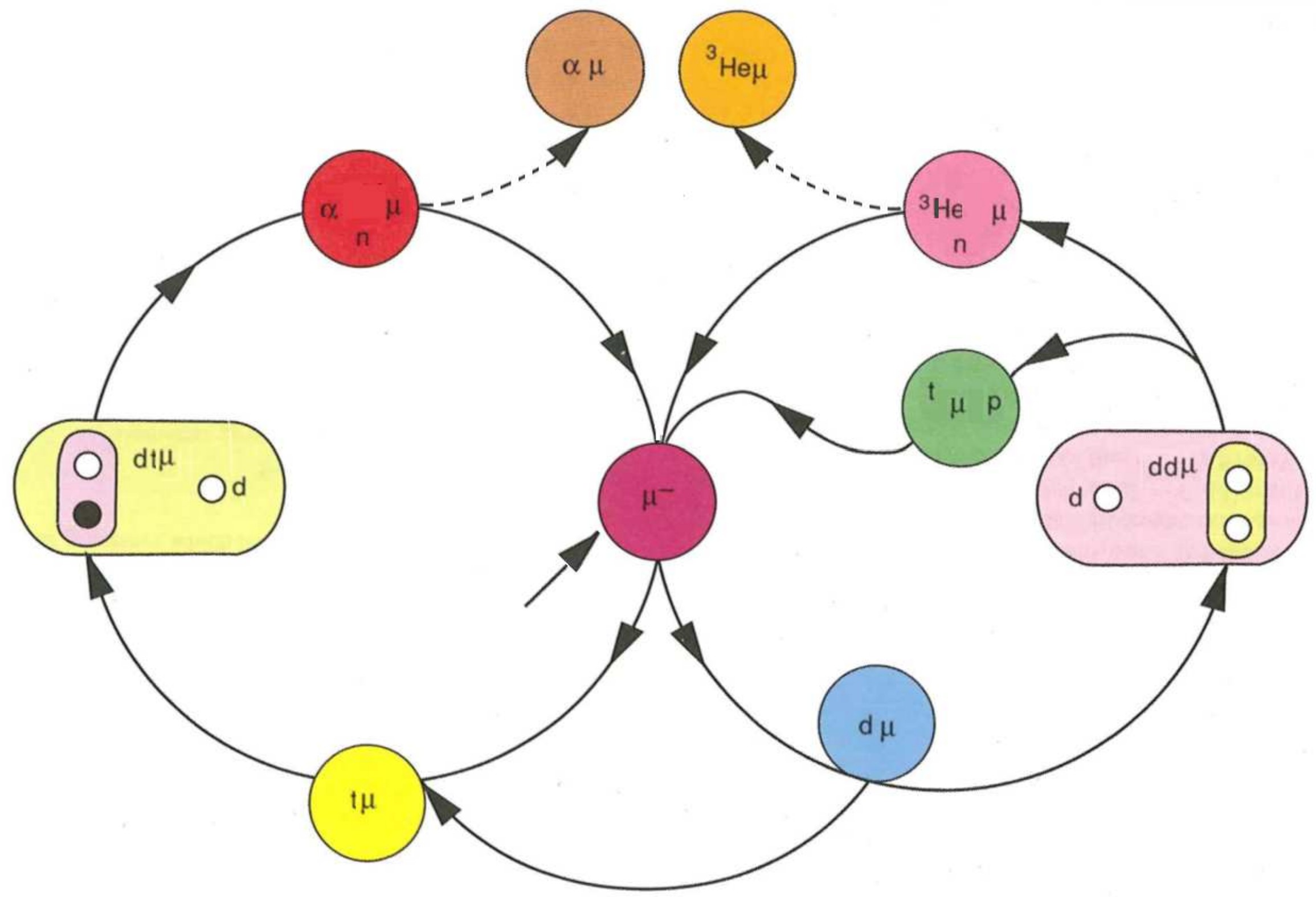}
\caption{Schematic of the muon-catalyzed fusion cycle. A negative muon binds to deuterium or tritium nuclei to form compact muonic atoms and molecules, drastically reducing the Coulomb barrier. The cycle is terminated when the muon becomes bound to the fusion $\alpha$ particle (``alpha sticking''), limiting the number of catalytic events. The isolated arrow marks the entry of an external muon into the cycle. Adapted from Ref.~\cite{Rafelski:1990in}.}
\label{fig:MuonFusion}
\end{figure}

The catalytic cycle of $\mu$CF, shown in Fig.~\ref{fig:MuonFusion}, proceeds as follows~\cite{Rafelski:1987Jones,Rafelski:1990in}:
\begin{enumerate}[label=(\arabic*),nosep]
\item A stopped $\mu^-$ is captured by a deuterium or tritium atom, forming a compact muonic atom $d\mu$ or $t\mu$ with principal quantum number $n=1$.
\item The muonic atom collides with another hydrogen isotope molecule, forming a muonic molecular ion ($dt\mu^+$, $dd\mu^+$, or $tt\mu^+$). In these molecules, the inter-nuclear distance is $\sim$\,500 times smaller than in the ordinary $H_2^+$ ion.
\item The nuclei in the muonic molecule fuse at a rate $\lambda_f\sim 10^{12}$~s${}^{-1}$ for $dt\mu^+$, roughly $10^6$ times faster than the muon's natural decay rate ($\tau_\mu = 2.197\,\mu\mathrm{s}$).
\item The muon is released after the fusion event and is available for the next catalytic cycle, unless it sticks to the $\alpha$-particle product (alpha sticking, probability $\omega_s$).
\end{enumerate}
The actual detailed chain needs to consider spin-states, and the probability of occupation of ground versus excited states in transient muonic atoms prior to molecule formation.

\paragraph*{\bf Reactions of the $dt$ cycle and alpha sticking}
The key fusion reactions in the $\mu$CF cycle operating in a $d$--$t$ mixed environment are:
\begin{align}
(dt\mu)^+ &\to \alpha + n + 17.59\,\mathrm{MeV} + \mu^- \qquad \text{(principal channel)}\,,\\
(dd\mu)^+ &\to {}^3\mathrm{He} + n + 3.27\,\mathrm{MeV} + \mu^- \qquad \text{(side channel)}\,,\\
(dd\mu)^+ &\to t + p + 4.03\,\mathrm{MeV} + \mu^- \qquad \text{(side channel)}\,,\\
(tt\mu)^+ &\to \alpha + n + n + 11.33\,\mathrm{MeV} + \mu^- \qquad \text{(side channel)}\,.
\end{align}
The $dt\mu^+$ principal channel is favored by strongest molecular formation resonance and benefits from fastest fusion rate, the larger energy yield, smallest predicted muon loss due to sticking to helium. Even so, it is generally believed that $\mu$CF yield is limited by alpha-sticking. With the sticking probability $\omega_s$ reduced from $\simeq 1.2$~\cite{Jackson:1957zza} to $\simeq 0.6\%$~\cite{jones1993evaluation,Rafelski:1988wq,Jones1997}, the muon remains bound to the $\alpha$-particle produced in the $d+t$ fusion (forming $\alpha\mu^+$), and is believed to be (mostly) lost from the catalytic cycle. If true, this caps the effective number of catalytic events at $\bar{n}\lesssim 1/\omega_s\simeq 80$--$170$, consistent with the $\sim 150$ fusions per muon achieved experimentally~\cite{Jones:1983pw,Jones:1986kk}. These experiments carried out at Los Alamos in the mid-1980s showed that $\mu$CF in dense $dt$ liquid approaches or slightly exceeds the physics break-even condition, the point where the fusion energy released per muon exceeds the muon rest-mass energy~\cite{Rafelski:1990in}. Moreover, reaction cycle analysis did obtain a very low and unexplained sticking probability in the presence of high T$_2$ abundance~\cite{jones1993evaluation}.

\paragraph*{\bf Muonic excimers and nuclear regeneration}
At that time, more than 30 years ago, it was not fully appreciated that $\alpha\mu$ could enter into chemical noble gas `excimer' bond with both $d$ and $t$, given that only one binding muon was available to form excimer muon molecule. Moreover, should a resonant formation of the muonic excimer in $\alpha\mu$ collision with T$_2$ be possible, the data presented by Jones~\cite{jones1993evaluation,Jones1997} could be interpreted by sticking free molecular muon regenerating fusion reaction $(\alpha\mu t)^{++}\to{}^7\mathrm{Li}+\mu+2.467\,\mathrm{MeV}$; we further note the sticking free molecular excimer fusion $(\alpha\mu d)^{++}\to{}^6\mathrm{Li}+\mu+1.474\,\mathrm{MeV}$. The sticking excimer reactions such as $(\alpha\mu t)^{++} \to{}^6\mathrm{Li}\mu+n -4.783\mathrm{MeV}$ are all endothermic and thus impossible. We conclude that nuclear regeneration in He-muonic-excimer molecules deserves to be further explored as a potential agent for explanation of $\mu$CF path to fusion. This remark applies also to the analog situation in $dd\mu$ fusion branch where the muon regenerating excimers of interest are $({}^3\mathrm{He}\mu d)^{++}$ and $({}^3\mathrm{He}\mu t)^{++}$.

\subsubsection{Towards 1,000 fusions per muon}\label{sssec:1000muCF}
The fundamental obstacle to $\mu$CF as a net energy source is the alpha-sticking probability $\omega_s$ which is the fraction of fusion events after which the muon remains bound to the product $\alpha$ particle and is lost from the catalytic cycle. The effective number of fusions per muon is approximately
\begin{equation}
\bar{n} \simeq \frac{\mu_{\rm live}}{\tau_\mu} \cdot \frac{1}{\omega_s + \tau_\mu/\tau_c}\,,
\end{equation}
where $\tau_\mu = 2.2\,\mu\mathrm{s}$ is the muon lifetime, $\tau_c$ is the mean cycling time for one catalytic event, and $\omega_s \simeq 0.006$--$0.012$ in $dt$ targets~\cite{jones1993evaluation}. The experimental record of 150 fusions per muon achieved by Jones and collaborators in dense, cold $dt$-liquid~\cite{Jones:1983pw,Jones:1986kk,Jones:1986Na} was a landmark result, since it exceeded early theoretical predictions~\cite{Jackson:2010Rem} that set the ceiling near 85.

The path toward $\bar{n}\sim 1000$ fusions per muon requires reducing $\omega_s$ by at least one order of magnitude. Several mechanisms have been discussed in the literature co-authored by one of us~\cite{Rafelski:1987Jones,Rafelski:1990gg,Rafelski:1990in}: 

\begin{enumerate}
\item {\bf Resonant reactivation.} In a dense, hot $dt$ target the $\alpha\mu$ ion produced after sticking can be re-stripped of the muon via collisions or through Auger processes in multi-electron environments. Elevated target temperature ($T \gtrsim 1\,\mathrm{keV}$) enhances the rate of collisional stripping. 
\item {\bf High tritium fraction.} Jones et al.\ showed that increasing the tritium fraction toward 100\% in a liquid $dt$ mixture significantly increases the fusion yield~\cite{jones1993evaluation}. This effect is attributed to the suppression of the $(dt\mu)$ Vesman resonance~\cite{Vesman:1967} formation time and to the favorable kinematics of $dt\mu$ vs.\ $dd\mu$ formation in the tritium-enriched environment. However, it could also be due to nuclear reactivation driven by muonic excimers we described above.
\item {\bf Density effects.} At liquid and above-liquid hydrogen densities $\phi \gtrsim 1$, the muon cycling time $\tau_c$ decreases while the muon survival fraction improves. Jones et al.\ reported unexpected density enhancements in muon catalyzed $dt$ fusion~\cite{Jones:1986kk} that pointed to cooperative effects not captured in dilute-gas models. 
\item {\bf Laser-assisted de-sticking.} Some proposals made more recently explore using tuned laser pulses to photo-ionize the $\alpha\mu$ bound state, returning the muon to the free state before it decays. However, the binding energy of the 1s-$\alpha\mu$ atomic level is $4\times 13.6\,\mathrm{eV}\times 208\sim 11.3\,\mathrm{keV}$, requiring state of the art X-ray laser capability. It is uncertain that such system could reach economical viability, let alone be experimentally tested.
\item {\bf High-density degenerate plasma.} At densities $\phi\equiv\rho/\rho_0\sim 10^3$ and temperatures $T\sim 10$--$100$~eV, a hydrogen plasma becomes partially electron-degenerate. In this regime the stopping power of the $(\alpha\mu)^+$ ion is substantially reduced, because the Fermi energy of the degenerate electrons ($E_F\sim 50$~eV at $\phi=10^3$) suppresses the ionization losses that dominate in liquid hydrogen at standard temperature and pressure (STP). Since muon reactivation (stripping of the $\mu$ from the $\alpha$) depends on the ratio of the stripping cross-section to the stopping power, a reduced stopping power directly enhances the reactivation probability $R$, potentially approaching unity in the deeply degenerate limit~\cite{Rafelski:1988wq,Rafelski:1991DOE}. However, in a dense plasma the resonant Vesman molecular formation mechanism~\cite{Vesman:1967} is suppressed, requiring in flight fusion processes which in general are a bit slower.
\end{enumerate}

\subsubsection{Many possible (catalyzed) fusion reactions}\label{sssec:ManyFusions}
One little explored particle catalyzed path involves reactions beyond pure hydrogen isotopes mix. In general there is a presumption that a catalytic chain is impossible as any $Z>1$ nucleus will bind the catalyst deeply and immobilize it. We now discuss escape routes from this hypothesis, which are possible in the presence of near threshold nuclear resonances of target $Z$ element with incoming neutral, neutron-like object that in terms of nuclear activity is actually a hydrogen isotope, for example a muonic hydrogen isotope exotic atom.

Such threshold resonances are difficult to recognize and thus their discovery is very recent and very probably largely incomplete~\cite{Aprahamian:2025AA}. In our opinion a viable discovery path for systems involving hydrogen isotopes is involving muonic hydrogen (exotic) atoms. The role of these threshold resonances in catalyzed fusion cannot be overstated. This is so for two reasons: (a) Without a heavy electron catalyst it is practically not possible to bring together at relative zero (positive) kinetic energy two nuclei allowing these literally to stick with each other; (b) In many cases such states are nuclear molecules with mutual nuclear attraction not sufficient to bind, but the Coulomb barrier is preventing their separation, making the resonance very sharp: Unbound nuclear molecules often have vanishing reorganization coupling into other nuclear forms; professional language refers here to `spectroscopic factor' being unity for the nuclear molecular state. For this reason the search for such narrow resonant states in alternate nuclear collision channels is often not feasible and even impossible as there is no measurable strength reaction available.

The reader will notice that above we specifically considered the situation in which nuclear attraction is just not enough to bind the nuclear molecule. This allows the collision of two nuclei to form the heavier object compared to the sum of their masses. The new reaction path involving truly bound nuclear molecules is, however, allowed when a (catalytic) third object is present, in our case the catalyzing particle, most commonly a muon. The muon not only allows the nuclear approach but also can be ejected while the bound nuclear molecule is formed, carrying out the binding energy. Moreover, the muon also carries a quantum number which could allow an otherwise forbidden nuclear threshold reaction to proceed. Given these novel aspects we think that a potential explanation of the large muon catalyzed fusion yields we discussed in a previous section, which remains in our opinion unexplained, could be the resonant nuclear stripping or conversion of muons combined with resonant threshold resonances in particular in the $tt$ channel considered recently~\cite{Aprahamian:2025AA}.

\paragraph*{\bf Catalyzed fusion on $Z>1$ targets}
Any of these threshold nuclear molecules will, after a relatively long lifespan, transition radiatively into another state -- of course nuclear processes if energetically allowed can also proceed. For known compound nuclear near threshold resonances (non-molecular), one of us (JR) found several $Z>1$ catalyzed nuclear fusions paths allowing several fusion cycles before a muon is immobilized -- this implies a much more drastic increase in more suitable conditions. We present in \rt{tab:ReactionTable} exothermic reactions of hydrogen isotopes $p,d,t$ with nuclei of up to $Z = 7$, $A = 15$, as presented by Harley, M{\"u}ller, and Rafelski~\cite{Harley:1988wp}. We note a recent study of the system $p\mu+\mathrm{B}$~\cite{Wang:2026zuj}. As we will note, especially for protons, one can extend this table considerably.

\begin{table}[ht]
\caption{Survey of light-element nuclear combination energy balance $Q$ involving the three hydrogen isotopes $p,d,t$ combining with select isotopes through atomic number $Z=7$ (nitrogen) and mass number 15. Both $Q$ and the reduced mass $\mu_\mathrm{r}$ are given in units of MeV. The (logarithm) of tunneling $D_{1\sigma}$ and logarithm of fusion rate $\tilde{\lambda}_{t}$ for muon catalyzed fusion collisions are shown as well, for technical details see text. Data adapted from Harley, M{\"u}ller, and Rafelski~\cite{Harley:1988wp}.}
\label{tab:ReactionTable}
\centering
\small
\begin{tabular}{lcccc|lcccc}
Reaction & $Q$ & $\mu_\mathrm{r}$ & $\log(D_{1s\sigma})$ & $\log(\tilde{\lambda}_{t})$ &
Reaction & $Q$ & $\mu_\mathrm{r}$ & $\log(D_{1s\sigma})$ & $\log(\tilde{\lambda}_{t})$ \\
\hline
\ $d$ + $p$ & 6 & 625 & $-5\,(-3)$ & 13 (15) & ${}^{10}$B + $p$ & 9 & 852 & $-6\,(-6)$ & 12 (13) \\
\ $d$ + $d$ & 24 & 938 & $-6\,(-4)$ & 12 (14) & ${}^{10}$B + $d$ & 25 & 1562 & $-8\,(-8)$ & 10 (11) \\
\ $t$ + $p$ & 20 & 703 & $-5\,(-3)$ & 12 (15) & ${}^{10}$B + $t$ & 24 & 2159 & $-10\,(-9)$ & 8 (9) \\
\ $t$ + $d$ & 17 & 1125 & $-7\,(-4)$ & 11 (14) & ${}^{11}$B + $p$ & 16 & 860 & $-6\,(-6)$ & 12 (13) \\
\ $t$ + $t$ & 12 & 1404 & $-8\,(-5)$ & 10 (13) & ${}^{11}$B + $d$ & 19 & 1586 & $-8\,(-8)$ & 10 (11) \\
${}^{3}$He + $d$ & 17 & 1125 & $-7\,(-5)$ & 11 (13) & ${}^{11}$B + $t$ & 21 & 2205 & $-10\,(-9)$ & 8 (9) \\
${}^{3}$He + $t$ & 16 & 1404 & $-7\,(-6)$ & 11 (12) & ${}^{12}$C + $p$ & 2 & 866 & $-6\,(-6)$ & 12 (13) \\
${}^{4}$He + $d$ & 1.5 & 1248 & $-7\,(-5)$ & 11 (13) & ${}^{12}$C + $d$ & 10 & 1606 & $-9\,(-8)$ & 10 (10) \\
${}^{4}$He + $t$ & 2.5 & 1602 & $-8\,(-6)$ & 10 (12) & ${}^{12}$C + $t$ & 15 & 2245 & $-10\,(-9)$ & 8 (9) \\
${}^{6}$Li + $p$ & 6 & 804 & $-6\,(-5)$ & 12 (13) & ${}^{13}$C + $p$ & 8 & 871 & $-6\,(-6)$ & 12 (13) \\
${}^{6}$Li + $d$ & 22 & 1405 & $-8\,(-6)$ & 10 (12) & ${}^{13}$C + $d$ & 16 & 1624 & $-9\,(-8)$ & 10 (10) \\
${}^{6}$Li + $t$ & 18 & 1871 & $-9\,(-8)$ & 9 (11) & ${}^{13}$C + $t$ & 13 & 2280 & $-10\,(-10)$ & 8 (9) \\
${}^{7}$Li + $p$ & 17 & 820 & $-6\,(-5)$ & 12 (13) & ${}^{14}$C + $p$ & 10 & 875 & $-6\,(-6)$ & 12 (13) \\
${}^{7}$Li + $d$ & 17 & 1457 & $-8\,(-7)$ & 10 (12) & ${}^{14}$C + $d$ & 11 & 1640 & $-9\,(-8)$ & 10 (10) \\
${}^{7}$Li + $t$ & 17 & 1964 & $-9\,(-8)$ & 9 (10) & ${}^{14}$C + $t$ & 10 & 2311 & $-10\,(-10)$ & 8 (9) \\
${}^{9}$Be + $p$ & 6.6 & 844 & $-6\,(-5)$ & 12 (13) & ${}^{14}$N + $p$ & 7 & 875 & $-6\,(-6)$ & 12 (13) \\
${}^{9}$Be + $d$ & 16 & 1533 & $-8\,(-7)$ & 10 (11) & ${}^{14}$N + $d$ & 21 & 1640 & $-9\,(-8)$ & 10 (10) \\
${}^{9}$Be + $t$ & 13 & 2105 & $-10\,(-9)$ & 8 (10) & ${}^{14}$N + $t$ & 19 & 2311 & $-11\,(-10)$ & 8 (8) \\
${}^{10}$Be + $p$ & 11 & 853 & $-6\,(-5)$ & 12 (13) & ${}^{15}$N + $p$ & 12 & 879 & $-6\,(-6)$ & 12 (13) \\
${}^{10}$Be + $d$ & 13 & 1562 & $-8\,(-7)$ & 10 (11) & ${}^{15}$N + $d$ & 14 & 1654 & $-9\,(-8)$ & 10 (10) \\
${}^{10}$Be + $t$ & 11 & 2159 & $-10\,(-9)$ & 8 (9) & ${}^{15}$N + $t$ & 16 & 2339 & $-11\,(-10)$ & 8 (8) \\
\hline
\end{tabular}
\end{table}

All of the displayed reactions involve isotopes that are stable, or relatively long lived, and are readily available in nature; this statement, however, could be contested for artificially made tritium, long lived mostly radiogenic ${}^{10}\mathrm{Be}$, and limited inventory ${}^3\mathrm{He}$. In \rt{tab:ReactionTable} the presented $Q$-value (in MeV) shows the net energy release where the largest values associated with energetically favorable cycles were chosen, often unreachable without a catalyst. The nuclear colliding pair reduced mass $\mu_\mathrm{r}$ (in MeV), not to be confused with the muon, controls the tunneling probability through the not fully screened by the muon Coulomb wall.

\paragraph*{\bf Tunneling probability}
This tunneling probability in semiclassical WKB-approximation is given by 
\begin{align}
D=2\pi\eta e^{-2\pi\eta}\,,\qquad \eta=\frac{1}{\pi}\int_{R_0}^{R_1}\sqrt{2\mu_\mathrm{r}(V(R)-E^*)}dR\,,
\end{align}
which reduces to $\eta\to\eta_{\rm S}$ (cf.~\req{eq:Sfactor}) for the Coulomb potential, and setting $R_0\to0$ and $R_1$ to the classical turning-point. The relative energy of colliding nuclei $E^*$ captures much of the effect of muon screening of the Coulomb repulsion: $E^*$ increases as nuclear separation $R$ decreases in view of energy conservation and the significantly increasing binding of the muon in the attractive combined Coulomb field of two nuclei, for details we refer to Ref.\,\cite{Harley:1988wp}. The results shown are for the most bound quasi-muon-molecular $1s\sigma$ state giving the logarithm of the tunneling probability $D_{1s\sigma}$ in \rt{tab:ReactionTable}. 

In muon-catalyzed hydrogen fusion reactions, the surrounding electron cloud has minimal influence, compare \rsec{ssec:PlasmaScreening}. This is so since muon molecular dynamics is occurring deep within the electron cloud. However, processes leading to the formation of muonic molecule, not described here, are coupled to the electron dynamics as the scale differences are bridged by highly excited transient muonic orbitals.

A finite nuclear range of interaction was allowed for in Ref.\,\cite{Harley:1988wp}, but not the possibility of relatively large in size nuclear molecules. Therefore even the `optimistic' estimate seen in the parenthesis could be an underestimate. The reason that a range for penetrability is presented in \rt{tab:ReactionTable} is due to the difficulty of fully understanding the three-body dynamics as the gain in binding of the muon is significantly impacting the nuclear part of the quantum wave function. A true three-body study should follow for the most interesting cases once these are identified.

Checking the table entries for the penetrability results, we observe that the reduced-mass $\mu_\mathrm{r}$ impact on penetrability becomes increasingly significant as we go to larger $Z$. For proton reactions there could actually be an advantage for the $Z > 1$ case compared even to $dt$, also in view of the nuclear relative energy $E^*$ increase $\propto Z^2$, which in principle can overcompensate the impact of Coulomb repulsion $\propto Z$. Further advantage is provided by the smaller system size, which decreases with increasing $Z$, and by the larger nuclear size, which increases as $A^{1/3}$. Therefore for protons, the table created soon 40 years ago does not exhaust all the $Z>1$ opportunities.

\paragraph*{\bf Fusion rate estimate}
To obtain nuclear fusion reaction rate $\lambda_f$ (per second) consider the product of the intrinsic nuclear reaction strength with the probability of finding the nuclei in the interaction range which contains the penetration probability already seen in the penetration factor $D_{1s\sigma}$. To obtain in qualitative manner across many systems the rate of nuclear fusion reaction $\lambda_f$ we apply semiclassical quantitative ideas combining Gamow's `attempt' frequency $\nu$ with the chance of achieving intrinsically the desired outcome, and the penetration probability of the Coulomb wall. This approach is validated in a study of $\alpha$ decay of heavy nuclei.

The rate seen in \rt{tab:ReactionTable} assumes for $\nu$ a muon molecular vibrational constant $\nu= 10^{18}\,\mathrm{s}^{-1}$. The tilde reminds us that to obtain full rate for the $1s\sigma$ molecular state we must further multiply the shown rate by the probability that the combined $Z$-hydrogen system is found in this configuration, for $Z=1$ this is the only state possible but for high values of $Z$ this needs to be allowed for, but can be expected to be not of importance compared to other qualitative approximations that entered and the range of values with the optimistic rate again indicated in parenthesis.

\subsubsection{Muon production}\label{sssec:muon_production}
The energy economy of $\mu$CF is determined by two largely independent components: (i) the fusion yield $\bar{n}$ per muon, limited by sticking; and (ii) the energy cost $E_\mu$ per usable negative muon delivered to the fusion target. Both must be optimized simultaneously before $\mu$CF can be evaluated as a potential energy source.

A break-even analysis~\cite{Rafelski:1990in} shows that the energy released by $\mu$CF exceeds the energy cost of muon production (roughly $5\,\mathrm{GeV}$ per muon via pion production at an accelerator) only if $\bar{n} \gtrsim 300$, assuming a $dt$ $Q$-value of $17.6\,\mathrm{MeV}$ and a wall-plug efficiency of $\sim 30$\%. Achieving $\bar{n}\sim 1000$ would provide a comfortable margin even accounting for thermal cycle inefficiencies. The required reduction in sticking has not yet been demonstrated experimentally, and represents the central open problem of the field. Nevertheless, the non-thermal character of $\mu$CF (and its complete independence from plasma confinement and its operation at room temperature) makes it conceptually unique among all fusion approaches surveyed in this work, and justifies continued theoretical and experimental attention.

\paragraph*{\bf Standard muon production chain}
Negative muons are produced as decay products of negative pions, $\pi^-\to\mu^-+\bar\nu_\mu$ (lifetime $\tau_{\pi^-}=26.0$~ns). In classic discussion of muon production for fusion, pions are in turn generated by hadronic interactions of an energetic proton beam on a nuclear target, $p+\mathrm{nucleus}\to \pi^-+X$. The overall production chain is therefore
\begin{equation}
p_{\rm beam}\;\xrightarrow{\;\rm hadronic\;}\;\pi^-\;\xrightarrow{\;26\,\mathrm{ns}\;}\;\mu^-\;(\text{stopped in fusion target})\,.
\end{equation}
The energy cost of this standard approach was analyzed in detail in the 1991 DOE final report~\cite{Rafelski:1991DOE}. Monte Carlo hadronic cascade simulations of several metallic internal targets (Be, Pb, Ag, Cu, Al) found a production cost of approximately 11~GeV of beam energy per first-interaction negative pion. However, by exploiting the secondary shower particles emitted from the thin internal target at azimuthal angles $>36^\circ$ and redirecting them to an external converter target, the effective $\pi^-$ cost was reduced to $\sim 2.7$~GeV per negative pion at a primary beam momentum of 5~GeV/$c$ using a neutron rich target such as beryllium as both internal and external target material.

Two production geometries were studied~\cite{Rafelski:1991DOE}:
\begin{enumerate}[label=(\arabic*),nosep]
\item \textbf{Active target scheme.} The pion production target is placed immediately adjacent to the fusion vessel. This minimizes pion transport losses but restricts target geometry and neutron shielding options.
\item \textbf{Storage ring with internal target.} A proton beam circulates in a storage ring with a thin internal beryllium target. Secondary shower beams from this target are collected at an external second target to maximize $\pi^-$ yield per unit beam energy. This scheme
achieves the factor-of-four improvement in muon cost noted above by harvesting shower particles that would otherwise be lost.
\end{enumerate}
Accounting for pion transport and muon stopping in the fusion vessel, the overall energetic cost is $\sim 10$~GeV of thermal power per usable stopped $\mu^-$, or $\sim 5$~GeV of electrical energy per muon when accelerator wall-plug efficiency is included~\cite{Rafelski:1990in,Rafelski:1991DOE}.

\paragraph*{\bf $e^+e^-$ resonance as a muon source: why $\phi(1020)$ wins}
The pion-based muon production chain requires a multi-GeV proton accelerator, imposing substantial capital cost. An alternative exploits low-energy $e^+e^-$ resonances to produce mesons that subsequently decay to muons. Three light vector mesons are candidates: the $\rho(770)^0$, the $\omega(782)$, and the $\phi(1020)$. We show in the following how the three resonances differ in their practical utility for $\mu$CF. 

The peak cross-section for a spin-1 resonance is
\begin{equation}
\sigma_\mathrm{peak} = \frac{12\pi}{M_R^2}\,\frac{\Gamma_{ee}\,\Gamma_\mathrm{had}}{\Gamma_\mathrm{tot}^2}\,,
\end{equation}
with Table~\ref{tab:ee_muon} summarizing the key parameters.
\begin{table}[ht]
\centering
\caption{Comparison of light $e^+e^-$ vector resonances considered as potential muon sources. $\sigma_\mathrm{peak}$ is the peak hadronic cross-section; $\Gamma_\mathrm{tot}$ the total width; $\sigma({\mu^-})$ the effective cross-section for negative muon production accounting for decay branching fractions. Data from PDG~2024~\cite{ParticleDataGroup:2024cfk}.}
\label{tab:ee_muon}
\begin{tabular}{lccccl}
\toprule
Resonance & $M$ (MeV) & $\Gamma_\mathrm{tot}$ (MeV) & $\sigma_\mathrm{peak}$ ($\mu$b) & $\sigma(\mu^-)$ ($\mu$b) & Muon path \\
\midrule
$\rho(770)^0$ & 775 & 147 & 1.15 & $\sim 1.15$ & $\rho\to\pi^+\pi^-\to\mu$ \\
$\omega(782)$ & 783 & 8.68 & 1.66 & $\sim 1.47$ & $\omega\to\pi^+\pi^-\pi^0\to\mu$ \\
$\phi(1020)$ & 1020 & 4.25 & 4.22 & $\sim 1.70$ & $\phi\to K^+K^-\to\mu$ \\
\bottomrule
\end{tabular}
\end{table}

This idea to consider resonant muon production arises adapting the results of the existing DA$\Phi$NE $\phi$-factory at the INFN Laboratori Nazionali di Frascati (LNF), Italy~\cite{Vignola:1994DAFNE,Zobov:2007xw}, which has operated since 1999 at $\sqrt{s}=1020$~MeV delivering low-momentum, nearly monochromatic $K^-$ beams at a rate of $\sim 300$~$\phi$ decays per second to experiments including KLOE, FINUDA, DEAR, SIDDHARTA, and SIDDHARTA-2~\cite{Sirghi:2023wok,Iliescu:2016zgx}. These experiments exploit the $K^-$ beam for kaonic atom X-ray spectroscopy and kaon-nucleon scattering length measurements. The same $K^-$ beam can be redirected allowing to use $K^-\to\mu^-\bar\nu_\mu$ decay as an alternative muon source for $\mu$CF.

We believe that $\phi(1020)$ is the preferred candidate for the following reasons:
\begin{enumerate}[label=(\arabic*),nosep]
\item 
\textbf{Resonance sharpness.} The $\rho(770)$ has a total width of 147~MeV, about 20\% of its mass $m_\rho=775$\,MeV/$c^2$. A collider operating at $\sqrt{s}\simeq 775$~MeV cannot exploit the resonance peak efficiently; the beam energy spread required to scan the resonance is $\sim 150$~MeV. Operationally, this is little different from simply running a continuous $\pi^-$ production target, which is precisely the strategy of the existing pion-based muon sources analyzed in~\cite{Rafelski:1991DOE}. The $\phi(1020)$, of mass $m_\phi=1019.5$\,MeV/$c^2$, has a width well below half a percent of its mass, $\Gamma_{\phi\mathrm{tot}} = 4.25$~MeV. This can be matched by a standard collider beam energy spread, allowing to take advantage of resonant production at a well-defined center-of-mass energy of $\sqrt{s}\simeq1020\pm 5$~MeV.
\item 
\textbf{Kaon decay chain.} The $\phi$ decays to $K^+K^-$ (49.1\%), $K_LK_S$ (33.9\%), and the remainder goes to the three-pion channel. Most kaons decay directly into a muon e.g. $K^-\to\mu^-\bar\nu_\mu$ at a branching fraction of 63.6\%. One can estimate that about 10\% of $\phi$ chain decay does not produce a muon; this is the decay path that takes neutral kaons into neutral pions. Said differently a vast majority of produced $\phi$ produces in the decay chain one muon pair. We note that this is also the case for the $\rho$ produced $\pi^+\pi^-$ pairs. However, unlike formation of two Kaons in $\phi$ decay which are easily slowed down as their kinetic energy is in 10-15\,MeV range, pions from $\rho$-decay are relativistic with $\sim 200$~MeV kinetic energy, similar to the conventional pion source. It is clear that the $\phi/K$ path provides a distinct intermediate step with better-defined kinematics.
\item 
\textbf{Boost geometry.} In a geometrically asymmetric $e^+e^-$ collider where the beam energies and angles are carefully chosen, all produced $\phi$ can be directed at chosen speed towards reaction vessel allowing the cascade decay muons to be well localized and thus assure a large acceptance of muons for fusion. Here the weak decay $K^-\to\mu^-\bar\nu_\mu$ is convenient since the neutrino removes 236 MeV of available decay energy due to its vanishingly small mass. One could indeed use a highly asymmetric $e^+e^-$ collider such that about as many as 25\% of produced $\mu^-$ are created at rest in the reaction chamber. A full estimate requires in depth consideration for both $\phi$ and $\rho$ production.
\item 
\textbf{Existing infrastructure.} The DA$\Phi$NE $e^+e^-$ collider at Frascati (Italy) was built and operated expressly as a $\phi$ factory, demonstrating the technical feasibility of this approach at a modest scale. No equivalent dedicated $\rho$-factory infrastructure exists or would be straightforward to justify. There is apparently already a project underway to explore muon production at DA$\Phi$NE.
\end{enumerate}

The effective $\mu^-$ cross-section from the $e^+e^-\to \phi$ resonance is $\sigma(\mu^-)\simeq 1700$~nb (combining $K^+K^-$ and $K_LK_S$ contributions), competitive with or exceeding the other resonances on a per-event basis while retaining the sharp-resonance advantage. Whether the full system-level $e^+e^-\to \phi$ energy cost per stopped $\mu^-$ improves over the optimized storage-ring pion-production scheme of~\cite{Rafelski:1991DOE} remains an open quantitative question: There is need to consider collider wall-plug power, kaon capture efficiency, and muon transport losses. The resonant production mechanism and well-defined kaon decay kinematics provide qualitative advantages; the $\phi$-factory concept thus deserves serious consideration and energy cost evaluation as a muon source for $\mu$CF systems.

\paragraph*{\bf To summarize $\mu$CF theoretical and commercial viability}
The lifespan bound on number of fusions per muon is counted in 1000s of fusions due to long life span $\tau_\mu=2.2\mu$s and sub-nanoscale timing of all reaction processes in dense D-T target. Such a number of fusions per muon vastly exceeds the muon cost estimate we developed here based on resonant $\phi(1020)$ production. However, $\mu$CF is limited by muon sticking to $\alpha$ or to ${}^3\mathrm{He}$. This caps the effective yield of fusions per muon to below about $150$ fusion events observed. However, experiments have shown an intriguing departure from theoretical sticking estimates especially so within analysis that accounts for muon cycling and sticking in a high tritium environment. We considered unexplored nuclear muon regeneration mechanisms which justify continued interest in the $\mu$CF path to fusion.

As of 2026 at least two private companies: Muons, Inc.\ and Acceleron Fusion are pursuing $\mu$CF-based reactor and neutron-source concepts, supported in part by ARPA-E funding. Near-term applications exploiting the high-flux 14.1~MeV neutrons produced by $dt\mu$ fusion (isotope transmutation, medical radioisotope production) are viable well before energy break-even, since the outcome requirements are several orders of magnitude less demanding than for net electricity generation~\cite{Kamimura:2021msf,Yamashita:2022rtu}. 

\section{Evaluation, discussion and outlook}\label{sec:Dis}
\subsection{Fusion pathways}
In this work we have shown why study of the science of nuclear fusion with ultimate application to energy generation should not be viewed today to have some clear cut unique path leading to near term technological realization. We have argued that there is a broad class of nuclear pathways whose feasibility depends on multitude of unexplored scientific factors often requiring novel technological realizations. It should be remembered that this remark applies strongly to the `traditional' $dt$ fusion path irrespective of realization model: laser implosion (direct or indirect), dynamic EM field implosion, a variant not further discussed here, or plasma (stellarator, tokamak). We clearly suggested that ICF holds a significant advantage due to spatially tiny neutron source allowing development of viable neutron control. 

The preceding sections have surveyed fusion pathways across an unusually broad physics range, from stellar quasi-equilibrium reactors, to laboratory-scale thermal plasmas, and beam driven non-equilibrium ignition and nano-structured environments. We attempt to illustrate the width of considered environments in~\rf{fig:ProcessTimeVolume}, where in space size-time plane different systems are displayed in qualitative terms. The Big Bang Nucleosynthesis (BBN) is present in a homogeneous thermally equilibrated expanding Universe, spatially as big as it can be. However, fusion reactions end within an `hour'. Stellar nucleosynthesis is opposite, a modest sized but very long duration. Large manmade fusion experiments are designed or capable to operate for rather short microsecond to minute duration burns and are a large building sized. Beam driven approaches are aiming to be `table top' sized but those requiring giant lasers have some miniaturization to achieve. Resonant antenna for light systems are metaphorical future ``space planes'' where ignition size and burn time is the smallest. 

\begin{figure}[ht]
\centering
\includegraphics[width=0.75\linewidth]{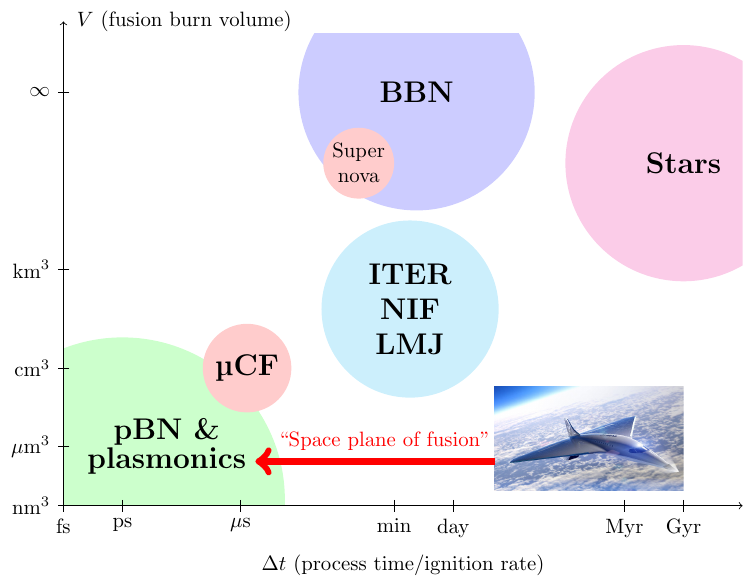}
\caption{Comparison of time and scale of fusion environments. Big Bang Nucleosynthesis (BBN) is present in a homogeneous thermally equilibrated Universe filling plasma which cools in expanding Universe and fusion reactions end within an `hour'. Stellar nucleosynthesis is a near equilibrium in plasma process with reactor size in front of our eyes. Large manmade fusion experiments are designed or capable to operate for short microsecond to minute duration burns. Aneutronic approaches (such as proton-boron and plasmonics) are metaphorical future ``space planes.'' Nuclear burn is limited in both time and space to below microscale. The space plane art credit to \textcopyright\ Virgin Galactic~\cite{VirginGalactic:2020}.
}
\label{fig:ProcessTimeVolume}
\end{figure}

\subsection{Fusion fuel}
The light-element abundances summarized in~\rf{fig:NuclearAsh} and \rf{fig:NuclearAshUniverse} imply that the relevant fuels are not equally available: deuterium, lithium, boron, carbon, nitrogen, and helium isotopes each occupy different positions in the terrestrial and cosmic nuclear inventory. The phrase ``limitless fusion energy'' must be qualified by isotope supply, the reaction cycle chosen and the ability of the chosen realization to close any required breeding or fuel loop, while developing the transition of the society to nuclear fusion energy.

\paragraph*{\bf Fusion self heating}
For thermal plasmas, the central constraint remains the requirement that charged-particle heating, see \req{eq:powerFusion}, exceed radiative losses, dominated at high temperature by Bremsstrahlung, \req{eq:powerBrem}. As illustrated in~\rf{fig:PowerBalance_AllFuels}, the favorable reactivity of the $dt$ channel lowers the ignition threshold relative to $dd$, which itself is worse in terms of neutron flux but does not need tritium. An intermediate strategy is to ``spike'' a deuterium plasma with a modest fraction of ${}^3$He, as in~\rf{fig:He3Spike}. In this case the additional $d+{}^3$He channel increases charged-particle heating. However, bremsstrahlung rises through the corresponding increase in $Z_{\rm eff}$. For suitable temperatures and mixture fractions, a net gain in $P_{\rm fus}/n_e^2$ relative to pure deuterium can be achieved without introducing thus production of a proliferation agent, tritium.
 
More generally, controlled doping of light hydrogen plasma with light nuclei (including ${}^3$He, Be, B, etc) offers a pathway to tailor reaction networks and enhance secondary heating, albeit potentially at the cost of increased radiative losses and more complex burn dynamics. A self-consistent treatment of doped plasmas, including full reaction networks, secondary burn fractions, evolving composition, and radiative feedback, remains today an important and yet entirely open nuclear science problem arguably with little if any presence in literature.

\paragraph*{\bf Conventional hydrogen burn}
The most technologically accessible high energy gain reaction remains the traditional fusion channel
\begin{align}
 d+t\to \alpha+n\,,
\end{align}
because of its large reactivity and lowest thermal ignition temperature, as illustrated by the thermal power-balance comparison in~\rf{fig:PowerBalance_AllFuels}. However, this accessibility comes at the cost of a $14~\mathrm{MeV}$ neutron carrying most of the released energy (which itself presents a radiation hazard; see~\rf{fig:NeutronSafety}) and the need for tritium-breeding to maintain and expand the fuel cycle, as discussed in~\rsec{sssec:ICFtritium}. 

In this sense, application of $dt$ fusion is not governed only by a solution to the reaction domain confinement problem (such as direct or indirect ICF; see~\rsec{sssec:picoICF}). It also requires development of science and technology addressing the fuel-cycle (tritium breeding) and fast neutron control, combining tritium breeding and shielding from interaction with conventional and newly developed stress and radiation wear resistant materials. These challenges remain whether or not the $dt$ burn is realized in magnetic confinement, inertial confinement, or muon-catalyzed fusion. We believe that the smaller is the source of neutrons, the more approachable is the technology of neutron energy harvesting and tritium breeding.

The pure deuterium fusion solution
\begin{align}
d+d\to p+t,\qquad d+d\to n+{}^3{\rm He}
\end{align}
is sometimes advanced as a cleaner alternative, but more complete reaction chain analysis, discussed in~\rsec{ssubsec:SecondHeating}, shows that this argument is incomplete. The primary $dd$ fusion reactions naturally generate secondary generation tritium and ${}^3\mathrm{He}$. This allows secondary reactions
\begin{align}
 d+t\to \alpha+n\,,\qquad
 d+{}^3{\rm He}\to \alpha+p\,.
\end{align}
In conditions that allow $dd$-burn, it is difficult to control produced tritium burn, which is typically a 1000 times more likely. One can therefore say that produced tritium secondary reactions are unavoidable and at near 100\% level in magnetic or inertial confinement devices. Thus heavy hydrogen limitless fusion capability in thermal environment is always accompanied by high energy neutron production within the coupled reaction network, which is replacing naive single-step fuel burn. On the other hand, these secondary reactions can substantially enhance plasma heating, as illustrated in~\rf{fig:DD_boosted}.

Complete assessment of $dd$ fusion therefore requires tracking the burn rates of produced $t$ and harder to burn ${}^3\mathrm{He}$ in dependence on device temperature. Since the secondary $t$-reactions introduce energetic neutrons, we can ask if there is any advantage of solving $dd$ fusion technological challenge: Back of envelope estimates suffice to show that the number of fast neutrons one must handle per unit of energy production can be worse than in the technologically much less difficult $dt$ burn, while the required plasma volume to achieve required energy production given the reduced burn rate is about 1000 times greater. We therefore believe that thermal fusion burn is best achieved with $dt$ fuel, and we need to address the related neutron and tritium issues with some urgency today.
 
Considering on first look a clean ${d}^{3\!}\mathrm{He}$-fusion burn, we note appearance of undesired $dd$ burn. The problem: The neutron producing $dd$ burn dominates, up to temperatures around 20 keV, the desired ${d}^{3\!}\mathrm{He}$-fusion burn. Reduction of fractional abundance of $d$ in the ${}^3\mathrm{He}$ plasma will limit chance of $dd$ fusion but greatly increase bremsstrahlung radiation losses, considering higher operational temperature required in $d+{}^3\mathrm{He}$ burn channel, see~\rsec{sssec:RadLosses}. We believe that the technological challenges of realizing \underline{thermal} ${d}^{3\!}\mathrm{He}$-fusion limiting neutron production are well above our present day technological capabilities. 

\subsection{Future fusion solutions}
Therefore, any practical approach to a clean $d+{}^3\mathrm{He}$ burn requires, as we have discussed, the injection of a high flux resonant reaction energy ${}^3\mathrm{He}$-beam into a $d$ target. Such a system is clearly not a thermal plasma device. We remember here that larger $Z>2$ thermal plasma enhances Bremsstrahlung losses beyond the fusion gain. We have also addressed another alternative, the use of a proton beam to create $d$ in reactions on beryllium and accompany this by sequel $d+{}^3\mathrm{He}$ reaction, which is re-creating the proton needed to continue a fusion cycle. This proposal is a conceptual idea; the realization could require harvesting the high energy of produced protons before recycling these to produce deuterons on Be.

{\it In summary of the above we thus conclude that a \underline{thermal plasma} and radiation clean energy producing fusion device that can be built on Earth has not been conceptually discovered, implying that large, costly, advanced in construction magnetic confinement devices are a classic example of premature rush to engineering solution without a demonstrated science case. A thermal fusion energy source requires breakthrough novel ideas for alternate to $dt$ fusion plasmas or the development of fast neutron handling technologies with energy harvesting and tritium breeding. This conclusion applies in general, thus involving both magnetic and inertial confinement when and if most of energy produced is through thermal burn reactions. It applies to muon catalyzed fusion as we know it today. However, we anticipate that smaller nuclear burn environments and thus smaller neutron sources will be easier to handle.}

\paragraph*{\bf Search for aneutronic fusion method}
This understanding motivates the recent interest in aneutronic reactions. Searches for laboratory aneutronic beam-target approaches and environments are motivated at least in part by the fully aneutronic stellar fusion burn cycles, discussed in~\rsec{sssub:sunN}. In general aneutronic processes require much higher inter-particle collision energy. This in turn has stimulated interest in non-equilibrium reaction pathways involving use of particle beams to either fully drive a fusion reactor or to ignite in spark manner a thermal burn front. 

In such directed particle motion systems, the central control variable is no longer a thermal quasi-equilibrium temperature. Instead, fusion yield depends on the ion energy spectrum, target density, stopping power of produced fusion secondaries, and pointlike fusion inducement by meta-material field-enhancement. Examples include laser-driven $p+{}^{11}\mathrm{B}$ fusion seen in~\rf{fig:PBFusion} and~\rf{fig:YieldpB}, requiring boron-nitride reaction cycles in~\rf{fig:BNcycles} to assure proton production for fusion cycling, potentially implemented in plasmonic/nano-structured target concepts discussed in~\rsec{sssec:nanoF}. Another similar aneutronic system involving a mix of ${}^9\mathrm{Be}$ with ${}^{3}\mathrm{He}$, see \rsec{ssec:Beryl}, has a vastly different realization method needing to account for the high energy of protons.

Nuclear fusion reactions driven by high contrast high intensity short pulsed lasers have now been explored for more than two decades. The advantage is the ability to use isotopes which leads to a rate of neutron production that is small to start with and potentially reducible. This is the aneutronic fusion. One must also note research works with opposite objective, i.e. interest to cheaply produce neutrons on demand, needed for practical technological applications~\cite{Karsch2002Neutrons}.

\paragraph*{\bf Aneutronic fusion targets}
For energy production our objectives in context of laser fusion is to find paths to fully aneutronic energy production: There are numerous laser driven reactions where neutrons are absent in the primary step; some we discussed in this work, see for example Table~\ref{tab:aneutronic_reactions}. However, the fusion reaction chains have to be explored assuring secondary and unavoidable reactions are predominantly aneutronic too. Such secondary reactions are typically induced by energetic fusion products including $\alpha$-particles.

A mitigation strategy to achieve nearly fully aneutronic burn is the use of a mixed target such as boron-nitride~\rsec{sssec:BNcycles}, where energetic $\alpha$-particles from $p+{}^{11}\mathrm{B}$ fusion are more likely to react with nitrogen and produce secondary aneutronic reactions while enhancing the overall fusion yield per incoming beam particle and allowing cycling with production of secondary protons. The optimal aneutronic high energy yield fuel mix for (laser) particle beam driven fusion is little known today; there is much to discover!

\paragraph*{\bf Particle catalyzed fusion}
Our discussion of muon-catalyzed fusion in \rsec{ssec:INTmuCF} traced the historical path of $\mu$CF from Frank's 1947 theoretical idea~\cite{Frank:1947} through Zel'dovich's more detailed and insightful consideration~\cite{Zel:1954}, the all accidental Alvarez discovery~\cite{Alvarez:1957un}, and the decisive experimental breakthroughs of Jones and collaborators in the early 1980s~\cite{Jones:1983pw,Jones:1986kk,Jones:1986Na}. The detailed treatment in Sect.\,\ref{ssec:MuonCatalyzed} builds on these results and recognizes the specific physical bottleneck (the loss of muon from the cycle of reactions attributed to alpha sticking) that has prevented $\mu$CF from crossing the engineering break-even threshold originating in the muon production cost.

The historical significance of $\mu$CF in the broader nuclear fusion context cannot be overstated. $\mu$CF provided unambiguous experimental demonstration of true cold fusion: nuclear reactions proceeding at laboratory and even cryogenic temperatures without thermal plasma confinement~\cite{Rafelski:1987Jones}. The 150+ fusions per muon recorded by Jones et al.\ remains a benchmark for non-equilibrium catalytic processes. The interplay between atomic physics (muonic molecule formation), nuclear physics ($dt\mu$ fusion rates), and particle physics (muon production and decay) that accompanies $\mu$CF motivated conceptual development of the laser-driven non-equilibrium approaches described in Sects.~\ref{ssec:exploreLas}--\ref{ssec:Plasmonics}: both exploit the idea that nuclear reactions can be triggered in a controlled, non-thermal environment, bypassing diverse bottlenecks limiting thermal plasma schemes.

The question posed, whether $\bar{n}\sim 1000$ fusions per muon can be achieved, e.g., by density-enhanced resonant reactivation, as outlined in \rsec{sssec:1000muCF}, is not purely academic. Resolution of this question, and progress in the reduction of cost of muon production, will determine whether $\mu$CF remains a scientific curiosity, or finds a viable path toward energy applications. This said, $\mu$CF, rooted in foundational domain of particle physics, attracts the curious investigative minds needed to advance nuclear fusion science.

\subsection{Outlook}
The outlook for future progress in theoretical nuclear fusion is therefore twofold:
\begin{enumerate}
\item 
For thermal fusion, the key theoretical task is to replace single-channel ignition estimates with reaction-network modeling of advanced fuels, evaluating carefully the neutron load per energy produced and handling difficulties of (ultra) fast neutrons. Such models should include evolving fuel composition, secondary burn fractions, charged-particle deposition, neutron escape, radiative feedback, and transport losses. This is especially important for $dd$, $d{}^3\mathrm{He}$, and any deuterium doped plasmas, where secondary heating may move a fuel mixture closer to net power balance but may also introduce neutron-producing pathways.
\item 
For non-thermal fusion, the key task is to develop equally self-consistent models of interaction length and energy deposition in special cases (such as in proposed ${}^3\mathrm{He}+d$ fusion) for direct energy harnessing from produced charged particles. These models must connect microscopic fusion cross-sections to ion acceleration, stopping, target geometry, plasma formation, and/or heated matter fireball expansion. The particle beam driven, often also laser-driven, plasmonic, and nano-structured approaches benefit from bypassing many difficulties of thermal nuclear fusion. The challenge is to accumulate enough nuclear optical depth on ultrashort time scales to make the per particle fusion yield energetically significant and eventually, to harness direct electrical power.
\item
It should be noted that particle beam driven fusion space propulsion is by far the preferred solution given the natural directionality of generated thrust and electrical power harnessing at minimal level to keep system operational in deep space, while using solar power when available. We return to this matter in the closing of this work.
\end{enumerate}

Non-equilibrium approaches alter fusion landscape vastly by relaxing the assumption of Maxwellian ion distributions. In beam-target regimes which do not aim to use produced fast particles as propellant, a relevant metric is the (nuclear) interaction length, see \req{eq:Lint}, with the associated areal mass $m_{\rm areal}$ in \req{eq:m_areal}. As~\rf{fig:InteractionLength} makes clear, even near resonance peaks, aneutronic reactions require substantial target areal densities to reach unit optical depth.

The laser-driven $p+{}^{11}\mathrm{B}$ experiments (see~\rf{fig:PBFusion} and~\rf{fig:YieldpB}) and boron-nitride cycles (see \rf{fig:BNcycles}) demonstrate that non-equilibrium, picosecond-scale fusion is experimentally accessible. In these schemes, fusion is driven by transient ion acceleration and beam--target interactions rather than by a Maxwellian ion distribution, so the relevant control parameters shift from bulk temperature and confinement time to the ion energy spectrum, stopping length, target density, and the timing of resonant energy deposition relative to hydrodynamic expansion and thermalization.

\paragraph*{\bf Space propulsion}
Let us close this report with some futuristic consideration of the small scale (nano) fusion in space propulsion. On Earth, nuclear fusion solution must demonstrate a significant amplification factor in energy, allowing fusion power to compete economically. Moreover, for fusion power to be acceptable, it must be near to 100\% safe with radiation load much reduced compared to nuclear fission reactors. This makes deployment of fusion on Earth complicated, as one must achieve several somewhat contradictory objectives at once. 

In space all that matters is reliability, portability and absence of penetrating (neutron) radiation. These aspects permeate this article. In space there is no need for an amplification factor; as long as the Sun shines, there is electricity to drive lasers or particle beams for fusion. What matters is ability to achieve thrust, as measured in rocket community by specific impulse 
\begin{align}
I_\mathrm{sp}\equiv\frac{v_\mathrm{exhaust}}{g_0} \propto \sqrt{\frac{Q}{m_\mathrm{propellant}}}
\end{align}
where $g_0$ is Earth standard acceleration, $Q$ is the energy gained per reaction and $m_\mathrm{propellant}$ is the exhaust mass. $I_\mathrm{sp}$ defines how efficiently a rocket engine converts its propellant into usable thrust. A higher value of $I_\mathrm{sp}$ means the engine gets a larger push from a smaller amount of fuel. 

For the best chemical rockets $I_\mathrm{sp}\simeq 450$\,s. The NASA Dawn spacecraft had in interplanetary space a specific impulse $I_\mathrm{sp}=3,100$\,s for its xenon ion propulsion system ($v_\mathrm{exhaust}\simeq 30$\,km/s=$10^{-4}$c). The kinetic energy per nucleon of this ion drive was thus $E_\mathrm{kin}\simeq 5\times 10^{-9}\times935\,\mathrm{MeV}=4.65$\,eV, achieved by combining solar power with electrostatic grid at keV potential level for singly ionized xenon. We keep in mind that fusion energies are million times greater and thus $I_\mathrm{sp}$ can be up to 1000 times greater. The major difference to ion drive is the ultra short-pulse operation of the fusion assisted propulsion system. This is allowing to use the same solar power to reach in principle a much greater, up to factor 1000, specific impulse.

The technological requirements for fusion amplified ion drive are challenging but not forbidding. We need light-weight versions of:
\begin{enumerate}
\item 
A high repetition-rate intense laser capable to prepare the reaction target area by stripping electrons, allowing to reduce stopping power for incoming pulsed ion beam for fusion. 
\item
A conventional or laser driven pulsed 200\,keV protons (for proton-Boron drive), or $\sim600$\,keV ${}^3\mathrm{He}^{++}$ (for ${}^3\mathrm{He}\to d$ drive) ion beam. Note that particle speed for ${}^3\mathrm{He}^{++}$ is the same as for protons, while ${}^3\mathrm{He}^{++}$ has 33\% smaller charge to mass ratio $Q/M$. The required ${}^3\mathrm{He}^{++}$ pulsed ion beam could be easier to prepare.
\end{enumerate}

The space propulsion application is arguably the most near-term route to use in a real-world system the small scale fusion introduced in this article. Power requirements of fusion assisted system are comparable to electrostatic ion drive due to pulsed operation. The drive is just `another' ion drive realization capable of considerably fusion-amplifying the achieved ionic rocket thrust. The issue of energy gain, which hinders deployment of fusion on Earth, is of no relevance, as required power is provided, in the initial realization, entirely by the Sun. This said: none of conventional magnetic or inertial solutions for fusion could be today considered for space deployment due to intrinsic weight and/or inability to burn safe fusion fuel. 

\vspace{6pt} 
\authorcontributions{Both authors contributed to creation, computation and development of this manuscript as follows: JR assembled presentation material for Particles and Plasma 2025 meeting which lecture content was developed jointly with AJS, who later contributed several updates regarding isotope abundances, traditional fusion data and plasma burn rates. The wide-scoped presentation of different approaches to nuclear fusion, past, present, and future, was developed by JR. Both authors participated in writing, review, and editing. Both authors have read and agreed to the published version of the manuscript.} 

\funding{Wilhelm and Else Heraeus Foundation provided limited travel support in Summer 2025, partially covering travel cost of one of us (JR) facilitating a presentation at the WE-Heraeus Seminar Particles and Plasma 2025 meeting which motivated this article. Authors received no other external funding.}

\dataavailability{All figures and data sets are available in the ancillary files of \url{https://arxiv.org/abs/2609.01366} and in the GitHub repository \url{https://github.com/ajsteinmetz/fusion-insights}.}

\acknowledgments{JR thanks NAPLIFE project leader Tam{\'a}s Bir{\'o} for the kind invitation to participate online in group meetings with observer status; Vojt{\v e}ch Petr\'a{\v c}ek of Czech Technical University and CFP (Check Fusion Partners) for a careful reading of the manuscript and for valuable comments. AJS thanks Sandy Yousif for help in photographing the optical properties of the Lycurgus cup.} 

\conflictsofinterest{JR is a co-author on {US Patent App.~19/773,972}, {US6909764B2}, and {FR2988897B1} concerning fusion concepts briefly discussed in this article. The authors report no other conflicts.}

\abbreviations{Abbreviations}{
The following abbreviations are used in this manuscript:
\\

\noindent 
\begin{tabular}{@{}ll}
BBN & Big Bang nucleosynthesis \\
BN & Boron nitride \\
CNO & Carbon--nitrogen--oxygen (cycle) \\
HBT & Hanbury Brown--Twiss \\
IC & Inertial confinement \\
ICF & Inertial confinement fusion \\
ITER & International Thermonuclear Experimental Reactor \\
LIBS & Laser-induced breakdown spectroscopy \\
LULI & Laboratoire pour l'Utilisation des Lasers Intenses \\
LMJ & Laser M{\'e}gajoule \\
LSP & Localized surface plasmon \\
LSPP & Localized surface plasmon polariton \\
LLE & Laboratory for Laser Energetics \\
MJ & Megajoule \\
NAPLIFE & NAno-Plasmonic Laser Induced (or Ignited) Fusion Energy\\
NIF & National Ignition Facility \\
$pp$ & Proton--proton (chain) \\
$dt$ & Deuterium--tritium \\
$dd$ & Deuterium--deuterium \\
$p$BN & proton--boron-nitride \\
CF & Catalyzed fusion \\
$\mu$CF & Muon-catalyzed fusion \\
PIC & Particle in cell (computational simulation method) \\
WKB & Wentzel-Kramers-Brillouin
\end{tabular}
}


\reftitle{References}


\bibliography{aneutronic-fusion-v3.bib}
\PublishersNote{}

\end{document}